\documentclass{article}%
\usepackage{amsmath}
\usepackage{amsfonts}
\usepackage{amssymb}
\usepackage{graphicx}%
\newtheorem{theorem}{Theorem}

\newtheorem{definition}[theorem]{Definition}

\newtheorem{proposition}[theorem]{Proposition}
\newtheorem{remark}[theorem]{Remark}

\newenvironment{proof}[1][Proof]{\noindent\textbf{#1.} }{\ \rule{0.5em}{0.5em}}
\begin{document}

\title{An Ambiguous Statement Called \ `Tetrad Postulate' and the Correct Field
Equations Satisfied by the Tetrad Fields\thanks{published: \textit{Int. J.
Mod. Phys. D}. \textbf{14}, 2095-2150 (2005).}}
\author{Waldyr A. Rodrigues Jr. and Quintino A. G. de Souza\\Institute of Mathematics, Statistics and Scientific Computation\\IMECC-UNICAMP, CP 6065\\13083-970 Campinas, SP, Brazil\\walrod@ime.unicamp.br, quintino@ime.unicamp.br}
\date{November 29 2004\\
with misprints and typos corrections: January 2007}
\maketitle
\tableofcontents

\begin{abstract}
The names \textit{tetrad}, \textit{tetrads}, \textit{cotetrads}, have been
used with many different meanings in the physical literature, not all of them,
equivalent from the mathematical point of view. In this paper we introduce
unambiguous definitions for each one of those terms, and show how the old
miscellanea made many authors to introduce in their formalism an
\textit{ambiguous} statement called `tetrad postulate', which has been source
of many misunderstandings, as we show explicitly examining examples found in
the literature. Since formulating Einstein's field equations intrinsically in
terms of cotetrad fields \ $\theta^{\mathbf{a}}$, $\mathbf{a}=\mathbf{0,1,2,3}%
$ is an worth enterprise, we derive the equation of motion of each
$\theta^{\mathbf{a}}$ using modern mathematical tools (the Clifford bundle
formalism and the theory of the square of the Dirac operator). Indeed, we
identify (giving all details and theorems) from the square of the Dirac
operator some noticeable mathematical objects, namely, the Ricci, Einstein,
covariant D'Alembertian and the Hodge Laplacian operators, which permit to
show that each $\theta^{\mathbf{a}}$ satisfies a well defined wave equation.
Also, we present for completeness a detailed derivation of the cotetrad wave
equations from a variational principal. We compare the cotetrad wave equation
satisfied by each $\theta^{\mathbf{a}}$ with some others appearing in the
literature, and which are unfortunately in error.

\end{abstract}

\section{Introduction}

In what follows we identify an \textit{ambiguous} statement called `tetrad
postulate' (a better name, as we shall see would be `naive tetrad postulate')
that appears often in the Physics literature (see e.g.,
\cite{carroll,0,gsw,rovelli,11,12}, to quote only a few examples here). We
identify the genesis of the wording \ `tetrad postulate' as a result of a
\textit{deficient }identification of some mathematical objects of differential
geometry. Note that we used the word ambiguous, not the word wrong. This is
because, as we shall show, the equation dubbed 'tetrad postulate' can be
rigorously interpreted as meaning that the \textit{components }of a covariant
derivative in the direction of a vector field ${\mbox{\boldmath$\partial$}}%
_{\mu}$ of a \ certain tensor field $\mathbf{Q}$ (Eq.(\ref{Q})) are null (see
Eq.(\ref{TETRAD POSTULATE})). This equation is \textit{not }a postulate.
Indeed, it is nothing more than the intrinsic expression of an
\textit{obvious} identity of differential geometry that we dubbed the freshman
identity (Eq.(\ref{18})). However, if the freshman identity is used naively as
if meaning a `tetrad postulate' misunderstandings may arise, and in what
follows we present some of them, by examining some examples that we found in
the literature. We comment also on a result called \ `Evans Lemma' of
differential geometry and claimed in \cite{0} to be as important as the
Poincar\'{e} lemma. We show that \ `Evans Lemma' as presented in \cite{0} is a
false statement, the proof offered by that author being invalid because in
trying to use the naive tetrad postulate he did incorrect use of some
fundamental concepts of differential geometry, as, e.g.,\footnote{In order to
not confuse the numeration of equations in \textbf{\cite{0} }with the
numeration of the equations in the present report we denote in what follows an
equation numered Eq.(x) in \textbf{\cite{0}} by Eq.(xE).} \ his (wrong) Eq.(41E).

We explain all that in details in what follows. We observe also that in
\cite{0,1,2,3,4,5} it is claimed that `Evans Lemma' is the basic pillar of a
(supposed) generally covariant unified field theory developed there. So, once
we prove that `Evans Lemma' is a \textit{wrong }premise, all the theory
developed in \cite{0,1,2,3,4,5} is automatically disproved.

Using modern mathematical tools (namely the theory of \textit{Clifford
bundles} and the theory of the square of the Dirac operator\footnote{The Dirac
operator used in this paper acts on sections of a Clifford bundle. So, it is
not to be confused with the (spin) Dirac operator that acts on section of a
spin-Clifford bundle. Details can be found in \cite{10}. In particular the
square of the Dirac operator is different form the square of the spin-Dirac
operator, first calculated by Lichnerowicz \cite{lichi}. The difference of
these squares will be presented in another publication.}), we present
\textit{two} derivations (which includes all the necessary mathematical
theorems)\footnote{These equations already appeared in \cite{14,17}, but the
necessary theorems (proved in this report) needed to prove them have not been
given there.} of the correct differential equations satisfied by the
\textit{cotetrad} fields $\theta^{\mathbf{a}}=q_{\mu}^{\mathbf{a}}dx^{\mu}$ on
a Lorentzian manifold, modelling a gravitational field in General Relativity.
The first derivation find the tetrad field equations directly from Einstein's
field equations, once we identify, playing with the square of the Dirac
operator acting on sections of the Clifford bundle, the existence of some
remarkable mathematical objects, namely, the \textit{Ricci}, \textit{Einstein}%
, \textit{covariant D'Alembertian}, and \textit{Hodge Laplacian} operators
\cite{17}. The second derivation (presented here for completeness) is
achieved\ using a variational principle, after expressing the Einstein-Hilbert
Lagragian in terms of the tetrad fields\footnote{For the best of our
knwoledge, the Einstein-Hilbert Lagrangian write explicitly in terms of the
tetrad fields appears in \cite{wallner}. See also \cite{rumpf} and related
material in \cite{mughe,thirring}.}. Our objective in presenting those
derivations was the one of comparing the correct equations with the ones
presented, e.g., in \cite{1,2,3,4,5,kaniel} and which appears as Eq.(49E) in
\textbf{\cite{0}}.

The functions $q_{\mu}^{\mathbf{a}}$ \ appearing as components of the cotetrad
fields $\theta^{\mathbf{a}}$ in a coordinate basis can be used to define a
tensor $\mathbf{Q}=q_{\mu}^{\mathbf{a}}\mathbf{e}_{\mathbf{a}}\otimes dx^{\mu
}$ (see Eq.(\ref{sachs 1})). $\mathbf{Q}$ satisfies \textit{trivially} in any
general Riemann-Cartan spacetime a second order differential equation which in
intrinsic form is \ $g^{\nu\mu}\mathbf{\nabla}_{{\mbox{\boldmath$\partial$}}%
_{\nu}}\mathbf{\nabla}_{{\mbox{\boldmath$\partial$}}_{\mu}}\mathbf{Q}=0$. From
that equation (numbered Eq.(\ref{q rc1}) below.) we can, of course, write \ a
wave equation for the each one of the functions $q_{\mu}^{\mathbf{a}}$ in
\textit{any} Riemann-Cartan spacetime. It is this equation that author of
\cite{0} attempted to obtain and that he called \textit{'}Evans lemma'.
\ However, as we already said, his \textit{final} result is not correct. In
what follows, we \ use the intrinsic Eq.(\ref{q rc1}) to write wave equations
for the functions $q_{\mu}^{\mathbf{a}}$ only in the particular case of a
general Lorentzian spacetime\footnote{If necessary these equations can be also
written for a Riemman-Cartan spacetime.}. This restriction is done here for
the following reason. Wave equations for the functions $q_{\mu}^{\mathbf{a}}$
can also be derived from the \textit{correct} equations satisfied by the
$\theta^{\mathbf{a}}$\ in General Relativity (see Eq.(\ref{11.1}) below).
Then, by comparing both equations we obtain a constrain equation, involving
these functions, the components of the Ricci tensor and the components of the
energy-momentum tensor and its trace (Eq.(\ref{l4})). The paper has three
appendices. In Appendix A we give a very simple example of the many
misunderstandings that the use of the naive `tetrad postulate' may produce. We
hope that this example may be understood even by readers with only a small
knowledge of differential geometry. In appendix B we give the details of the
calculations needed for deriving the equations for the tetrad fields in
General Relativity from a variational principle. Those calculations require a
working knowledge of the Clifford bundle formalism. Finally, in Appendix C we
give a brief answer to a comment of the author of \cite{0} who just posted a
refutation to a preliminary version of this paper. We show that all his claims
in his refutation are without valid foundation.

\section{Recall of Some Basic Results}

In what follows $M$ is a real differential manifold \cite{8} with $\dim M=4$
which will be made part of the definition of a spacetime (whose points are
\textit{events}) of General Relativity, or of a general Riemann-Cartan type
theory. As usual we denote the tangent and cotangent spaces at $e\in M$ by
$T_{e}M$ and $T_{e}^{\ast}M$. Elements of $T_{e}M$ are called \textit{vectors}
and elements of $T_{e}^{\ast}M$ are called \textit{covectors}. The structures
$TM=\cup_{e}T_{e}M$ and $T^{\ast}M=\cup_{e\in M}T_{e}^{\ast}M$ are
\textit{vector bundles} called respectively the \textit{tangent} and
\textit{cotangent} bundles. Sections of $TM=\cup_{e\in M}T_{e}M$ are called
\textit{vector fields} and sections of $T^{\ast}M=\cup_{e\in M}T_{e}^{\ast}M$
are called \textit{covector fields} (or $1$-form fields). We denote moreover
by $T^{r,s}M$ the bundle of $r$-covariant and $s$-contravariant tensor fields
and by $\tau M=\bigoplus\nolimits_{r,s=0}^{\infty}T^{r,s}M$, the tensor bundle
of $M$. Also, $\bigwedge TM=\bigoplus_{i=0}^{4}\bigwedge^{i}TM$ and $\bigwedge
T^{\ast}M=\bigoplus_{i=0}^{4}\bigwedge^{i}T^{\ast}M$, denote respectively the
bundles of (nonhomogeneous) \textit{multivector fields} and \textit{multiform
fields}.

\begin{remark}
\label{elementar}It is important to keep in mind, in order to appreciate some
of the comments presented in the next section, that $T_{e}M$ and $T_{e}^{\ast
}M$ are $4$-dimensional vector spaces over the real field $\mathbb{R}$, i.e.,
\textrm{dim} $T_{e}M$ = \textrm{dim} $T_{e}^{\ast}M=4$. Also note the
identifications $\bigwedge^{0}T_{e}M$ $=\bigwedge^{0}T_{e}^{\ast}M$ $=$
$\mathbb{R}$, $\bigwedge^{1}T_{e}M$ $=T_{e}M$ and $\bigwedge^{1}T_{e}^{\ast}M$
$=T_{e}^{\ast}M$. Keep also in mind that $\dim\bigwedge^{i}T_{e}%
M=\dim\bigwedge^{i}T_{e}^{\ast}M=\binom{4}{i}$. More details on these
structures will be given in Section 6, where they are to be used.
\end{remark}

To proceed we suppose that $M$ is a connected, paracompact and noncompact
manifold. We give the following standard definitions.

\subsection{Spacetimes}

\begin{definition}
\label{lorentz manif}A Lorentzian manifold is a pair $(M,\mathbf{g})$, where
$\mathbf{g}\in\sec T^{2,0}M$ is a Lorentzian metric of signature $(1,3)$,
i.e., for all $e\in M$, $T_{e}M\simeq T_{e}^{\ast}M\simeq\mathbb{R}^{4}$. For
each $e\in M$ the pair $(\mathbb{R}^{4},\mathbf{g}_{e})\equiv$ $\mathbb{R}%
^{1,3}=$ $(\mathbb{R}^{4},\mathbf{\eta})$ is a Minkowski vector
space\footnote{$\mathbf{\eta}$ is a metric of Lorentzian signature $-2$ in
$\mathbb{R}^{4}$.} \cite{8}.
\end{definition}

\begin{remark}
\label{elementar 1}We shall always suppose that the tangent space at $e\in M$
is equipped with the metric $\mathbf{g}_{e}$ and so, we eventually write by
abuse of notation $T_{e}M\simeq T_{e}^{\ast}M\simeq\mathbb{R}^{1,3}$. Take
into account also, that in general the tangent spaces at different points of
the manifold $M$ cannot be identified, unless the manifold possess some
additional \ appropriate structure \cite{choquet}.
\end{remark}

\begin{definition}
\label{spacetime}A spacetime $\mathfrak{M}$ is a pentuple $(M,\mathbf{g}%
,$\textbf{$\nabla$}$\mathbf{,\tau}_{\mathbf{g}},\mathbf{\uparrow})$ where
$(M,\mathbf{g},\mathbf{\tau}_{\mathbf{g}},\mathbf{\uparrow})$ is an oriented
Lorentzian manifold (oriented by $\mathbf{\tau}_{\mathbf{g}}$) \ and time
oriented by an appropriate equivalence relation\footnote{See \cite{8} for
details.} (denoted $\uparrow$) for the timelike vectors at the tangent space
$T_{e}M$, $\forall e\in M$. \textbf{$\nabla$\ }is a linear
connection\footnote{More precisely, \textbf{$\nabla$} is a covariant
derivative operator associated to a connection $\mathbf{\omega}$, which is a
section of a principal bundle called the frame bundle of $M$. \textbf{$\nabla
$} acts on sections of the tensor bundle \cite{choquet}. We will need to
specify with more details the precise nature of \textbf{$\nabla$} in order to
present in an inteligible way the ambiguities associated with the \ \ `tetrad
postulate'. This will be done in Section 4.} for $M$ such that \textbf{$\nabla
$}$\mathbf{g}=0$.
\end{definition}

\begin{remark}
In General Relativity, Lorentzian spacetimes are models of gravitational
fields \cite{8}.
\end{remark}

\begin{definition}
Let $\mathbf{T}$ and $\mathbf{R}$ be respectively the torsion and curvature
tensors of \textbf{$\nabla$}. If in addition to the requirements of the
previous definitions, $\mathbf{T}($\textbf{$\nabla$}$)=0$, then $\mathfrak{M}$
is said to be a Lorentzian spacetime. The particular Lorentzian spacetime
where $M\simeq\mathbb{R}^{4}$ and such that $\mathbf{R}($\textbf{$\nabla$%
}$)=0$ is called Minkowski spacetime\footnote{It is important to not confound
Minkowski spacetime with $\mathbb{R}^{1,3}$, the Minkowski vector space.} and
will be denoted by $\mathcal{M}$. When $\mathbf{T}($\textbf{$\nabla$}$)$ is
possibly nonzero, $\mathfrak{M}$ is said to be a Riemann-Cartan spacetime
(RCST). A particular RCST such that $\mathbf{R}($\textbf{$\nabla$}$)=0$ is
called a teleparallel spacetime.
\end{definition}

We will also denote by $F(M)$ the frame bundle of $M$ and by $P_{\mathrm{SO}%
_{1,3}^{e}}(M)$ the principal bundle of oriented \textit{Lorentz tetrads}.
Those bundles will be used in Section 4 to give some additional details
concerning the nature of the tangent, cotangent and tensor bundles, as
associated vector bundles to $F(M)$ or $P_{\mathrm{SO}_{1,3}^{e}}(M)$, which
are necessary to clarify misunderstandings related to the naive \ `tetrad postulate'.

\subsection{On the Nature of Tangent and Cotangent Fields I}

Let $U\subset M$ be an open set and let $(U,\varphi)$ be a coordinate chart of
the maximal atlas of $M$. We recall that $\varphi$ is a differentiable mapping
from $U$ to an open set of $\mathbb{R}^{4}$. The coordinate functions of the
chart are denoted by $x^{\mu}:U\rightarrow\mathbb{R}$, $\mu=0,1,2,3$.

Consider the subbundles $TU\subset TM$ and $T^{\ast}U\subset T^{\ast}M$. There
are two types of vector fields (respectively covector fields) in $TU$
(respectively $T^{\ast}U$) which are such that at each point (event) $e\in U$
define interesting bases for $T_{e}U$ (respectively $T_{e}^{\ast}U$).

\begin{definition}
\textbf{coordinate basis }for\textbf{ }$TU$\textbf{. \ }A set\footnote{Also we
say that $\{e_{\mu}\}\in\sec F(U)\subset\sec F(M)$, i.e., is a section of the
frame bundle.} $\{e_{\mu}\}$, $e_{\mu}\in\sec TU$, $\mu=0,1,2,3$ is called a
coordinate basis for $TU$ if there exists a coordinate chart $(U,\varphi)$ and
coordinate functions $x^{\mu}:U\rightarrow\mathbb{R}$, $\mu=0,1,2,3$, such
that for each (differentiable) function $f:M\rightarrow\mathbb{R}$ we have
$(\varphi(e)\equiv x)$%
\begin{equation}
e_{\mu}(f)|_{e}=\left.  \frac{\partial}{\partial x^{\mu}}(f\circ\varphi
^{-1})\right\vert _{x} \label{1}%
\end{equation}

\end{definition}

\begin{remark}
Due to this equation mathematicians often write \ $e_{\mu}%
={\mbox{\boldmath$\partial$}}_{\mu}$ and sometimes even $e_{\mu}%
=\frac{\partial}{\partial x^{\mu}}=\partial_{\mu}$. Also by abuse of notation
it is usual to see (in physics texts) $f\circ\varphi^{-1}$ written simply as
$f$ or $f(x)$, and here we eventually use such sloppy notation, when no
confusion arises.
\end{remark}

\begin{definition}
\textbf{coordinate basis }for\textbf{ }$T^{\ast}U$\textbf{. \ }A set
$\{\theta^{\mu}\}$, $\theta^{\mu}\in\sec T^{\ast}U$, $\mu=0,1,2,3$ is called a
coordinate basis for $T^{\ast}U$ if there exists a coordinate chart
$(U,\varphi)$ and coordinate functions $x^{\mu}:U\rightarrow\mathbb{R}$,
$\mu=0,1,2,3$, such that $\theta^{\mu}=dx^{\mu}$.
\end{definition}

Recall that the basis $\{\theta^{\mu}\}$ is the dual basis of
$\{{\mbox{\boldmath$\partial$}}_{\mu}\}$ and we have $\theta^{\mu
}({\mbox{\boldmath$\partial$}}_{\nu})=\delta_{\nu}^{\mu}$.

Now, in general the coordinate basis $\{{\mbox{\boldmath$\partial$}}_{\mu}\}$
is not orthonormal, this means that if the \textit{pullback} of $\mathbf{g}$
in $T^{2,0}\varphi(U)$ is written as usual (with abuse of notation) as
$\mathbf{g}=g_{\mu\nu}(x)dx^{\mu}\otimes dx^{\nu}$ then,%
\begin{equation}
\mathbf{g}({\mbox{\boldmath$\partial$}}_{\mu},{\mbox{\boldmath$\partial$}}%
_{\nu})|_{x}=\mathbf{g}({\mbox{\boldmath$\partial$}}_{\nu}%
,{\mbox{\boldmath$\partial$}}_{\mu})|_{x}=g_{\mu\nu}(x) \label{2'}%
\end{equation}
and in general the real functions $g_{\mu\nu}:\varphi(U)\rightarrow\mathbb{R}$
are not constant functions.

Also, if $g\in\sec T^{0,2}M$ is the metric of the cotangent bundle, we have
(writing for the \textit{pullback} of $g$ in $T^{0,2}\varphi(U)$, $g=g^{\mu
\nu}(x){\mbox{\boldmath$\partial$}}_{\mu}\otimes{\mbox{\boldmath$\partial$}}%
_{\nu}$)%
\begin{equation}
g(dx^{\mu},dx^{\nu})|_{x}=g^{\mu\nu}(x), \label{3}%
\end{equation}
and the real functions $g^{\mu\nu}:\varphi(U)\rightarrow\mathbb{R}$ satisfy
\begin{equation}
g^{\mu\nu}(x)g_{\mu\alpha}(x)=\delta_{\alpha}^{\nu},\text{ }\forall
x\in\varphi(U). \label{4}%
\end{equation}

\subsection{Tetrads and Cotetrads}

\begin{definition}
\textbf{orthonormal basis} for $TU$. A set $\{\mathbf{e}_{\mathbf{a}%
}\},\mathbf{e}_{\mathbf{a}}\in\sec TU$, with $\mathbf{a=}0,1,2,3$ \ is said to
be an orthonormal basis for $TU$ if and only if for any $x\in\varphi(U),$
\begin{equation}
\mathbf{g}(\mathbf{e}_{\mathbf{a}},\mathbf{e}_{\mathbf{b}})|_{x}%
=\eta_{\mathbf{ab}} \label{5}%
\end{equation}
where the $4\times4$ matrix with entries $\eta_{\mathbf{ab}}$ is the diagonal
matrix $\mathrm{diag}(1,-1,-1,-1)$. When no confusion arises we shall use the
sloppy (but very much used) notation $\eta_{\mathbf{ab}}=\mathrm{diag}%
(1,-1,-1,-1)$.
\end{definition}

\begin{definition}
\textbf{orthonormal basis} for $T^{\ast}U$. A set $\{\theta^{\mathbf{a}%
}\},\theta^{\mathbf{a}}\in\sec T^{\ast}U$, with $\mathbf{a=}0,1,2,3$ is said
to be an orthonormal basis for $T^{\ast}U$ if and only if for any $x\in
\varphi(U),$
\begin{equation}
\mathbf{g}(\theta^{\mathbf{a}},\theta^{\mathbf{b}})|_{x}=\eta^{\mathbf{ab}%
}=\mathrm{diag}(1,-1,-1,-1). \label{6}%
\end{equation}

\end{definition}

Recall that the basis $\{\theta^{\mathbf{a}}\}$ is the dual basis of the
basis$\ \{\mathbf{e}_{\mathbf{a}}\}$, i.e., $\theta^{\mathbf{a}}%
(\mathbf{e}_{\mathbf{b}})=\delta_{\mathbf{b}}^{\mathbf{a}}$

\begin{definition}
\label{tetrad}The set $\{\mathbf{e}_{\mathbf{a}}\}$ considered as a section of
the orthonormal frame bundle $P_{\mathrm{SO}_{1,3}^{e}}(U)\subset
P_{\mathrm{SO}_{1,3}^{e}}(M)$ is called a tetrad basis for $TU$. The set
$\{\theta^{\mathbf{a}}\}$ is called a cotetrad basis for $T^{\ast}U$.
\end{definition}

\begin{remark}
We recall that a\ global (i.e., defined for all $e\in M)$ tetrad
$($cotetrad$)$ basis for $TM$ $(T^{\ast}M)$ exists if and only if $M$ in
Definition \ref{spacetime} is a spin manifold $($see, e.g.,\cite{9,10}$)$.
This result is the famous Geroch theorem \cite{geroch}.
\end{remark}

\begin{remark}
Besides that bases, it is also convenient to define reciprocal bases. So, the
reciprocal basis of $\{{\mbox{\boldmath$\partial$}}_{\mu}\}\in\sec F(U)$ is
the basis of $\{{\mbox{\boldmath$\partial$}}^{\mu}\}\in\sec F(U)$ such that
$\mathbf{g}({\mbox{\boldmath$\partial$}}_{\mu},{\mbox{\boldmath$\partial$}}%
^{\nu})=\delta_{v}^{\mu}$. Also, the reciprocal basis of the basis
$\{\theta^{\mu}=dx^{\mu}\}$ of $T^{\ast}U$, $\theta^{\mu}\in\sec T^{\ast}U$,
$\mu=0,1,2,3$ is the basis $\{\theta_{\mu}\}$ of $T^{\ast}U$, $\theta_{\mu}%
\in\sec T^{\ast}U$, $\mu=0,1,2,3$ such that $g(\theta_{\mu},\theta^{\nu
})=\delta_{v}^{\mu}$. Also $\{\mathbf{e}^{\mathbf{a}}\},\mathbf{e}%
^{\mathbf{a}}\in\sec TU$, $\mathbf{a}=0,1,2,3$ with $\mathbf{g(e}^{\mathbf{a}%
},\mathbf{e}_{\mathbf{b}}\mathbf{)=\delta}_{\mathbf{b}}^{\mathbf{a}}$ is
called the reciprocal basis of the basis $\{\mathbf{e}_{\mathbf{a}}\}$.
Finally, $\{\theta_{\mathbf{a}}\},\theta_{\mathbf{a}}\in\sec T^{\ast}U$,
$\mathbf{a}=0,1,2,3$ with $g\mathbf{(}\theta_{\mathbf{a}},\theta^{\mathbf{b}%
}\mathbf{)=\delta}_{\mathbf{a}}^{\mathbf{b}}$ is called the reciprocal basis
of $\{\theta^{\mathbf{a}}\}$.
\end{remark}

Now, consider a vector field $V\in\sec TU$ and a covector field $C\in\sec
T^{\ast}U$. We can express $V$ and $C$ in the coordinate basis
$\{{\mbox{\boldmath$\partial$}}_{\mu}\},\{{\mbox{\boldmath$\partial$}}^{\mu
}\}$ and $\{\theta^{\mu}=dx^{\mu}\},\{\theta_{\mu}\}$ by
\begin{equation}
V=V^{\mu}{\mbox{\boldmath$\partial$}}_{\mu}=V_{\mu}%
{\mbox{\boldmath$\partial$}}^{\mu}\text{, }\qquad C=C_{\mu}dx^{\mu}=C^{\mu
}\theta_{\mu} \label{7}%
\end{equation}
and in the tetrad basis $\{\mathbf{e}_{\mathbf{a}}\},\{\mathbf{e}^{\mathbf{a}%
}\}$ and $\{\theta^{\mathbf{a}}\}$, $\{\theta_{\mathbf{a}}\}$by
\begin{equation}
V=V^{\mathbf{a}}\mathbf{e}_{\mathbf{a}}=V_{\mathbf{a}}\theta^{\mathbf{a}%
}\text{, }\qquad C=C_{\mathbf{a}}\theta^{\mathbf{a}}=C^{\mathbf{a}}%
\theta_{\mathbf{a}}. \label{8}%
\end{equation}

\section{Some Misconceptions and Misunderstandings Involving Tetrads}

In this section we analyze some statements found in section 1 \textbf{\cite{0}
}which is said to be dedicated to give many distinct definitions of
\ `tetrads'. Unfortunately that section is full of misconceptions and
misunderstandings, which are the origin of many errors in papers signed by
author of \cite{0}. In order to appreciate that statement, let us recall some facts.

First, recall that each one of the tetrad fields (as defined in the previous
Section, Definition \ref{tetrad} ), $\mathbf{e}_{\mathbf{a}}\in\sec TU$,
$\mathbf{a}=0,1,2,3$, as any vector field, can be expanded using Eq.(\ref{7})
in the coordinate basis $\{{\mbox{\boldmath$\partial$}}_{\mu}\}$, as%
\begin{equation}
\mathbf{e}_{\mathbf{a}}=q_{\mathbf{a}}^{\mu}{\mbox{\boldmath$\partial$}}_{\mu
}. \label{9}%
\end{equation}

Also, each one of the cotetrad fields $\{\theta^{\mathbf{a}}\},\theta
^{\mathbf{a}}\in\sec TU$, $\mathbf{a=}0,1,2,3$, as any covector field, can be
written as%

\begin{equation}
\theta^{\mathbf{a}}=q_{\mu}^{\mathbf{a}}dx^{\mu}. \label{10}%
\end{equation}

\begin{remark}
The functions $q_{\mathbf{a}}^{\mu},q_{\mu}^{\mathbf{a}}:\varphi
(U)\rightarrow\mathbb{R}$ are \textbf{real} functions and satisfy%
\begin{equation}
q_{\mathbf{a}}^{\mu}q_{\mu}^{\mathbf{b}}=\mathbf{\delta}_{\mathbf{a}%
}^{\mathbf{b}}\text{, }\qquad q_{\mathbf{a}}^{\mu}q_{\nu}^{\mathbf{a}}%
=\delta_{\nu}^{\mu}\text{ .} \label{11}%
\end{equation}

\end{remark}

It is trivial to verify the formulas
\begin{align}
g_{\mu\nu}  &  =q_{\mu}^{\mathbf{a}}q_{\nu}^{\mathbf{b}}\eta_{\mathbf{ab}%
}\text{, }\qquad g^{\mu\nu}=q_{\mathbf{a}}^{\mu}q_{\mathbf{b}}^{\nu}%
\eta^{\mathbf{ab}},\nonumber\\
\eta_{\mathbf{ab}}  &  =q_{\mathbf{a}}^{\mu}q_{\mathbf{b}}^{\nu}g_{\mu\nu
},\qquad\eta^{\mathbf{ab}}=q_{\mu}^{\mathbf{a}}q_{\nu}^{\mathbf{b}}g^{\mu\nu}.
\label{12}%
\end{align}

Now to some comments.

(c1) In Eq.(9E) and Eq.(10E) it is written\footnote{In \textbf{\cite{0}}
\ instead of the symbol \ $\curlywedge$ the symbol $\wedge$ has been used for
the exterior product. This distinction is necessary here because the
convention for the exterior product that we used in the second part of the
paper is different from the one used in \textbf{\cite{0}.}}
\begin{align}
&
\begin{tabular}
[c]{|c|}\hline
$q_{\mu\nu}^{c(A)}=q_{\mu}^{\mathbf{a}}\curlywedge q_{\nu}^{\mathbf{b}}%
,$\\\hline
\end{tabular}
\tag{9E}\\
&
\begin{tabular}
[c]{|c|}\hline
$q_{\mu\nu}^{\mathbf{ab}}=q_{\mu}^{\mathbf{a}}q_{\nu}^{\mathbf{b}}=q_{\mu
}^{\mathbf{a}}\otimes q_{\nu}^{\mathbf{b}}.$\\\hline
\end{tabular}
\ \ \ \ \tag{10E}%
\end{align}

Of course, these \textit{unusual} notations used to multiply \textit{scalar
functions} in the above equations, if they are to have any meaning at all,
must be understood as a notation suggested from the result of correct
mathematical operations. The problem is that in \textbf{\cite{0}} they are not
well specified and we have some ambiguity. Indeed, we have the possibilities:%
\begin{align}
\theta^{\mathbf{a}}\otimes\theta^{\mathbf{b}}  &  =q_{\mu}^{\mathbf{a}}q_{\nu
}^{\mathbf{b}}\eta_{\mathbf{ab}}dx^{\mu}\otimes dx^{\nu}\label{definition}\\
=  &  \theta^{\mathbf{a}}\wedge\theta^{\mathbf{b}}+\theta^{\mathbf{a}}%
\overset{s}{\otimes}\theta^{\mathbf{b}},
\end{align}
where the algebraists definitions \cite{bourbaki,crumeyrolle} of
$\theta^{\mathbf{a}}\wedge\theta^{\mathbf{b}}$ and $\theta^{\mathbf{a}%
}\overset{s}{\otimes}\theta^{\mathbf{b}}$ are:%

\begin{align}
\theta^{\mathbf{a}}\wedge\theta^{\mathbf{b}}  &  =\frac{1}{2}\left(
\theta^{\mathbf{a}}\otimes\theta^{\mathbf{b}}-\theta^{\mathbf{b}}\otimes
\theta^{\mathbf{a}}\right) \nonumber\\
&  =\frac{1}{2}\left(  q_{\mu}^{\mathbf{a}}q_{\nu}^{\mathbf{b}}-q_{\mu
}^{\mathbf{b}}q_{\nu}^{\mathbf{a}}\right)  dx^{\mu}\otimes dx^{\nu
}\label{algebra definition}\\
&  =q_{\mu}^{\mathbf{a}}dx^{\mu}\wedge q_{\nu}^{\mathbf{b}}dx^{\nu}=q_{\mu
}^{\mathbf{a}}q_{\nu}^{\mathbf{b}}dx^{\mu}\wedge dx^{\nu}\label{13 0}\\
&  =\frac{1}{2}\left(  q_{\mu}^{\mathbf{a}}q_{\nu}^{\mathbf{b}}-q_{\nu
}^{\mathbf{a}}q_{\mu}^{\mathbf{b}}\right)  dx^{\mu}\wedge dx^{\nu}
\label{13 0 bis}%
\end{align}

\begin{align}
\theta^{\mathbf{a}}\overset{s}{\otimes}\theta^{\mathbf{b}}  &  =\frac{1}%
{2}\left(  \theta^{\mathbf{a}}\otimes\theta^{\mathbf{b}}+\theta^{\mathbf{b}%
}\otimes\theta^{\mathbf{a}}\right) \nonumber\\
&  =\frac{1}{2}\left(  q_{\mu}^{\mathbf{a}}q_{\nu}^{\mathbf{b}}+q_{\mu
}^{\mathbf{b}}q_{\nu}^{\mathbf{a}}\right)  dx^{\mu}\otimes dx^{\nu
}\label{13 0bis}\\
&  =q_{\mu}^{\mathbf{a}}dx^{\mu}\overset{s}{\otimes}q_{\nu}^{\mathbf{b}%
}dx^{\nu}=q_{\mu}^{\mathbf{a}}q_{\nu}^{\mathbf{b}}dx^{\mu}\overset{s}{\otimes
}dx^{\nu}\\
&  =\frac{1}{2}\left(  q_{\mu}^{\mathbf{a}}q_{\nu}^{\mathbf{b}}+q_{\nu
}^{\mathbf{b}}q_{\mu}^{\mathbf{a}}\right)  dx^{\mu}\overset{s}{\otimes}%
dx^{\nu}. \label{13 0 bisss}%
\end{align}

So, we have the following possibilities for identification of symbols:

(a) Use Eq.(\ref{13 0 bis}) and Eq.(\ref{13 0 bisss}). This results in
\begin{align}
q_{\mu}^{\mathbf{a}}\wedge q_{\nu}^{\mathbf{b}}  &  =\frac{1}{2}\left(
q_{\mu}^{\mathbf{a}}q_{\nu}^{\mathbf{b}}-q_{\nu}^{\mathbf{a}}q_{\mu
}^{\mathbf{b}}\right)  ,\label{13}\\
\bar{q}_{\mu\nu}^{\mathbf{ab}}  &  =q_{\mu}^{\mathbf{a}}\overset{}{\otimes
}q_{\nu}^{\mathbf{b}}=\frac{1}{2}\left(  q_{\mu}^{\mathbf{a}}q_{\nu
}^{\mathbf{b}}+q_{\nu}^{\mathbf{b}}q_{\mu}^{\mathbf{a}}\right)  ,
\label{13bis}\\
q_{\mu}^{\mathbf{a}}\otimes q_{\nu}^{\mathbf{b}}  &  =q_{\mu}^{\mathbf{a}%
}\overset{}{\otimes}q_{\nu}^{\mathbf{b}}+q_{\mu}^{\mathbf{a}}\overset{}%
{\wedge}q_{\nu}^{\mathbf{b}}.
\end{align}

(b) Use now Eq.(\ref{algebra definition}) \ and Eq.(\ref{13 0bis}). This
results in the alternative possibility%

\begin{align}
q_{\mu}^{\mathbf{a}}\overset{a}{\wedge}q_{\nu}^{\mathbf{b}}  &  =\frac{1}%
{2}\left(  q_{\mu}^{\mathbf{a}}q_{\nu}^{\mathbf{b}}-q_{\nu}^{\mathbf{a}}%
q_{\mu}^{\mathbf{b}}\right)  ,\label{last 13}\\
\mathbf{q}_{\mu\nu}^{\mathbf{ab}}  &  =q_{\mu}^{\mathbf{a}}\overset
{as}{\otimes}q_{\nu}^{\mathbf{b}}=\frac{1}{2}\left(  q_{\mu}^{\mathbf{a}%
}q_{\nu}^{\mathbf{b}}+q_{\mu}^{\mathbf{b}}q_{\nu}^{\mathbf{a}}\right)
,\label{13 BIS}\\
q_{\mu}^{\mathbf{a}}\otimes q_{\nu}^{\mathbf{b}}  &  =q_{\mu}^{\mathbf{a}%
}\overset{as}{\otimes}q_{\nu}^{\mathbf{b}}+q_{\mu}^{\mathbf{a}}\overset
{a}{\wedge}q_{\nu}^{\mathbf{b}}.
\end{align}

To decide what the author of \textbf{\cite{0}} had in mind, we need to look at
line 19 in Table 1 of \textbf{\cite{0}}.\textbf{ }There, we learn that the
definition of the exterior product ($\curlywedge$) used there is\footnote{The
definition given in Eq.(\ref{algebra definition}) is used mainly (for very
good reasons, that we refrain to discuss here) by algebraists
\cite{bourbaki,crumeyrolle}, However, many physcisits working in General
Relativity use it, as, e.g, \cite{benntu}.\ \ The definition given by
Eq.(\ref{alternative definition}) is eventually more popular among authors
working on differential geometry, (see, e.g. \cite{aubin}) and some authors
working in General Relativiy. \ In particular, this definition is also the one
used in \cite{carroll} (see his Eq.(1.79)), and also the one used, e.g., in
\cite{frankel,9}. Both may be used, each one has its merits, but it is a good
idea for a reader to first knows what the author means. We have discussed this
issue in details in \cite{fmr}.}: given $A,B\in\sec T^{\ast}M,$%
\begin{equation}
A\curlywedge B=A\otimes B-B\otimes A, \label{alternative definition}%
\end{equation}
since line 19 of Table 1 in ME reads
\begin{equation}%
\begin{tabular}
[c]{|c|}\hline
$\left(  A\curlywedge B\right)  _{\mu\nu}=A_{\mu}B_{\nu}-A_{\nu}B_{\mu}%
$\\\hline
\end{tabular}
\ \ \ \ \ . \label{genuine}%
\end{equation}
\ \ 

But, the author of\textbf{ \cite{0} }forgot to inform his readers that from
the genuine notation given by Eq.(\ref{genuine}) he starting using that
$\left(  A\curlywedge B\right)  _{\mu\nu}:=A_{\mu}\curlywedge B_{\nu}$.
Without that explanation the symbols $A_{\mu}\curlywedge B_{\nu}$ look as a
product of scalars, and as we just showed that symbols can be interpreted in
the alternative ways given above, which are different from the one eventually
intended by author of \textbf{\cite{0}}. Indeed, he should write
\[%
\begin{tabular}
[c]{|c|}\hline
$(\theta^{\mathbf{a}}\curlywedge\theta^{\mathbf{b}})_{\mu\nu}=(\theta
^{\mathbf{a}}\curlywedge\theta^{\mathbf{b}})({\mbox{\boldmath$\partial$}}%
_{\mu},{\mbox{\boldmath$\partial$}}_{\mu})=q_{\mu}^{\mathbf{a}}q_{\nu
}^{\mathbf{b}}-q_{\nu}^{\mathbf{a}}q_{\mu}^{\mathbf{b}}$\\\hline
\end{tabular}
\ \ \ \ ,
\]
and then advise his readers that he was going to represent $\left(
\theta^{\mathbf{a}}\curlywedge\theta^{\mathbf{b}}\right)  _{\mu\nu}$ by the
symbol $q_{\mu}^{\mathbf{a}}\curlywedge q_{\nu}^{\mathbf{b}}$ , i.e., $\left(
\theta^{\mathbf{a}}\curlywedge\theta^{\mathbf{b}}\right)  _{\mu\nu}:=q_{\mu
}^{\mathbf{a}}\curlywedge q_{\nu}^{\mathbf{b}}$.

At first sight it may seem that we are being very \textit{pedantic}. But if we
insist in notational issues, it is because as we are going to see in the
following sections, if the exact meaning of the symbols used are not precise,
ambiguities may appear in calculations a lit bit more sophisticated than the
ones above, resulting inevitably in nonsense.

(c2) Consider the statement following Eq.(22E) in page 437 of \textbf{\cite{0}%
}, namely:

"...The dimensionality of the tetrad matrix depends on the way it is defined:
for example, using Eqs.(6E) (7E), (11E) or (12E), the tetrad is a $4\times4$
matrix; using Eq.(13E), it is a $2\times2$ complex matrix."

This is a very misleading statement, which is a source in \cite{0,1,2,3,4,5}
of confusion. That statement has probably origin in some statements appearing
\cite{11,12}. Indeed, suppose that we consider Clifford valued differential
forms. i.e., objects that are sections of the bundle $\mathcal{C\ell
(}TM)\otimes\bigwedge T^{\ast}M$, where $\mathcal{C\ell(}TM)$ is the Clifford
bundle of nonhomogeneous multivector fields.\footnote{Note that in section 8
and the following ones we work with $\mathcal{C\ell(}T^{\ast}M)$, the bundle
of nonhomogeneous multiforms fields.} We consider as usual that $TM=\bigwedge
\nolimits^{1}TM\hookrightarrow\mathcal{C\ell(}TM)$ (details may be found in
\cite{13,14}) Consider the object
\begin{equation}
\mathbf{Q}=\mathbf{e}_{\mathbf{a}}\otimes\theta^{\mathbf{a}}=e_{\mu}\otimes
dx^{\mu}\in\sec\bigwedge\nolimits^{1}TM\otimes\bigwedge\nolimits^{1}T^{\ast
}M\hookrightarrow\mathcal{C\ell(}TM)\otimes\bigwedge T^{\ast}M \label{sachs 1}%
\end{equation}
We define now the object $S\in\sec$\ $\mathcal{C\ell(}TM)\otimes\bigwedge
T^{\ast}M$ by%
\begin{align}
S  &  =\mathbf{Qe}_{\mathbf{0}}:=e_{\mu}\mathbf{e}_{\mathbf{0}}\otimes
dx^{\mu}=\mathbf{q}_{\mathbf{\mu}}\otimes dx^{\mu}\label{sachs 2}\\
&  =\mathbf{e}_{\mathbf{a}}\mathbf{e}_{\mathbf{0}}\otimes\theta^{\mathbf{a}%
}=\mathbf{q}_{\mathbf{a}}\otimes\theta^{\mathbf{a}}%
\end{align}

As showed in details in \cite{13}, the objects
\begin{equation}
\mathbf{q}_{\mathbf{\mu}}=e_{\mu}\mathbf{e}_{\mathbf{0}}\in\sec\mathcal{C\ell
}^{(0)}\mathcal{(}TM), \label{paravector}%
\end{equation}
where $\mathcal{C\ell}^{(0)}\mathcal{(}TM)$ is the even subalgebra of
$\mathcal{C\ell(}TM)$. As it is well known, for each $e\in M$, $\mathcal{C\ell
}^{(0)}\mathcal{(}T_{e}M)=\mathbb{R}_{3,0}$, a Clifford algebra also known as
Pauli algebra, the reason being the fact that as a matrix algebra,
$\mathbb{R}_{3,0}\simeq\mathbb{C(}2\mathbb{)}$, the algebra of $2\times2$
complex matrices. Sachs thought that the $\mathbf{q}_{\mathbf{\mu}}$ would be
quaternion fields, but indeed they are not. They are paravector fields.
Important for our comments is the fact that the matrix representation of the
$\mathbf{q}_{\mathbf{\mu}}$ are $2\times2$ complex matrices that \ as well
known may be expanded in terms of the identity matrix and the Pauli matrices.
Now, having in mind that we can write $\mathbf{q}_{\mathbf{\nu}}=e_{\nu
}\mathbf{e}_{\mathbf{0}}=q_{\nu}^{\mathbf{a}}\mathbf{e}_{\mathbf{a}}%
\mathbf{e}_{\mathbf{0}}=q_{\nu}^{\mathbf{a}}\mathbf{q}_{\mathbf{a}}$, we can
understand that the real functions $q_{\nu}^{\mathbf{a}}$ appears as
components of complex functions in the matrix representations of the
$\mathbf{q}_{\mathbf{\nu}}$. But this, of course, does not mean that the
tetrads are complex matrices, as stated in \cite{0}. We can define a covariant
derivative $\mathbf{\nabla}^{c}$ operator (see details in \cite{10}) acting on
sections of the Clifford bundle of multivectors $\mathcal{C\ell(}TM)$. Then,
we can define the covariant derivative of the paravector fields $\mathbf{q}%
_{\mathbf{\nu}}$ (or their matrix representations) in the direction of the
coordinate vector field $e_{\mu}={\mbox{\boldmath$\partial$}}_{\mu}$. This
would be written as $\mathbf{\nabla}_{{\mbox{\boldmath$\partial$}}_{\mu}}%
^{c}\mathbf{q}_{\mathbf{\nu}}=\mathbf{\nabla}_{{\mbox{\boldmath$\partial$}}%
_{\mu}}^{c}(q_{\nu}^{\mathbf{a}}\mathbf{q}_{\mathbf{a}}):=(\mathbf{\nabla
}_{{\mu}}^{c}q_{\nu}^{\mathbf{a}})\mathbf{q}_{\mathbf{a}}$, thereby defining
unambiguously the symbols $\mathbf{\nabla}_{{\mu}}^{c}q_{\nu}^{\mathbf{a}}$ as
the components of the covariant derivatives of the paravector fields
$\mathbf{q}_{\mathbf{\nu}}$ in the paravector field basis $\{\mathbf{q}%
_{\mathbf{a}}\}.$

Of course, it is possible to think of another matrix \ involving the real
functions $q_{\mu}^{\mathbf{a}}.$ Indeed, \ forget \ for a while the bundle
$\mathcal{C\ell(}TM)\otimes\bigwedge T^{\ast}M$ and consider an object
$\mathbf{P}\in\sec T^{1,1}M.$ Such object (sometimes called a vector valued
1-form) can be written in the \ `hybrid' basis $\{\mathbf{e}_{\mathbf{a}%
}\otimes dx^{\mu}\}$ of $T^{1,1}U$ as
\begin{equation}
\mathbf{P}=P_{\nu}^{\mathbf{a}}\mathbf{e}_{\mathbf{a}}\otimes dx^{\nu}
\label{P}%
\end{equation}
and can, of course be represented by a $4\times4$ \textit{real }matrix in the
standard way. In particular, we can imagine a $\mathbf{Q}\in\sec T^{1,1}M$
such that $Q_{\mu}^{\mathbf{a}}=q_{\mu}^{\mathbf{a}}$. This
\begin{equation}
\mathbf{Q=}q_{\mu}^{\mathbf{a}}\mathbf{e}_{\mathbf{a}}\otimes dx^{\nu}
\label{Q}%
\end{equation}
can be, of course, be appropriately identified in an obvious way with the
\textbf{Q }$\ $defined in Eq.(\ref{sachs 1}), this being the reason that we
used the same symbol. As we shall show below we \textit{cannot} identify the
components of the covariant derivative of $\mathbf{Q}$ in the direction of the
vector field ${\mbox{\boldmath$\partial$}}_{\mu}$, which we will denote by
$\mathbf{\nabla}_{\mu}q_{\mu}^{\mathbf{a}}=($\textbf{$\nabla$}%
$_{{\mbox{\boldmath$\partial$}}_{\mu}}\mathbf{Q})_{\nu}^{\mathbf{a}}$ with the
components of the covariant derivative of the $%
\mbox{\boldmath{$\theta$}}%
^{\mathbf{a}}$ in the direction of the vector field
${\mbox{\boldmath$\partial$}}_{\mu}$, which will be denote by \textbf{$\nabla
$}$_{\mu}^{-}q_{\nu}^{\mathbf{a}}$, which is given by Eq.(\ref{19}) below. It
is also not licit to identify $\mathbf{\nabla}_{\mu}q_{\mu}^{\mathbf{a}}$ with
$\mathbf{\nabla}_{\mu}^{c}q_{\mu}^{\mathbf{a}}$.

As we shall see, it is this wrong identification that leads to the ambiguous
statement called \ `tetrad postulate'.

\begin{remark}
Any how, before proceeding we have an observation concerning the symbols
$q_{\mu}^{\mathbf{a}}\overset{a}{\wedge}q_{\nu}^{\mathbf{b}}$ and
$\mathbf{q}_{\mu\nu}^{\mathbf{ab}}$. The idea of associating a linear
combination of $\mathbf{q}_{\mu\nu}^{\mathbf{ab}}$, as defined in
Eq.(\ref{13 BIS}) with a gravitational field and a multiple of $q_{\mu
}^{\mathbf{a}}\overset{a}{\wedge}q_{\nu}^{\mathbf{b}}$ \ as defined by
Eq.(\ref{last 13}) with an electromagnetic field already appeared in the old
Sachs book \cite{11} (see also Sachs recent book \cite{12}). The only
difference is that Sachs introduces the fields $q_{\mathbf{a}}^{\mu},q_{\mu
}^{\mathbf{a}}:\varphi(U)\rightarrow\mathbb{R}$ as coefficients of the matrix
representations \ of the paravector vector fields $\mathbf{q}_{\mu}$ defined
in Eq.(\ref{paravector}) (which he incorrectly identified with quaternion
fields). Unfortunately that idea does not work as proved in \cite{13,14}, and
much the same arguments can be given for the theory proposed in
\cite{0,20,1,2,3,4,5} and will not be repeated here.
\end{remark}

For what follows we need to keep in mind that---as explained in the previous
section--- the functions $q_{\mathbf{a}}^{\mu},q_{\mu}^{\mathbf{a}}%
:\varphi(U)\rightarrow\mathbb{R}$ are always \textit{real functions}, and that
set $\{q_{\mu}^{\mathbf{a}}\}$\textit{, }can appear as components of very,
\textit{distinct }objects, e.g.\textit{,} for each fixed $\mathbf{a}$,
$\{q_{\mu}^{\mathbf{a}}\}$ can be interpreted as the components of a covector
field (namely $%
\mbox{\boldmath{$\theta$}}%
^{\mathbf{a}}$) in the basis $\{dx^{\mu}\}$ or for fixed $\mu$, $\{q_{\mu
}^{\mathbf{a}}\}$ as the components of the vector field
${\mbox{\boldmath$\partial$}}_{\mu}$ in the basis $\{\mathbf{e}_{\mathbf{a}%
}\}.$ Also, the set $\{q_{\mathbf{a}}^{\mu}\}$ for each fixed $\mathbf{a}$ can
be interpreted as the components of the vector field $\mathbf{e}_{\mathbf{a}}$
in the basis ${\mbox{\boldmath$\partial$}}_{\mu}$. Also, for varying
\ $\mathbf{a}$ and $\mu$ the $\{q_{\mu}^{\mathbf{a}}\}$ can be thought as the
components of the tensor $\mathbf{Q}$ given by Eq.(\ref{Q}), etc. So, it is
crucial to distinguish without ambiguity in what context the set of real
functions $\{q_{\mu}^{\mathbf{a}}\}$ (or $\{q_{\mathbf{a}}^{\mu}\}$) is being used.

(c3) Consider the statement before Eq.(23E) of \cite{0}:

"The tetrad is a vector-valued one-form, i.e., is a one-form $q_{\mu}$ with
labels $\mathbf{a}$. If $\mathbf{a}$ takes values 1,2 or 3 of a Cartesian
representation of the tangent space, for example, the vector%
\begin{equation}
\mathbf{q}_{\mu}=q_{\mu}^{1}\mathbf{i+}q_{\mu}^{2}\mathbf{j+}q_{\mu}%
^{3}\mathbf{k} \tag{23E}%
\end{equation}
can be defined in this space. Each of the components $q_{\mu}^{1}%
\mathbf{,}q_{\mu}^{2}$ or\textbf{ }$q_{\mu}^{3}$ are scalar-valued one-forms
of differential geometry [2], and each of the $q_{\mu}^{1}\mathbf{,}q_{\mu
}^{2}$, and\textbf{ }$q_{\mu}^{3}$ is therefore a covariant four vector in the
base manifold. The three scalar-valued one-forms are therefore the three
components of the vector-valued one-forms $q_{\mu}^{\mathbf{a}}$, the tetrad form."

Well, that sentence contains a \textit{sequence} of misconceptions.

The first part of the statement \ namely \ `The tetrad is a vector-valued
one-form, i.e., is a one-form $q_{\mu}$ with labels $\mathbf{a}$' only has
meaning if the functions $q_{\mu}^{\mathbf{a}}$ are interpreted as the
components of the tensor $\mathbf{Q}$ defined by Eq.(\ref{Q}). So, the next
part of the statement, namely Eq.(23E) is meaningless.

First, the tangent space to each $e\in M,$ where $M$ is the manifold where the
theory was supposed to be developed is a real $4$-dimensional space. So, as we
observed in Remark \ref{elementar}, $\mathbf{a}$ must take the values
$0,1,2,3.$ More, as observed in Remark \ref{elementar 1} the tangent spaces at
different points of a general manifold $M$ in general \textit{cannot} be
identified, unless the manifold possess some additional appropriate structure,
which is not the case in Evans paper. As such, the objects defined in Eq.(23E)
have nothing to do with the concept of tangent vectors, as Evans would like
for future use in some identifications that he used in \textbf{\cite{0}
(}and\textbf{ }\cite{20,1,2,3,4,5}\textbf{ }and also in some old papers that
he signed alone or with the AIAS group and that were published in \textbf{FPL
}and other journals\footnote{A very detailed discussion of the many non
sequitur results of those papers is given in \cite{carro}. A replic by Evans
to that paper is to be found in Evans
website.:http://www.aias.us/pub/rebutal/finalrebutaldocument.pdf. A treplic to
Evans note can be found in:
http://www.ime.unicamp.br/rel\_pesq/2003/ps/rp28-03.pdf. The reading of those
documents is important \ for any reader that \ eventually wants to know some
details of the reason we get involved with Evans theories. A complement to the
previous paper can be found at
http://arxiv.org/PS\_cache/math-ph/pdf/0311/0311001.pdf.}) \textbf{ }to
justify some (wrong) calculations of his \textbf{B}(3) theory. This means also
that $\mathbf{q}_{\mu}$ in Eq.(23E) cannot be identified with the basis
vectors ${\mbox{\boldmath$\partial$}}_{\mu}$. They are simply mappings
$U\rightarrow$ \ $\mathcal{F(U)}\otimes\mathbb{R}^{3}$, where $\mathcal{F(U)}$
is a subset of the set of (smooth) functions in $U$. We emphasize again: The
vectors in set $(\mathbf{i,j,k)}$ as introduced by Evans are \textit{not}
tangent vector fields to the manifold $M$, i.e., they are not sections of
$TU$. The set $(\mathbf{i,j,k)}$ is simply a basis of the real
three-dimensional vector space $\mathbb{R}^{3}$, which has been introduced by
Evans without any clear mathematical motivation.

\section{ Some Results from the Theory of Connections}

(i) In what follows we denote by $F(M)$ the \textit{principal bundle }of
linear frames. The structural group of this bundle is $Gl(4,\mathbb{R})$, the
general linear group on 4-dimensions.\footnote{For details, the reader may
consult as an introduction the books \cite{choquet,frankel}. A more advanced
view of the subject can be acquired studying, e.g.,\cite{konomi,palais}.}

(ii) the elements of $F(M)$ are called frame fields (or simply frames). A
frame $\{e_{\alpha}\}\in\sec F(M)$ can be identified with a basis of $TM$, the
tangent bundle.

(iii) We suppose that the manifold $M$ is equipped with a Lorentz metric
$\mathbf{g}\in\sec T^{(2.0)}M.$ We denote by $P_{\mathrm{SO}_{1,3}^{e}}(M)$
the bundle of \textit{orthonormal frames}. Its structural group is
\textrm{SO}$_{1,3}^{e}$, the homogeneous orthochronous Lorentz group.
$P_{\mathrm{SO}_{1,3}^{e}}(M)$ is said to be a reduction of $F(M).$ A frame
$\{\mathbf{e}_{\mathbf{a}}\}\in\sec$ $P_{\mathrm{SO}_{1,3}^{e}}(M)$ is called
an orthonormal frame.

(iv) A \ linear connection on $F(M)$ is a $1$-form with values in the Lie
algebra $gl(4,\mathbb{R)}$, which needs to satisfy a set of well specified
properties, which we are not going to specify here, since they will be not
necessary in what follows.

(v) It is a theorem of the theory of connections that each connection defined
in $P_{\mathrm{SO}_{1,3}^{e}}(M)$ \ determines a connection in $F(M)$ ( a
linear connection)

(vi) Given the pair $(M,\mathbf{g}),$ \ a linear connection on $F(M)$, which
is determined by a connection on the bundle of orthonormal frames
$P_{\mathrm{SO}_{1,3}^{e}}(M)$ is called \textit{metric compatible}.

(vii) Any connection in a principal bundle determines a connection in each
associated vector bundle to it.

\subsection{On the Nature of Tangent and Cotangent Fields II}

(viii) We are going to work exclusively with spacetime structures in this
paper which have the pair $(M,\mathbf{g})$ as substructure. Under this
condition we recall that the tangent and cotangent bundles $TM$ and $T^{\ast
}M$ (already introduced in section 2) can also be written as the associated
vector bundles
\begin{equation}
TM=F(M)\times_{\rho^{+}(Gl(4,\mathbb{R}^{4}))}\mathbb{R}^{4}=P_{\mathrm{SO}%
_{1,3}^{e}}(M)\times_{\rho^{+}(\mathrm{SO}_{1,3}^{e})}\mathbb{R}^{4},
\label{t bundle}%
\end{equation}
and the cotangent bundle is
\begin{equation}
T^{\ast}M=F(M)\times_{\rho^{-}(Gl(4,\mathbb{R}^{4}))}\mathbb{R}^{4}%
=P_{\mathrm{SO}_{1,3}^{e}}(M)\times_{\rho^{-}(\mathrm{SO}_{1,3}^{e}%
)}\mathbb{R}^{4}. \label{cot bundle}%
\end{equation}
In the above equations, $\rho^{+}(Gl(4,\mathbb{R}^{4}))$ ($\rho^{+}%
(\mathrm{SO}_{1,3}^{e})$) refers to the \textit{standard} representations of
the groups $Gl(4,\mathbb{R}^{4})$ (\textrm{SO}$_{1,3}^{e}$) and $\rho
^{-}(Gl(4,\mathbb{R}^{4}))$ ($\rho^{-}(\mathrm{SO}_{1,3}^{e})$) refers to the
\textit{dual} representations. Given these results the bundle of $(r,s)$
tensors is (in obvious notation)%
\begin{align}
T^{(r,s)}(M)  &  =\bigotimes\nolimits_{s}^{r}F(M)\times_{\bigotimes
\nolimits_{s}^{r}\rho^{+}(Gl(4,\mathbb{R}^{4}))}(\bigotimes\nolimits_{s}%
^{r}\mathbb{R}^{4})\nonumber\\
&  =\bigotimes\nolimits_{s}^{r}P_{\mathrm{SO}_{1,3}^{e}}(M)\times
_{\bigotimes\nolimits_{s}^{r}\rho^{+}(\mathrm{SO}_{1,3}^{e})}(\bigotimes
\nolimits_{s}^{r}\mathbb{R}^{4}) \label{rs te bundle}%
\end{align}

(ix) The tensor bundle is denoted here, as in Section 2 by $\mathbf{\tau
}M=\bigoplus\nolimits_{r,s=0}^{\infty}T^{(r,s)}(M).$

(x) Any connection in a principal bundle determines a connection in each
associated vector bundle to it. A connection on a vector bundle is also called
a \textit{covariant derivative. }

\subsection{\textbf{$\nabla$}$^{+},$\textbf{$\nabla$}$^{-}$ and
\textbf{$\nabla$}}

Let $X,Y\in\sec TM$, any vector fields, $\alpha\in\sec T^{\ast}M$ any covector
(also called a 1-form field) and \ $\mathbf{P}\in\sec\mathbf{\tau}M$ \ any
general tensor. Then, we have the following three covariant derivatives
operators, \textbf{$\nabla$}$^{+},$\textbf{$\nabla$}$^{-}$ and \textbf{$\nabla
$}, defined as follows:%
\begin{align}
\mathbf{\nabla}^{+}  &  :\sec TM\times\sec TM\rightarrow\sec TM,\nonumber\\
(X,Y)  &  \mapsto\mathbf{\nabla}_{X}^{+}Y, \label{CO VETOR}%
\end{align}

\begin{align}
\mathbf{\nabla}^{-}  &  :\sec TM\times\sec T^{\ast}M\rightarrow\sec
TM,\nonumber\\
(X,\alpha)  &  \mapsto\mathbf{\nabla}_{X}^{-}\alpha, \label{CO FORMA}%
\end{align}%
\begin{align}
\mathbf{\nabla}  &  :\sec TM\times\sec\mathbf{\tau}M\rightarrow\sec\tau
M,\nonumber\\
(X,\mathbf{P})  &  \mapsto\mathbf{\nabla}_{X}\mathbf{P}, \label{CO TENSOR}%
\end{align}

(xi) Each one of the covariant derivative operators introduced above satisfy
the following properties: Given, differentiable functions $f,g:M\rightarrow
\mathbb{R}$, vector fields $X,Y\in\sec TM$ and $\mathbf{P,Q}\in\sec
\mathbf{\tau}M$ we have%

\begin{align}
\mathbf{\nabla}_{fX+gY}\mathbf{P}  &  =f\mathbf{\nabla}_{X}\mathbf{P+}%
g\mathbf{\nabla}_{Y}\mathbf{P},\nonumber\\
\mathbf{\nabla}_{X}(\mathbf{P+Q})  &  =\mathbf{\nabla}_{X}\mathbf{P}%
+\mathbf{\nabla}_{X}\mathbf{Q},\nonumber\\
\mathbf{\nabla}_{X}(f\mathbf{P)}  &  =f\mathbf{\nabla}_{X}(\mathbf{P)+}%
X(f\mathbf{)P},\nonumber\\
\mathbf{\nabla}_{X}(\mathbf{P\otimes Q})  &  =\mathbf{\nabla}_{X}%
\mathbf{P\otimes Q}+\mathbf{P\otimes\nabla}_{X}\mathbf{Q}.
\end{align}

(xii) The \textit{absolute differential} of \ $\mathbf{P\in\sec}T^{(r,s)}(M)$
is given by the mapping%
\begin{align}
\mathbf{\nabla}  &  \mathbf{:}\sec T^{(r,s)}(M)\rightarrow\sec T^{(r,s+1)}%
(M),\label{absolute dif}\\
&  \mathbf{\nabla P(}X\mathbf{,}X_{1},...,X_{s},\alpha_{1},...,\alpha_{r})\\
&  =\mathbf{\nabla}_{X}\mathbf{P(}X_{1},...,X_{s},\alpha_{1},...,\alpha
_{r}),\\
X_{1},...,X_{s}  &  \in\sec TM,\alpha_{1},...\alpha_{r}\in\sec T^{\ast}M.
\end{align}

(xiii) To continue we must give the relationship between \textbf{$\nabla$%
}$^{+},$\textbf{$\nabla$}$^{-}$ and \textbf{$\nabla$}. So, let us suppose that
a connection has been chosen according to what have been said in (vi) above.

Then, given the coordinate bases $\{{\mbox{\boldmath$\partial$}}_{\mu
}\},\{{\mbox{\boldmath$\partial$}}^{\mu}\},\{\theta^{\mu}=dx^{\mu}%
\},\{\theta_{\mu}\}$ and the orthonormal bases$\{\mathbf{e}_{\mathbf{a}%
}\},\{\mathbf{e}^{\mathbf{a}}\},\{\theta_{\mathbf{a}}\},\{\theta^{\mathbf{a}%
}\}$ defined in Section 2, we have the definitions of the connection
coefficients associated to the respective covariant derivatives in the
respective basis, e.g.,%
\begin{align}
\mathbf{\nabla}_{{\mbox{\boldmath$\partial$}}_{\mu}}^{+}%
{\mbox{\boldmath$\partial$}}_{\nu}  &  =\Gamma_{\mu\nu}^{\rho}%
{\mbox{\boldmath$\partial$}}_{\rho},\;\;\;\mathbf{\nabla}%
_{{\mbox{\boldmath$\partial$}}_{\sigma}}^{-}{\mbox{\boldmath$\partial$}}^{\mu
}=-\Gamma_{\sigma\alpha}^{\mu}{\mbox{\boldmath$\partial$}}^{\alpha},\\
\mathbf{\nabla}_{\mathbf{e}_{\mathbf{a}}}^{+}\mathbf{e}_{\mathbf{b}}  &
=\omega_{\mathbf{ab}}^{\mathbf{c}}\mathbf{e}_{\mathbf{c}},\qquad
\mathbf{\nabla}_{\mathbf{e}_{\mathbf{a}}}^{+}\mathbf{e}^{\mathbf{b}}%
=-\omega_{\mathbf{ac}}^{\mathbf{b}}\mathbf{e}^{\mathbf{c}},\;\nonumber\\
\hspace{0.15cm}\mathbf{\nabla}_{\mathbf{e}_{\mathbf{a}}}^{-}\theta
_{\mathbf{b}}  &  =\omega_{\mathbf{ba}}^{\mathbf{c}}\theta_{\mathbf{c}%
}=-\omega_{\mathbf{bac}}\theta^{\mathbf{c}}\label{14 esp}\\
\;\;\text{ }\omega_{\mathbf{abc}}  &  =\eta_{\mathbf{ad}}\omega_{\mathbf{bc}%
}^{\mathbf{d}}=-\omega_{\mathbf{cba}},\text{ }\omega_{\mathbf{a}}%
^{\mathbf{bc}}=\eta^{\mathbf{bk}}\omega_{\mathbf{kal}}\eta^{\mathbf{cl}%
},\text{ }\omega_{\mathbf{a}}^{\mathbf{bc}}=-\omega_{\mathbf{a}}^{\mathbf{cb}%
}\\
\mathbf{\nabla}_{{\mbox{\boldmath$\partial$}}_{\mu}}^{+}\mathbf{e}%
_{\mathbf{b}}  &  =\omega_{\mu\mathbf{b}}^{\mathbf{c}}\mathbf{e}_{\mathbf{c}%
},\nonumber\\
\mathbf{\nabla}_{{\mbox{\boldmath$\partial$}}_{\mu}}^{-}dx^{\nu}  &
=-\Gamma_{\mu\alpha}^{\nu}dx^{\alpha},\;\;\;\mathbf{\nabla}%
_{{\mbox{\boldmath$\partial$}}_{\mu}}^{-}\theta_{\nu}=\Gamma_{\mu\nu}^{\rho
}\theta_{\rho},\nonumber\\
\mathbf{\nabla}_{\mathbf{e}_{\mathbf{a}}}^{-}\theta^{\mathbf{b}}%
=-\omega_{\mathbf{ac}}^{\mathbf{b}}\theta^{\mathbf{c}}  &
,\;\;\;\mathbf{\nabla}_{{\mbox{\boldmath$\partial$}}_{\mu}}^{-}\theta
^{\mathbf{b}}=-\omega_{\mu\mathbf{a}}^{\mathbf{b}}\theta^{\mathbf{a}}.
\end{align}

To understood how \textbf{$\nabla$} works, consider its action, e.g., on the
sections of $T^{(1,1)}M=TM\otimes T^{\ast}M$. For that case, if $X\in\sec TM$,
$\alpha\in\sec T^{\ast}M$, we have that
\begin{equation}
\mathbf{\nabla}=\mathbf{\nabla}^{+}\otimes\mathrm{Id}_{T^{\ast}M}%
+\mathrm{Id}_{TM}\otimes\mathbf{\nabla}^{-}, \label{cov d}%
\end{equation}
and
\begin{equation}
\mathbf{\nabla}(X\otimes\alpha)=(\mathbf{\nabla}^{+}X)\otimes\alpha
+X\otimes\mathbf{\nabla}^{-}\alpha. \label{COV D1}%
\end{equation}

The general case, of $\mathbf{\nabla}$ acting on sections of $\mathbf{\tau}M$
is an obvious generalization of the precedent one, and details are left to the reader.

(xiv) We said that a connection determined under the \ conditions given in
(vi) above is \textit{metric compatible}. This is given explicitly by the
condition that for the metric tensor $\mathbf{g}\in\sec(T^{\ast}M\otimes
T^{\ast}M)$ we have
\begin{equation}
\mathbf{\nabla g}=0 \label{metric compatible}%
\end{equation}

(xv) It is supposed in what follows that \textbf{$\nabla$} is \textit{not}
necessarily torsion free.\footnote{Note that the metric compatibility
condition $\mathbf{\nabla g}=0$, does not necessarily imply that the torsion
tensor is null, $\mathbf{T}=0$. When \ $\nabla\mathbf{g}=0$ and $\mathbf{T}%
=0$, $\mathbf{\nabla}$ is called the Levi-Civita connection, and it is unique.
In that case, the connection coefficients (Cristoffel symbols) in a coordinate
basis are \textit{symmetric}. But, the connection coefficients in a tetrad
basis can be written in a very useful way for computations as
\textit{antisymmetric}. See, e.g., Eq.(\ref{14 esp}).}

(xvi) Also, we assume that we are studying a connection which is not
\textit{teleparallel}, i.e., there is \textit{no} orthonormal basis for
$TU\subset TM$ such that \textbf{$\nabla$}$_{\mathbf{e}_{\mathbf{a}}%
}\mathbf{e}_{\mathbf{b}}=0$, for all $\mathbf{a},\mathbf{b}$ $=0,1,2,3$. So,
in general, $\omega_{\mathbf{ab}}^{\mathbf{c}}\neq0$ and%
\begin{equation}
\mathbf{\nabla}_{\mathbf{e}_{\mathbf{a}}}^{-}\theta^{\mathbf{b}}%
=-\omega_{\mathbf{ac}}^{\mathbf{b}}\theta^{\mathbf{c}}\neq0\text{.}
\label{14a}%
\end{equation}

(xvii) For every vector field $V\in\sec TU$ and a covector field $C\in\sec
T^{\ast}U$ we have%

\begin{equation}
\mathbf{\nabla}_{{\mbox{\boldmath$\partial$}}_{\mu}}^{+}V=\mathbf{\nabla
}_{{\mbox{\boldmath$\partial$}}_{\mu}}^{+}(V^{\alpha}%
{\mbox{\boldmath$\partial$}}_{\alpha}),\quad\text{ }\mathbf{\nabla
}_{{\mbox{\boldmath$\partial$}}_{\mu}}^{-}C=\mathbf{\nabla}%
_{{\mbox{\boldmath$\partial$}}_{\mu}}^{-}(C_{\alpha}\theta^{\alpha})
\label{15}%
\end{equation}

and using the properties introduced in (xi) above, \textbf{$\nabla$%
}$_{{\mbox{\boldmath$\partial$}}_{\mu}}^{+}V$ can be written as:%
\begin{align}
\mathbf{\nabla}_{{\mbox{\boldmath$\partial$}}_{\mu}}^{+}V  &  =\mathbf{\nabla
}_{{\mbox{\boldmath$\partial$}}_{\mu}}^{+}(V^{\alpha}%
{\mbox{\boldmath$\partial$}}_{\alpha})=(\mathbf{\nabla}%
_{{\mbox{\boldmath$\partial$}}_{\mu}}^{+}V)^{\alpha}%
{\mbox{\boldmath$\partial$}}_{\alpha}\nonumber\\
&  =({\mbox{\boldmath$\partial$}}_{\mu}V^{\alpha}){\mbox{\boldmath$\partial$}}%
_{\alpha}+V^{\alpha}\mathbf{\nabla}_{{\mbox{\boldmath$\partial$}}_{\mu}}%
^{+}{\mbox{\boldmath$\partial$}}_{\alpha}\nonumber\\
&  =\left(  \frac{\partial V^{\alpha}}{\partial x^{\mu}}+V^{\rho}\Gamma
_{\mu\rho}^{\alpha}\right)  {\mbox{\boldmath$\partial$}}_{\alpha
}=(\mathbf{\nabla}_{\mu}^{+}V^{\alpha}){\mbox{\boldmath$\partial$}}_{\alpha},
\label{16}%
\end{align}
where it is to be keeped in mind that
\begin{equation}
(\mathbf{\nabla}_{{\mbox{\boldmath$\partial$}}_{\mu}}^{+}V)^{\alpha}%
\equiv\mathbf{\nabla}_{\mu}^{+}V^{\alpha}. \label{16bis}%
\end{equation}

Also, we have
\begin{align}
\mathbf{\nabla}_{{\mbox{\boldmath$\partial$}}_{\mu}}^{-}C  &  =\mathbf{\nabla
}_{{\mbox{\boldmath$\partial$}}_{\mu}}^{-}(C_{\alpha}\theta^{\alpha
})=(\mathbf{\nabla}_{{\mbox{\boldmath$\partial$}}_{\mu}}^{-}C)_{\alpha}%
\theta^{\alpha}\nonumber\\
&  =\left(  \frac{\partial C_{\alpha}}{\partial x^{\mu}}-C_{\beta}\Gamma
_{\mu\alpha}^{\beta}\right)  \theta^{\alpha},\nonumber\\
&  \equiv(\mathbf{\nabla}_{\mu}^{-}C_{\alpha})\theta^{\alpha} \label{17}%
\end{align}
where it is to be keeped in mind that \footnote{Recall that other authors
prefer the notations $(\mathbf{\nabla}_{{\mbox{\boldmath$\partial$}}_{\mu}%
}V)^{\alpha}=V_{:\mu}^{\alpha}$ and $(\mathbf{\nabla}%
_{{\mbox{\boldmath$\partial$}}_{\mu}}C)_{\alpha}\equiv C_{\alpha:\mu}$. What
is important is always to have in mind the meaning of the symbols.}
\begin{equation}
(\mathbf{\nabla}_{{\mbox{\boldmath$\partial$}}_{\mu}}^{-}C)_{\alpha}%
\equiv\mathbf{\nabla}_{\mu}^{-}C_{\alpha}, \label{17new}%
\end{equation}

\begin{remark}
\label{dv}Eqs.(\ref{16}) and (\ref{17}) \textit{define} the \textit{symbols}
\textbf{$\nabla$}$_{\mu}^{+}V^{\alpha}$ and \textbf{$\nabla$}$_{\mu}%
^{-}C_{\alpha}$. The symbols \textbf{$\nabla$}$_{\mu}^{+}V^{\alpha}%
:\varphi(U)\rightarrow\mathbb{R}$ are real functions, which are the components
of the vector field \ \textbf{$\nabla$}$_{{\mbox{\boldmath$\partial$}}_{\mu}%
}^{+}V$ in the basis $\{{\mbox{\boldmath$\partial$}}_{\alpha}\}$. Also,
\textbf{$\nabla$}$_{\mu}^{-}C_{\alpha}:\varphi(U)\rightarrow\mathbb{R}$ are
the components of the covector field $C$ in the basis $\{\theta^{\alpha}\}$.
\end{remark}

\begin{remark}
The standard practice of many Physics textbooks of representing,
\textbf{$\nabla$}$_{\mu}^{+}V^{\alpha}$ \ and \textbf{$\nabla$}$_{\mu}%
^{+}V^{\alpha}$ by\textbf{ $\nabla$}$_{\mu}V^{\alpha}$ will be avoided here.
This is no pedantism, as we are going to see. Moreover, we observe that the
standard practice of calling \ \textbf{$\nabla$}$_{\mu}^{+}V^{\alpha}$ the
covariant derivative of the "vector" field $V^{\alpha}$ generates a lot of
confusion, for many people, confounds the symbol \textbf{$\nabla$}$_{\mu}^{+}$
(appearing in \textbf{$\nabla$}$_{\mu}^{+}V^{\alpha}$) with the real covariant
derivative operator, which is \textbf{$\nabla$}$_{{\mbox{\boldmath$\partial$}}%
_{\mu}}^{+}$.\footnote{An explicit warning concerning this observation can be
found at page 210 of \cite{9}.
\par
{}} Also, in many Physics textbooks the symbol \textbf{$\nabla$}$_{\mu}^{+}$
is sometimes also used as a sloppy notation for the symbol \textbf{$\nabla$%
}$_{{\mbox{\boldmath$\partial$}}_{\mu}}^{+}$, something that generates yet
more confusion. The author of \cite{0,1,2,3,4,5}. e.g., has not escaped from
that confusion, and generated more confusion yet.
\end{remark}

\begin{remark}
In analyzing Eqs. (\ref{16}) and (\ref{17}) we see that in the process of
taking the covariant derivative the action\ of the basis vector fields
${\mbox{\boldmath$\partial$}}_{\alpha}$ on a vector field $V$ and on a
covector field $C$ are
\begin{align}
{\mbox{\boldmath$\partial$}}_{\mu}V  &  ={\mbox{\boldmath$\partial$}}_{\mu
}(V^{\alpha}{\mbox{\boldmath$\partial$}}_{\alpha})=\frac{\partial V^{\alpha}%
}{\partial x^{\mu}}{\mbox{\boldmath$\partial$}}_{\alpha},\label{17bis}\\
{\mbox{\boldmath$\partial$}}_{\mu}C  &  ={\mbox{\boldmath$\partial$}}_{\mu
}(C_{\alpha}\theta^{\alpha})=\frac{\partial C_{\alpha}}{\partial x^{\mu}%
}\theta^{\alpha},
\end{align}
from where we infer the rules\footnote{These rules are crucial for the writing
of the covariant derivative operator on the Clifford bundles $\mathcal{C\ell
(}TM)$ and $\mathcal{C\ell(}T^{\ast}M)$. See Eq.(\ref{der1}).} $($to be used
with care$)$
\begin{align}
{\mbox{\boldmath$\partial$}}_{\mu}\left(  {\mbox{\boldmath$\partial$}}_{\nu
}\right)   &  =0,\nonumber\\
{\mbox{\boldmath$\partial$}}_{\mu}\left(  \theta^{\alpha}\right)   &  =0.
\label{17biss}%
\end{align}

\end{remark}

Next we recall that our given connection has been\ assumed to be \textit{not}
teleparallel, a statement that implies also
\begin{equation}
\mathbf{\nabla}_{\mathbf{e}_{\mathbf{b}}}^{-}%
\mbox{\boldmath{$\theta$}}%
^{\mathbf{a}}\neq0,\text{ }\mathbf{a},\mathbf{b}=0,1,2,3\mathbf{.}
\label{17bisss}%
\end{equation}
Take notice also that\ in general the $q_{\mathbf{b}}^{\mu}\ $cannot be all
null \ (otherwise the $\mathbf{e}_{\mathbf{b}}=q_{\mathbf{b}}^{\mu}e_{\mu}$
would be null). Also in the more general case, $\partial_{\mu}q_{\nu
}^{\mathbf{b}}\neq0$. Moreover, $%
\mbox{\boldmath{$\theta$}}%
^{\mathbf{a}}=q_{\alpha}^{\mathbf{a}}\theta^{\alpha}=q_{\alpha}^{\mathbf{a}%
}dx^{\alpha}$, and in general $q_{\alpha}^{\mathbf{a}}\neq0$ and
$\partial_{\nu}q_{\alpha}^{\mathbf{a}}\neq0$. It is now a well-known freshman
exercise presented in many good textbooks to verify that the following
identity holds:
\begin{equation}
\partial_{\mu}q_{\nu}^{\mathbf{a}}+\omega_{\mu\mathbf{b}}^{\mathbf{a}}q_{\nu
}^{\mathbf{b}}-\Gamma_{\mu\mathbf{b}}^{\mathbf{a}}q_{\nu}^{\mathbf{b}}=0.
\label{18}%
\end{equation}

Indeed, from Eq.(\ref{17bisss}) we have,
\begin{equation}
\mathbf{\nabla}_{\mathbf{e}_{\mathbf{b}}}^{-}%
\mbox{\boldmath{$\theta$}}%
^{\mathbf{a}}=\omega_{\mathbf{bc}}^{\mathbf{a}}%
\mbox{\boldmath{$\theta$}}%
^{\mathbf{c}}=q_{\mathbf{b}}^{\mu}\mathbf{\nabla}%
_{{\mbox{\boldmath$\partial$}}_{\mu}}^{-}%
\mbox{\boldmath{$\theta$}}%
^{\mathbf{a}}=q_{\mathbf{b}}^{\mu}\omega_{\mu\nu}^{\mathbf{a}}\theta^{\nu}%
\neq0. \label{18bis}%
\end{equation}
\ 

\ Then, since in general \textbf{$\nabla$}$_{\mathbf{e}_{\mathbf{b}}}^{-}%
\mbox{\boldmath{$\theta$}}%
^{\mathbf{a}}\neq0$ and $q_{\mathbf{b}}^{\mu}\neq0,$ we must have in general,
$\omega_{\mu\nu}^{\mathbf{a}}\theta^{\nu}\neq0$ and thus%
\begin{equation}
\mathbf{\nabla}_{{\mbox{\boldmath$\partial$}}_{\nu}}^{-}%
\mbox{\boldmath{$\theta$}}%
^{\mathbf{a}}\neq0. \label{18a}%
\end{equation}
\ 

Now, using Eq.(\ref{17}) we can write
\begin{align}
\mathbf{\nabla}_{{\mbox{\boldmath$\partial$}}_{\mu}}^{-}%
\mbox{\boldmath{$\theta$}}%
^{\mathbf{a}}  &  =\mathbf{\nabla}_{{\mbox{\boldmath$\partial$}}_{\mu}}%
^{-}(q_{\alpha}^{\mathbf{a}}dx^{\alpha})=(\mathbf{\nabla}%
_{{\mbox{\boldmath$\partial$}}_{\mu}}^{-}%
\mbox{\boldmath{$\theta$}}%
^{\mathbf{a}})_{\alpha}dx^{\alpha}\nonumber\\
&  =(\mathbf{\nabla}_{\mu}^{-}q_{\nu}^{\mathbf{a}})dx^{\nu}%
=({\mbox{\boldmath$\partial$}}_{\mu}q_{\nu}^{\mathbf{a}}-\Gamma_{\mu\nu
}^{\beta}q_{\beta}^{\mathbf{a}})dx^{\nu} \label{19}%
\end{align}
Then, from Eq.(\ref{18a}) and \ Eq.(\ref{19}) it follows that (in general)
\begin{equation}
\mathbf{\nabla}_{\mu}^{-}q_{\nu}^{\mathbf{a}}\neq0. \label{20}%
\end{equation}
Having proved that crucial result for our purposes, recall that (see
Eq.(\ref{14 esp}))
\begin{equation}
\mathbf{\nabla}_{{\mbox{\boldmath$\partial$}}_{\mu}}^{-}%
\mbox{\boldmath{$\theta$}}%
^{\mathbf{a}}=-\omega_{\mu\mathbf{b}}^{\mathbf{a}}%
\mbox{\boldmath{$\theta$}}%
^{\mathbf{b}}=-q_{\nu}^{\mathbf{b}}\omega_{\mu\mathbf{b}}^{\mathbf{a}}%
\theta^{\mu}. \label{20bis}%
\end{equation}
Then from Eq.(\ref{19}) and Eq.(\ref{20bis}) we get the proof of
Eq.(\ref{18}), i.e.,
\begin{equation}
\partial_{\mu}q_{\nu}^{\mathbf{a}}-q_{\beta}^{\mathbf{a}}\Gamma_{\mu\nu
}^{\beta}=\partial_{\mu}q_{\nu}^{\mathbf{a}}-\Gamma_{\mu\mathbf{b}%
}^{\mathbf{a}}q_{\nu}^{\mathbf{b}}=-\omega_{\mu\mathbf{b}}^{\mathbf{a}}q_{\nu
}^{\mathbf{b}}\neq0. \label{20 bis}%
\end{equation}

\section{Comments on the \ `Tetrad Postulate'}

At page 438 of \textbf{\cite{0}} the following equation (that the author, says
to be known as the tetrad postulate)%
\begin{equation}
D_{\mu}q_{\nu}^{\mathbf{a}}=\partial_{\mu}q_{\nu}^{\mathbf{a}}+\omega
_{\mu\mathbf{b}}^{\mathbf{a}}q_{\nu}^{\mathbf{b}}-\Gamma_{\mu\mathbf{b}%
}^{\mathbf{a}}q_{\nu}^{\mathbf{b}}=0, \tag{24E}%
\end{equation}
is said to be the \textit{basis} for the \ `demonstration' of Evans Lemma. In
truth, that \ `demonstration' needs also his Eq.(41E), which as we shall see
is completely \textit{wrong}. Of course, several other authors (see below)
also call an equation \textit{like} Eq.(24E) \ `tetrad postulate'

So, we need to investigate if Eq.(24E) has any meaning within the theory of
covariant derivatives. This is absolutely necessary if someone is going to use
that equation as a basis for applications, in particular, in applications to
physical theories.

(a) To start, \ we immediately see that the statement contained in the first
member of Eq.(24E) cannot be identified with the statement \textbf{$\nabla$%
}$_{\mu}^{-}q_{\nu}^{\mathbf{a}}=0$. Indeed, to make such an identification is
simply \textit{wrong}, because we just showed that in general, \textbf{$\nabla
$}$_{\mu}^{-}q_{\nu}^{\mathbf{a}}\neq0$

(b) The freshman identity \ (Eq.(\ref{18})) is simply a compatibility
condition. It results from the condition introduced in (vi) in the previous
section. There is nothing of mysterious in it. However, for reasons that we
are going to explain below, such a compatibility condition generated a lot of
misunderstandings. To see this we need to do some other (almost trivial) calculations.

(c) So, let us next calculate \textbf{$\nabla$}$_{{\mbox{\boldmath$\partial$}}%
_{\mu}}^{+}{\mbox{\boldmath$\partial$}}_{\nu}$ in two different ways, as we
did for \textbf{$\nabla$}$_{{\mbox{\boldmath$\partial$}}_{\mu}}^{-}%
\mbox{\boldmath{$\theta$}}%
^{\mathbf{a}}$. Recalling that
\begin{equation}
{\mbox{\boldmath$\partial$}}_{\nu}=q_{\nu}^{\mathbf{a}}\mathbf{e}_{\mathbf{a}%
}, \label{will need}%
\end{equation}
we have:%
\begin{align}
\mathbf{\nabla}_{{\mbox{\boldmath$\partial$}}_{\mu}}^{+}%
{\mbox{\boldmath$\partial$}}_{\nu}  &  =\mathbf{\nabla}%
_{{\mbox{\boldmath$\partial$}}_{\mu}}^{+}(q_{\nu}^{\mathbf{a}}\mathbf{e}%
_{\mathbf{a}})\nonumber\\
&  ={\mbox{\boldmath$\partial$}}_{\mu}(q_{\nu}^{\mathbf{a}})\mathbf{e}%
_{\mathbf{a}}+q_{\nu}^{\mathbf{a}}(\mathbf{\nabla}%
_{{\mbox{\boldmath$\partial$}}_{\mu}}^{+}\mathbf{e}_{\mathbf{a}})\nonumber\\
&  =\left(  {\mbox{\boldmath$\partial$}}_{\mu}q_{\nu}^{\mathbf{a}}+q_{\nu
}^{\mathbf{b}}\omega_{\mu\mathbf{b}}^{\mathbf{a}}\right)  \mathbf{e}%
_{\mathbf{a}}\nonumber\\
&  =(\mathbf{\nabla}_{_{\mu}}^{+}q_{\nu}^{\mathbf{a}})\mathbf{e}_{\mathbf{a}}
\label{fres a1}%
\end{align}

Now, writing
\begin{equation}
\mathbf{\nabla}_{{\mbox{\boldmath$\partial$}}_{\mu}}^{+}%
{\mbox{\boldmath$\partial$}}_{\nu}=\Gamma_{\mu\nu}^{\rho}%
{\mbox{\boldmath$\partial$}}_{\rho}=\Gamma_{\mu\nu}^{\rho}q_{\rho}%
^{\mathbf{a}}\mathbf{e}_{a}, \label{fres a 2}%
\end{equation}
and from Eqs.(\ref{fres a1}) and (\ref{fres a 2}) we get again the freshman
identity,%
\begin{align}
&  {\mbox{\boldmath$\partial$}}_{\mu}q_{\nu}^{\mathbf{a}}+q_{\nu}^{\mathbf{b}%
}\omega_{\mu\mathbf{b}}^{\mathbf{a}}-\Gamma_{\mu\nu}^{\rho}q_{\rho
}^{\mathbf{a}}\label{frsh again}\\
&  =\partial_{\mu}q_{\nu}^{\mathbf{a}}+\omega_{\mu\mathbf{b}}^{\mathbf{a}%
}q_{\nu}^{\mathbf{b}}-\Gamma_{\mu\mathbf{b}}^{\mathbf{a}}q_{\nu}^{\mathbf{b}%
}=0\nonumber
\end{align}

\begin{remark}
It is very important before proceeding to keep in mind that \textbf{$\nabla$%
}$_{_{\mu}}^{-}q_{\nu}^{\mathbf{a}}$ given by Eq.(\ref{19}) and
\textbf{$\nabla$}$_{_{\mu}}^{+}q_{\nu}^{\mathbf{a}}$ given by
Eq.(\ref{fres a1}) are different functions, i.e., in general,%
\begin{equation}
\mathbf{\nabla}_{_{\mu}}^{+}q_{\nu}^{\mathbf{a}}\neq\mathbf{\nabla}_{_{\mu}%
}^{-}q_{\nu}^{\mathbf{a}} \label{ha ha ha 1}%
\end{equation}

\end{remark}

\begin{remark}
This shows that the statement contained in the first member of Eq.(24E) cannot
be identified with the statement \textbf{$\nabla$}$_{\mu}^{+}q_{\nu
}^{\mathbf{a}}=0$. Indeed, to make such an identification is simply
\textit{wrong}, since in general $\mathbf{\nabla}_{_{\mu}}^{+}q_{\nu
}^{\mathbf{a}}\neq0$
\end{remark}

(d) As our last exercise in this section we now calculate \textbf{$\nabla$%
}$_{{\mbox{\boldmath$\partial$}}_{\mu}}\mathbf{P}$, where $\mathbf{P}$
$\in\sec TU\otimes T^{\ast}U$, $U\subset M.$ Objects of this kind are, as we
already observed, often called vector valued differential forms. First taking
into account the structures of the associated vector bundles recalled in
Eq.(\ref{rs te bundle}), we expand $\mathbf{P}$ in the \ `hybrid`' basis
$\{\mathbf{e}_{\mathbf{a}}\otimes dx^{\nu}\}$ of $TU\otimes T^{\ast}U$, i.e.,
we write
\begin{equation}
\mathbf{P}=P_{\nu}^{\mathbf{a}}\mathbf{e}_{\mathbf{a}}\otimes dx^{\nu}.
\label{new error 1}%
\end{equation}
Then by definition, we have
\begin{align}
\mathbf{\nabla}_{{\mbox{\boldmath$\partial$}}_{\mu}}\mathbf{P}  &
=\mathbf{\nabla}_{{\mbox{\boldmath$\partial$}}_{\mu}}(P_{\nu}^{\mathbf{a}%
}\mathbf{e}_{\mathbf{a}}\otimes dx^{\nu})\label{new error 2}\\
=  &  (\mathbf{\nabla}_{{\mbox{\boldmath$\partial$}}_{\mu}}\mathbf{P})_{\nu
}^{\mathbf{a}}\mathbf{e}_{\mathbf{a}}\otimes dx^{\nu}\nonumber\\
&  =(\mathbf{\nabla}_{{\mu}}P_{\nu}^{\mathbf{a}}\mathbf{)e}_{\mathbf{a}%
}\otimes dx^{\nu}%
\end{align}

A standard computation yields%

\begin{equation}
\mathbf{\nabla}_{{\mu}}P_{\nu}^{\mathbf{a}}=\partial_{\mu}P_{\nu}^{\mathbf{a}%
}-\Gamma_{\mu\nu}^{\beta}P_{\beta}^{\mathbf{a}}+\omega_{\mu\mathbf{b}%
}^{\mathbf{a}}P_{\nu}^{\mathbf{a}}, \label{new error}%
\end{equation}
and in general, \textbf{$\nabla$}$_{{\mu}}P_{\nu}^{\mathbf{a}}\neq0$.

We have the

\begin{proposition}
Let%

\begin{equation}
\mathbf{Q}=q_{\nu}^{\mathbf{a}}\mathbf{e}_{\mathbf{a}}\otimes dx^{\nu}\in\sec
TU\otimes T^{\ast}U, \label{about Q}%
\end{equation}
where the\ functions $q_{\nu}^{\mathbf{a}}$ are the ones appearing in
Eqs.(\ref{10}) and (\ref{will need}). Then
\begin{equation}
\mathbf{\nabla Q=0,} \label{TETRAD POSTULATE}%
\end{equation}
\begin{equation}
\mathbf{\nabla}_{{\mu}}q_{\nu}^{\mathbf{a}}=(\mathbf{\nabla}%
_{{\mbox{\boldmath$\partial$}}_{\mu}}\mathbf{Q})_{\nu}^{\mathbf{a}}%
=\partial_{\mu}q_{\nu}^{\mathbf{a}}-\Gamma_{\mu\nu}^{\beta}q_{\beta
}^{\mathbf{a}}+\omega_{\mu\mathbf{b}}^{\mathbf{a}}q_{\nu}^{\mathbf{a}}=0.
\label{tetrad postulate}%
\end{equation}

\end{proposition}

\begin{proof}
Since \ the functions $q_{\nu}^{\mathbf{a}}$ are the ones appearing in
Eqs.(\ref{10}) and (\ref{will need}) then satisfy the true\ \ `freshman'
identity (Eq.(\ref{18})). Then from that equation and Eq.(\ref{new error}) the
proof follows.
\end{proof}

\begin{remark}
\ The tensor $\mathbf{Q}$\textbf{ }given by Eq.(\ref{about Q}) is for each
$e\in U\subset M$, $\left.  Q\right\vert _{e}$ simply the identity
(endomorphism) mapping $T_{e}U\rightarrow T_{e}U$, as any reader can easily
verify. We have more to say about $\mathbf{Q}$ in section 6.
\end{remark}

\subsection{Some Misunderstandings}

We just calculated the covariant derivatives in the direction of the vector
field ${\mbox{\boldmath$\partial$}}_{\nu}$ of the following tensor fields: the
1-forms $\theta^{\mathbf{a}}=q_{v}^{\mathbf{a}}dx^{v}$, the vector fields
${\mbox{\boldmath$\partial$}}_{v}=q_{v}^{\mathbf{a}}\mathbf{e}_{\mathbf{a}}$
and $\mathbf{Q}$, the identity tensor in $TU$. The components of those objects
in the basis above specified have been denoted respectively by \textbf{$\nabla
$}$_{{\mu}}^{-}q_{\nu}^{\mathbf{a}},$\textbf{$\nabla$}$_{{\mu}}^{+}q_{\nu
}^{\mathbf{a}}$ and \textbf{$\nabla$}$_{{\mu}}q_{\nu}^{\mathbf{a}}$ and we
arrived at the conclusion that in general,
\begin{align*}
\mathbf{\nabla}_{{\mu}}^{-}q_{\nu}^{\mathbf{a}}  &  \neq0,\\
\mathbf{\nabla}_{{\mu}}^{+}q_{\nu}^{\mathbf{a}}  &  \neq0,\\
\mathbf{\nabla}_{{\mu}}^{-}q_{\nu}^{\mathbf{a}}  &  \neq\mathbf{\nabla}_{{\mu
}}^{+}q_{\nu}^{\mathbf{a}},\\
\mathbf{\nabla}_{{\mu}}q_{\nu}^{\mathbf{a}}  &  =0.
\end{align*}

With this in mind we can now identify from where the ambiguities referred in
the introduction come from. Almost all physical authors use instead of the
three distinct symbols \textbf{$\nabla$}$_{{\partial}_{\mu}}^{-}%
,$\textbf{$\nabla$}$_{{\partial}_{\mu}}^{+}$ and \textbf{$\nabla$}%
$_{{\partial}_{\mu}}$ which represent, as we emphasized above three different
connections, the \textit{same} symbol, say $D_{{\partial}_{\mu}\text{ }}$ for
all of them. This clearly generated the absurd conclusions that we have
$D_{{\mu}}q_{\nu}^{\mathbf{a}}=0$ and $D_{{\mu}}q_{\nu}^{\mathbf{a}}\neq0$.

The reader at this point may be thinking: What you explained until now is so
\textit{trivial} that nobody will make such an stupidity of confusing symbols.
Are you sure, dear reader? Let us see.\textbf{ }

\subsubsection{Misunderstanding 1}

As we just observed the majority of physical textbooks and physical articles,
as, e.g., \cite{carroll,15,gsw,knapp,lebedev,pagels,rovelli, vollick} give
first rules for the covariant derivative (denoted in general by $\ D$%
\textit{)} for the components of tensors in a given basis, and when
introducing tetrads first state what they mean by using Eqs. (\ref{10}) and
(\ref{will need}). But then, immediately after that they say that the
`covariant derivative' of the tetrads $q_{\nu}^{\mathbf{a}}$ must be
calculated by
\begin{equation}
D_{{\mu}}q_{\nu}^{\mathbf{a}}=\partial_{\mu}q_{\nu}^{\mathbf{a}}-\Gamma
_{\mu\nu}^{\beta}q_{\beta}^{\mathbf{a}}+\omega_{\mu\mathbf{b}}^{\mathbf{a}%
}q_{\nu}^{\mathbf{b}}. \label{the texbook eq}%
\end{equation}
without specifying that $D_{{\mu}}q_{\nu}^{\mathbf{a}}$ \ must means the
$(\mathbf{\nabla}_{{\mbox{\boldmath$\partial$}}_{\mu}}\mathbf{Q})_{\nu
}^{\mathbf{a}}.$

After that they stated that we need a tetrad postulate\footnote{Since in
general no convincing expalnation is given for
Eq.(\ref{textbook tetrad postulate}), it should be better to call it the
\textit{naive tetrad postulate.}}, which is introduced as the statement%
\begin{equation}
D_{{\mu}}q_{\nu}^{\mathbf{a}}=\partial_{\mu}q_{\nu}^{\mathbf{a}}-\Gamma
_{\mu\nu}^{\beta}q_{\beta}^{\mathbf{a}}+\omega_{\mu\mathbf{b}}^{\mathbf{a}%
}q_{\nu}^{\mathbf{b}}=0. \label{textbook tetrad postulate}%
\end{equation}

Of course, this statement has meaning only if the $q_{\nu}^{\mathbf{a}}$ are
the components of the tensor $\mathbf{Q}$ (Eq.(\ref{about Q})), which is, as
we already recalled the identity endomorphism on $TU$. However, the statement
appears in many books and articles, e.g., in \cite{carroll,gsw} without the
crucial information and this generates inconsistencies. Let us show two of
them, using the physicists convention that all three operators \textbf{$\nabla
$}$_{{\mbox{\boldmath$\partial$}}_{v}}^{-},$\textbf{$\nabla$}%
$_{{\mbox{\boldmath$\partial$}}_{v}}^{+}$ and \textbf{$\nabla$}%
$_{{\mbox{\boldmath$\partial$}}_{v}}$ \ are to be represented by a unique
symbol, that we choose as being same symbol $D_{{\partial}_{\mu}}$.
In\footnote{Take care that some authors also use $D_{\mu}$ as meaning
$D_{{\partial}_{\mu}}.$} \cite{aldoper} the authors correctly calculate
\textbf{$\nabla$}$_{{\mbox{\boldmath$\partial$}}_{v}}^{-}%
{\mbox{\boldmath$\partial$}}_{v}$ (called there $D_{{\partial}_{\mu}%
}{\mbox{\boldmath$\partial$}}_{v}$) and correctly obtained the freshman
identity (their Eq.(5.32). After that they state in their comment 5.1.:
\ `Eq.(5.32) is frequently written as the vanishing of a `total \ covariant
derivative of the tetrad'. Then, they print the equivalent of
Eq.(\ref{textbook tetrad postulate}). This clearly means that they did not
grasp the meaning of the different symbols necessary to be used in an
unambiguous presentation the theory of connections, and they are not alone.
Statements of the same nature appears also, e.g., in
\cite{carroll,15,gsw,knapp,lebedev,pagels,rovelli,vollick} and also in
\cite{0,1,2,3,4,5}.

Specifically, we recall, e.g. that in \cite{15}, authors asserts that the true
identity given by Eq.(\ref{18}) is (unfortunately) written as $D_{\mu}q_{\nu
}^{\mathbf{a}}=0$ as in Eq.(\ref{textbook tetrad postulate}) (or Eq.(24E)) and
confused with the metric condition $D_{\nu}g_{\alpha\nu}=0.$

This old confusion of symbols, it seems, propagated also to papers and books
on supersymmetry, superfields and supergravity, as it will be clear for any
reader that has followed our discussion above and give a look, e.g., at pages
141-144 of \cite{sriva}. That author defines two different covariant
derivatives for the \ `tetrads' $q_{\nu}^{\mathbf{a}}$ without realizing that
in truth he was calculating the covariant derivative of different objects,
living in different vector bundles associated to $P_{\mathrm{SO}_{1,3}^{e}%
}(M)$. The fact is that unfortunately many authors use mathematical objects in
their papers without to know exactly their \textit{real} mathematical nature.
This generates many misunderstandings that propagate in the literature. For
example, in \cite{vollick}, where equations for the gravitational field are
derived from the Palatini method, which allows both the tetrads and the
connections to vary independently in the variation of the action, there are
two \ `tetrad postulates'. Both are expressions of the freshman identity. What
this author and many others forget to say is that the postulate' ( here a
better name would be, a constrain) is necessary to assure that a connection in
$P_{\mathrm{SO}_{1,3}^{e}}(M)$ determines a metric compatible connection in
$F(M)$, as we note in (vi) of section 4.

We now show the `tetrad postulate' generates even much more misunderstandings
than those already described above.

\subsubsection{Misunderstanding 2}

\ \ Observe that for a covector field $C$ we have from Eq.(\ref{17}) if the
symbol $D_{\mu}$ (without any comment) is used in place of the correct symbol
\textbf{$\nabla$}$_{\mu}^{-}$ that%

\begin{align}
D_{{\mbox{\boldmath$\partial$}}_{\mu}}C  &  =D_{{\mbox{\boldmath$\partial$}}%
_{\mu}}(C_{\nu}\theta^{\nu})=\left(  D_{{\mbox{\boldmath$\partial$}}_{\mu}%
}C\right)  _{\nu}\theta^{\nu}\nonumber\\
&  \equiv(D_{\mu}C_{\alpha})\theta^{\alpha}\nonumber\\
&  =\left(  {\partial}_{\mu}C_{\nu}-C_{\beta}\Gamma_{\mu\nu}^{\beta}\right)
\theta^{\nu}\\
&  =D_{{\mbox{\boldmath$\partial$}}_{\mu}}(C_{\mathbf{a}}\theta^{\mathbf{a}%
})=\left(  D_{{\mbox{\boldmath$\partial$}}_{\mu}}C\right)  _{\mathbf{a}}%
\theta^{\mathbf{a}}\\
&  \equiv(D_{{\mu}}C_{\mathbf{a}})\theta^{\mathbf{a}}\\
&  =\left(  {\partial}_{\mu}C_{\mathbf{a}}-C_{\mathbf{b}}\omega_{\mu
\mathbf{a}}^{\mathbf{b}}\right)  \theta^{\mathbf{a}}%
\end{align}

Now, since $C=C_{\nu}\theta^{\nu}=C_{\mathbf{a}}\theta^{\mathbf{a}}$, we have
that $C_{\nu}=q_{\nu}^{\mathbf{a}}C_{\mathbf{a}}$ and we can write%
\begin{align}
D_{\mu}C_{\alpha}  &  ={\partial}_{\mu}(q_{\nu}^{\mathbf{a}}C_{\mathbf{a}%
})-C_{\beta}\Gamma_{\mu\nu}^{\beta}\nonumber\\
&  =({\partial}_{\mu}q_{\nu}^{\mathbf{a}})C_{\mathbf{a}}+q_{\nu}^{\mathbf{a}%
}({\partial}_{\mu}C_{\mathbf{a}})-C_{\beta}\Gamma_{\mu\nu}^{\beta}\nonumber\\
&  =q_{\nu}^{\mathbf{a}}({\partial}_{\mu}C_{\mathbf{a}}-\omega_{\mu\mathbf{a}%
}^{\mathbf{b}}C_{\mathbf{b}})+C_{\mathbf{a}}\left(  {\partial}_{\mu}q_{\nu
}^{\mathbf{a}}-\Gamma_{\mu\nu}^{\beta}q_{\beta}^{\mathbf{a}}+\omega
_{\mu\mathbf{b}}^{\mathbf{a}}q_{\nu}^{\mathbf{b}}\right) \nonumber\\
&  =q_{\nu}^{\mathbf{a}}(D_{{\mu}}C_{\mathbf{a}}), \label{source of error}%
\end{align}
where in going to the last line we used the `freshman identity', i.e.,
Eq.(\ref{18}).

Now, if someone confounds the meaning of the symbols $D_{\mu}C_{\alpha}$ with
the covariant derivative of a vector field, taking into account that
$C_{\alpha}=q_{\nu}^{\mathbf{a}}C_{\mathbf{a}}$ he will use
Eq.(\ref{source of error}) to write the misleading equation

\ \ \ \ \ \ \ \ \ \ \ \ \ \ \ \ \ \ \ \ \ \ \ \ \ \ \ \ \ \ \ \ \ \ \ \ \ \ \ \ \ \ \ \ \ \ \ \ \ \
\begin{equation}%
\begin{tabular}
[c]{|c|}\hline
$D_{\mu}C_{\alpha}=D_{\mu}(q_{\nu}^{\mathbf{a}}C_{\mathbf{a}})=q_{\nu
}^{\mathbf{a}}(D_{{\mu}}C_{\mathbf{a}}),$\\\hline
\end{tabular}
\ \ \ \ \ \label{s error}%
\end{equation}
and someone must be tempted to think that the\ `tetrad postulate', i.e., the
statement that $D_{\mu}q_{\nu}^{\mathbf{a}}=0$ \ is necessary, for in that
case he could apply the Leibniz rule to the first member of Eq.(\ref{s error}%
), i.e., he could write
\begin{equation}
D_{\mu}(q_{\nu}^{\mathbf{a}}C_{\mathbf{a}})=(D_{\mu}q_{\nu}^{\mathbf{a}%
})C_{\mathbf{a}}+q_{\nu}^{\mathbf{a}}(D_{{\mu}}C_{\mathbf{a}})=q_{\nu
}^{\mathbf{a}}(D_{{\mu}}C_{\mathbf{a}}). \label{s error bis}%
\end{equation}

The fact is that:

(i) Whereas the symbols $D_{\mu}C_{\alpha}$ (meaning of course $\mathbf{\nabla
}_{\mu}^{-}C_{\alpha}$) are well defined, the symbol $D_{\mu}(q_{\nu
}^{\mathbf{a}}C_{\mathbf{a}})$ has not the meaning of being of being equal to
$D_{\mu}C_{\alpha}$

(ii) It is not licit to apply the Leibniz rule for the first member of
Eq.(\ref{s error bis}) The reason is the label \textbf{a} in each of the
factors have different ontology. In $q_{\nu}^{\mathbf{a}}$ , it is the $\nu$
component of the tetrad $\theta^{\mathbf{a}}$, i.e., $\theta^{\mathbf{a}%
}=q_{\nu}^{\mathbf{a}}dx^{v}$. In the second factor $\mathbf{a}$ labels the
components of the covector field $C$ in the tetrad basis, i.e.,
$C=C_{\mathbf{a}}\theta^{\mathbf{a}}$. In that way the term $q_{\nu
}^{\mathbf{a}}C_{\mathbf{a}}$ is \textit{not} the contraction of a vector with
a covector field and as such to apply the Leibniz rule to it, writing
Eq.(\ref{s error bis}) is a nonsequitur. Some authors, like in \cite{gsw} say
that $D_{\mu}q_{\nu}^{\mathbf{a}}=0$ \ in the sense of
Eq.(\ref{source of error}), i.e., $D_{\mu}C_{\alpha}=q_{\nu}^{\mathbf{a}%
}(D_{{\mu}}C_{\mathbf{a}})$ and say that this is a property of a \textit{spin
connection}. The fact is that $D$ must be understood in any \ case as the
appropriate connection acting on a well specified vector bundle, as discussed
in the previous section and satisfying the rules given in (vi) there.

\subsubsection{Misunderstanding 3.}

But what is a spin connection, an object said to be used, e.g., by \cite{gsw}?
Spin connection is the name that mathematicians give to a connection defined
on the covering bundle $SE(M)$ \ (called here \textit{spinor bundle
structure}) of $P_{\mathrm{SO}_{1,3}^{e}}(M)$, when $SE(M)$ exists, which
necessitates that ($M,$\texttt{g}$\mathtt{)}$ is a spin manifold. This imposes
constraints in the topology of the manifold $M$. For the particular case of a
manifold $M$ which is part of a spacetime structure,\ the constraint on the
topology of $M$ is given by the famous Geroch \cite{geroch} theorem, which
says that $P_{\mathrm{SO}_{1,3}^{e}}(M)$ must be trivial, i.e., has a global
section. Thus, in that case, global tetrad fields (and of course, cotetrad
fields) exist. Also, the wording spin connection can be used as meaning the
covariant derivatives acting on \ appropriate spinor bundles. A \textit{spinor
bundle} $S(M)$ is an associated vector bundle to the principal bundle $SE(M).$
Sections of a given $\ S(M)$ are called spinor fields. The spin connection
coefficients are related with the objects called $\omega_{\mu\mathbf{a}%
}^{\mathbf{b}}$ introduced, e.g., in Eq.(\ref{14 esp}) in a very
\textit{natural} way, but this will not be discussed here, because these
results will be not needed for what follows. Interested readers, may, e.g.,
consult \cite{10}.

Now, let us present one more serious misunderstanding in the next subsection.

\subsubsection{Misunderstanding 4}

Of course, we can introduce in $M$ many different connections
\cite{choquet,konomi,palais}. In particular, if $M$ is a spin manifold
\cite{10}, which as we just explained above means that $M$ has a global tetrad
$\{\mathbf{e}_{\mathbf{a}}\}$, $\mathbf{e}_{\mathbf{a}}\in\sec T^{\ast}M$,
$\mathbf{a}=0,1,2,3$ and has also a global cotetrad field $\{\theta
^{\mathbf{a}}\}$, $\theta^{\mathbf{a}}\in\sec T^{\ast}M$, $\mathbf{a}=0,1,2,3$
we can introduce a \textit{teleparallel connection---} call it $\mathbf{D}$---
such that
\begin{equation}
\mathbf{D}_{\mathbf{e}_{\mathbf{b}}}^{-}\theta^{\mathbf{a}}=0. \label{21}%
\end{equation}
From Eq.(\ref{21}) we get immediately after multiplying by $q_{\mu
}^{\mathbf{b}}$ and summing in the index $\mathbf{b}$ that%
\begin{equation}
q_{\mu}^{\mathbf{b}}\mathbf{D}_{\mathbf{e}_{\mathbf{b}}}^{-}(q_{\nu
}^{\mathbf{a}}dx^{\nu})\mathbf{=D}_{{\mbox{\boldmath$\partial$}}_{\mu}}%
^{-}(q_{\nu}^{\mathbf{a}}dx^{\nu})=(\mathbf{D}_{\mu}^{-}q_{\nu}^{\mathbf{a}%
})dx^{\nu}=0. \label{21bis}%
\end{equation}
\ \ 

Then, in this case we must have
\begin{equation}
\mathbf{D}_{\mu}^{-}q_{\nu}^{\mathbf{a}}=\left(  \mathbf{D}%
_{{\mbox{\boldmath$\partial$}}_{\mu}}^{-}\theta^{\mathbf{a}}\right)  _{\nu}=0
\label{22}%
\end{equation}

The important point here is that for the teleparallel connection, as it is
well-known the Riemann curvature tensor is \textit{null}, but the torsion
tensor is \textit{not null}. Indeed, given vector fields \ $X,Y\in\sec TM$,
the torsion operator is given by (see, e.g., \cite{choquet})%

\begin{equation}
\tau:(X,Y)\mapsto\tau(X,Y)=\mathbf{D}_{X}^{+}Y-\mathbf{D}_{Y}^{+}X-[X,Y].
\label{TORSION1}%
\end{equation}
First choose $X=\mathbf{e}_{\mathbf{a}}$, $Y=\mathbf{e}_{\mathbf{b}}$, with
$[\mathbf{e}_{\mathbf{a}}$, $\mathbf{e}_{\mathbf{b}}]=c_{\mathbf{ab}%
}^{\mathbf{d}}\mathbf{e}_{\mathbf{d}}$. Then since the $c_{\mathbf{ab}%
}^{\mathbf{d}}$ are not all null, we have
\begin{equation}
\tau(\mathbf{e}_{\mathbf{a}},\mathbf{e}_{\mathbf{b}})=T_{\mathbf{ab}%
}^{\mathbf{d}}\mathbf{e}_{\mathbf{d}}=c_{\mathbf{ab}}^{\mathbf{d}}%
\mathbf{e}_{\mathbf{d}}, \label{torsion1}%
\end{equation}
and the components $T_{\mathbf{ab}}^{\mathbf{d}}$ of the torsion tensor are
\textit{not} all null. Now, if we choose $X={\mbox{\boldmath$\partial$}}_{\mu
}$ and $Y={\mbox{\boldmath$\partial$}}_{\mu}$, then since
$[{\mbox{\boldmath$\partial$}}_{\mu},{\mbox{\boldmath$\partial$}}_{\nu}]=0$,
we can write
\begin{align}
\tau({\mbox{\boldmath$\partial$}}_{\mu},{\mbox{\boldmath$\partial$}}_{\nu})
&  =T_{\mu\nu}^{\mathbf{a}}\mathbf{e}_{\mathbf{a}}=\mathbf{D}%
_{{\mbox{\boldmath$\partial$}}_{\mu}}^{+}{\mbox{\boldmath$\partial$}}_{\nu
}-\mathbf{D}_{{\mbox{\boldmath$\partial$}}_{\nu}}^{+}%
{\mbox{\boldmath$\partial$}}_{\mu}=(\Gamma_{\mu\nu}^{\rho}-\Gamma_{\nu\mu
}^{\rho}){\mbox{\boldmath$\partial$}}_{\rho}\nonumber\\
&  =(\mathbf{D}_{{\mbox{\boldmath$\partial$}}_{\mu}}^{+}%
{\mbox{\boldmath$\partial$}}_{\nu})^{\mathbf{a}}\mathbf{e}_{\mathbf{a}%
}-(\mathbf{D}_{{\mbox{\boldmath$\partial$}}_{\nu}}^{+}%
{\mbox{\boldmath$\partial$}}_{\mu})^{\mathbf{a}}\mathbf{e}_{\mathbf{a}%
}\nonumber\\
&  =\mathbf{D}_{{\mbox{\boldmath$\partial$}}_{\mu}}^{+}\left(  q_{\nu
}^{\mathbf{a}}\mathbf{e}_{\mathbf{a}}\right)  -\mathbf{D}%
_{{\mbox{\boldmath$\partial$}}_{\nu}}^{+}\left(  q_{\mu}^{\mathbf{a}%
}\mathbf{e}_{\mathbf{a}}\right) \\
&  =\left(  \mathbf{D}_{\mu}^{+}q_{\nu}^{\mathbf{a}}\right)  \mathbf{e}%
_{\mathbf{a}}-\left(  \mathbf{D}_{\nu}^{+}q_{\mu}^{\mathbf{a}}\right)
\mathbf{e}_{\mathbf{a}}, \label{TORSION2}%
\end{align}
where
\begin{equation}
\mathbf{D}_{\mu}^{+}q_{\nu}^{\mathbf{a}}=(\mathbf{D}%
_{{\mbox{\boldmath$\partial$}}_{\mu}}^{+}{\mbox{\boldmath$\partial$}}_{\nu
})^{\mathbf{a}},\mathbf{D}_{\nu}^{+}q_{\mu}^{\mathbf{a}}=(\mathbf{D}%
_{{\mbox{\boldmath$\partial$}}_{\nu}}^{+}{\mbox{\boldmath$\partial$}}_{\mu
})^{\mathbf{a}}. \label{torsion3}%
\end{equation}
and $\left(  \mathbf{D}_{\mu}^{+}q_{\nu}^{\mathbf{a}}\right)  $ and
$(\mathbf{D}_{\nu}^{+}q_{\mu}^{\mathbf{a}})$ are in general non null. Indeed%
\begin{equation}
T_{\mu\nu}^{\mathbf{a}}=\mathbf{D}_{\mu}^{+}q_{\nu}^{\mathbf{a}}%
-\mathbf{D}_{\nu}^{+}q_{\mu}^{\mathbf{a}}, \label{torsion4}%
\end{equation}
and the $T_{\mu\nu}^{\mathbf{a}}\neq0$, as just proved. Now, e.g., in
\cite{gsw}, page 275, where the all the three distinct covariant derivatives
\textbf{$\nabla$}$_{{\partial}_{\mu}}^{-},$\textbf{$\nabla$}$_{{\partial}%
_{\mu}}^{+}$ and \textbf{$\nabla$}$_{{\partial}_{\mu}}$ introduced above \ are
represented by the same symbol $D_{\mu}$ we read: \textquotedblleft The
nonminimimality of a nonminimal spin connection is conveniently measured by
the so-called \ `torsion' $T_{\mu\nu}^{\mathbf{a}}$, defined by
\begin{equation}
T_{\mu\nu}^{\mathbf{a}}=D_{\mu}q_{\nu}^{\mathbf{a}}-D_{\nu}q_{\mu}%
^{\mathbf{a}}." \tag{(12.1.7 gsw)}%
\end{equation}

Now, application of Eq.(12.1.7gsw) to calculate the components of torsion
tensor, for the case of a teleparallel connection, instead of correct
Eq.(\ref{torsion4}) may generate a big confusion if as in \cite{gsw}, authors
adopt the tetrad postulate with the meaning given in
Eq.(\ref{textbook tetrad postulate}). Indeed, observe that if the \ `tetrad
postulate' is adopted then, the torsion tensor results null for a teleparallel
connection $\mathbf{D}$, and this is \textit{false}, as we just showed. The
use of Eq.(12.1.7gsw) may generate confusion also in the case of a Levi-Civita
connection a shown in Appendix A, if we compute compute the components of the
torsion tensor for the case of the structure $(\mathring{S}^{2},g,$%
\textbf{$\nabla$}$)$ using Eq.(\ref{torsion4}) and Eq.(12.1.7gsw).

\begin{remark}
It is very important to have in mind that author of \textbf{\cite{0}}
identified first the symbols $q_{\nu}^{\mathbf{a}}$ \ as the components of
$\theta^{\mathbf{a}}$ in a coordinate basis \ $\{dx^{v}\}$(line 55 in Table 1
of \cite{0} and as the components of of the coordinate basis vectors
$\mathbf{e}_{\nu}={\mbox{\boldmath$\partial$}}_{\nu}$ in the tetrad basis
$\theta^{\mathbf{a}}$ (line 53 in Table 1 of \textbf{\cite{0}}). He never
identified the $q_{\nu}^{\mathbf{a}}$, explicitly \ as the components of the
tensor $\mathbf{Q}$ given by our Eq.(\ref{new error 1}), since such a tensor
did not appear in his text. So, he can never claim that his \ `tetrad
postulate' has any meaning at all, but eventually he will do that after
reading our paper. With the choices given above, he can tell you that he was
just thinking about the tensor $\mathbf{Q}$. So, in Section 7 we shall
identify a crucial mathematical error in \cite{0} that invalidates completely
his supposed unified theory.
\end{remark}

\section{What is $\mathbf{Q}$?}

In the theory of connections\cite{konomi,rrv,coq} we introduce an affine
connection as a connection $\mathbf{\overset{\blacktriangle}{%
\mbox{\boldmath{$\omega$}}%
}}$ on the principal fiber bundle $F(M)$, with canonical projection
$\pi:F(M)\rightarrow M$. As already recalled $\mathbf{\overset{\blacktriangle
}{%
\mbox{\boldmath{$\omega$}}%
}}$ is a 1-form on $F(M)$ with values in the Lie algebra $gl(4,\mathbb{R)}$ of
the general linear group $Gl(4,\mathbb{R)}$. For each $p\in F(M)$ the tangent
space $T_{p}F(M)$ has a canonical decomposition $T_{p}F(M)=H_{p}F(M)\oplus
V_{p}F(M)$. Recall that each $p=(e,\{\left.  \mathbf{e}_{\mathbf{a}%
}\right\vert _{e}\})$, where $\{\left.  \mathbf{e}_{\mathbf{a}}\right\vert
_{e}\}$ is a frame for $e\in M.$ The derivative mapping $\pi_{\ast}$ of $\pi$
sends the tangent bundle $TF(M)$ to $TM$, i.e., $\pi_{\ast}:TF(M)\rightarrow
TM$. To continue, we need to know that if $T_{p}F(M)\ni v=$ $v_{\mathbf{h}%
}+v_{\mathbf{v}}$, $v_{\mathbf{h}}\in H_{p}F(M),$ $v_{\mathbf{v}}\in
V_{p}F(M)$, then $\mathbf{\overset{\blacktriangle}{%
\mbox{\boldmath{$\omega$}}%
}(}v_{h})=0$. Let $\mathbf{V}$ be some vector space and consider objects
$\mathbf{\phi\in}\sec\bigwedge\nolimits^{r}T^{\ast}F(M)\otimes\mathbf{V}$
called $r$-forms on $F(M)$ with values in $\mathbf{V}$. The \textit{exterior
covariant derivative} $\mathbf{D}^{\omega}$ of $\mathbf{\phi}$ is defined by
\begin{equation}
\mathbf{D}^{\omega}\mathbf{\phi(}v_{1},...,v_{r})=d\mathbf{\phi(}%
v_{\mathbf{h}1},...,v_{\mathbf{h}r}\mathbf{),}\label{q1}%
\end{equation}
where $d$ is the ordinary exterior derivative operator.

Take $\mathbf{V=}$\textbf{\ }$gl(4,\mathbb{R)}$ with basis $\{\mathbf{g}%
_{\mathbf{j}}^{\mathbf{i}}\}$, $\mathbf{i,j}=1,2,3,4$. Then $\mathbf{\overset
{\blacktriangle}{%
\mbox{\boldmath{$\omega$}}%
}}=\mathbf{\omega}_{\mathbf{j}}^{\mathbf{i}}\otimes\mathbf{g}_{\mathbf{i}%
}^{\mathbf{j}}$ and the curvature (2-form) of the connection is defined by
\begin{equation}
\overset{\blacktriangle}{\mathbf{\Omega}}=D^{\omega}\mathbf{\overset
{\blacktriangle}{%
\mbox{\boldmath{$\omega$}}%
}}. \label{q2}%
\end{equation}

Let $\mathbf{M}_{p}:T_{e}(M)\rightarrow\mathbb{R}^{4}$ be a mapping that sends
any vector $V\in T_{e}(M)$ into its components with respect to the basis
$\{\left.  \mathbf{e}_{\mathbf{a}}\right\vert _{e}\}$. Then,%

\begin{equation}
\mathbf{M}_{p}(V)=(\theta^{\mathbf{0}}(V),\theta^{\mathbf{1}}(V),\theta
^{\mathbf{2}}(V),\theta^{\mathbf{3}}(V)). \label{q3}%
\end{equation}

Now, take $\mathbf{V}=\mathbb{R}^{4}$, with canonical basis $\{\mathbf{E}%
_{\mathbf{a}}\mathbf{\}}$ consider the object $\overset{\blacktriangle
}{\mbox{\boldmath{$\theta$}}}\in\sec\bigwedge\nolimits^{1}T^{\ast}%
F(M)\otimes\mathbb{R}^{4}$ such that
\begin{equation}
\overset{\blacktriangle}{\mbox{\boldmath{$\theta$}}}(v) = M_{p}\pi_{\ast}(v),
v\in T_{p}F(M) \label{q4}%
\end{equation}
is called the \textit{soldering} form of the manifold. Unfortunately some
authors also call the soldering form, by the name of \textit{tetrad, }which
only serves the purpose of increasing even more the confusion involving the
issue under analysis.

The torsion of the connection $\overset{\blacktriangle}{%
\mbox{\boldmath{$\omega$}}%
}$ is defined as the 2-form
\begin{equation}
\overset{\blacktriangle}{\mathbf{\Theta}}=D^{\omega}\overset{\blacktriangle
}{\mbox{\boldmath{$\theta$}}}. \label{q5}%
\end{equation}
We can show that
\begin{equation}
\overset{\blacktriangle}{\mathbf{\Theta}\text{ }}=d\overset{\blacktriangle}{%
\mbox{\boldmath{$\theta$}}%
}\mathbf{+[\overset{\blacktriangle}{%
\mbox{\boldmath{$\omega$}}%
},\overset{\blacktriangle}{%
\mbox{\boldmath{$\theta$}}%
}],} \label{q6}%
\end{equation}
where $[\ ,\ ]$ denotes the commutator in the Lie algebra $\mathfrak{a}^{4}$
of the affine group \footnote{The symbol $\mathbb{\boxplus}$ means semi-direct
product.} $\mathbb{A}^{4}=Gl(4,\mathbb{R)\boxplus R}^{4}$. A basis of
$\mathfrak{a}^{4}$ is taken as $\{\mathbf{g}_{\mathbf{j}}^{\mathbf{i}%
},\mathbf{E}_{\mathbf{a}}\}$.

Let be $U\subset$ $M$ and $\varsigma:U\rightarrow\varsigma(U)\subset F(M)$. We
are interested in the pullbacks $%
\mbox{\boldmath{$\theta$}}%
=\varsigma^{\ast}$ $\overset{\blacktriangle}{%
\mbox{\boldmath{$\theta$}}%
}$ and $%
\mbox{\boldmath{$\omega$}}%
=\varsigma^{\ast}$ $\overset{\blacktriangle}{%
\mbox{\boldmath{$\omega$}}%
}$, once we give a local trivialization of the respective bundles. Now, $%
\mbox{\boldmath{$\omega$}}%
$ has components $\omega_{\mathbf{b}}^{\mathbf{a}}\in\sec T^{\ast}U$ which are
the connection 1-forms that we already introduced and used above. On the other
hand we can show that \cite{konomi}, chart $\langle x^{\mu}\rangle$ covering
$U$
\begin{equation}%
\mbox{\boldmath{$\theta$}}%
=\varsigma^{\ast}\overset{\blacktriangle}{%
\mbox{\boldmath{$\theta$}}%
}\text{ }=\theta^{\mathbf{a}}\otimes\mathbf{E}_{\mathbf{a}}=q_{\nu
}^{\mathbf{a}}dx^{\nu}\otimes\mathbf{E}_{\mathbf{a}}\in%
{\displaystyle\bigwedge\nolimits^{1}}
T^{\ast}M\otimes\mathbb{R}^{4}\text{.} \label{q7}%
\end{equation}
Now, $%
\mbox{\boldmath{$\theta$}}%
=\varsigma^{\ast}\overset{\blacktriangle}{%
\mbox{\boldmath{$\theta$}}%
}$, i.e., it is the pullback of the soldering form under a local
trivialization of the bundle $T^{\ast}F(M)\otimes\mathbb{R}^{4}$. $%
\mbox{\boldmath{$\theta$}}%
$ is called by some physical authors "\textit{tetrad}". We think that use of
this name is an unfortunate one.

We recall that if we calculate the pullback of the torsion tensor
$\overset{\blacktriangle}{\mathbf{\Theta}}$ we get the tensor $\mathbf{\Theta
}$ in the basis manifold $M$. Explicitly we have (taking into account
Eq.(\ref{q6}) and the fact that the operator $d$ commutes with pullbacks) that%
\begin{equation}
\mathbf{\Theta=(}d\theta^{a}+\omega_{\mathbf{b}}^{\mathbf{a}}\wedge
\theta^{\mathbf{b}}\mathbf{)\otimes E}_{\mathbf{a}}. \label{q8}%
\end{equation}

The objects $T^{\mathbf{a}}=d\theta^{a}+\omega_{\mathbf{b}}^{\mathbf{a}}%
\wedge\theta^{\mathbf{b}}$, $T^{\mathbf{a}}\in%
{\displaystyle\bigwedge\nolimits^{2}}
T^{\ast}M$ are called the torsion $2$-forms.

Now, given an orthonormal basis $\{\mathbf{e}_{\mathbf{a}}\}$ for $TU$ any
vector field $v=v^{\mathbf{a}}\mathbf{e}_{a}\in\sec TU$ we have%
\begin{equation}%
\mbox{\boldmath{$\theta$}}%
(v)=v^{\mathbf{a}}\mathbf{E}_{\mathbf{a}} \label{q9}%
\end{equation}
On the other hand recalling the definition of $\mathbf{Q=e}_{\mathbf{a}%
}\otimes\theta^{\mathbf{a}}\mathbf{\in}\sec%
{\displaystyle\bigwedge\nolimits^{1}}
TM\otimes%
{\displaystyle\bigwedge\nolimits^{1}}
T^{\ast}M,$ we have%
\begin{equation}
\mathbf{Q(}v\mathbf{)=}v^{\mathbf{a}}\mathbf{e}_{\mathbf{a}} \label{q10}%
\end{equation}
and we see that $%
\mbox{\boldmath{$\theta$}}%
$ is a kind of representation of $\mathbf{Q.}$ On the other hand the exterior
covariant derivative, denoted $\mathbf{d}^{\omega}$ of the vector valued
1-form $\mathbf{Q}$ is (see, e.g., ~\cite{frankel}) the \textit{torsion
tensor} of the connection in the basis manifold.
\begin{equation}
\mathbf{T=d}^{\omega}\mathbf{Q:=e_{\mathbf{a}}\otimes(}d\theta^{a}%
+\omega_{\mathbf{b}}^{\mathbf{a}}\wedge\theta^{\mathbf{b}}\mathbf{)}
\label{q11}%
\end{equation}
and we see that $\mathbf{\Theta}$ is a representation of $\mathbf{T}$.

It is very important to keep in mind that for a general Riemann-Cartan
manifold $\mathbf{T=d}^{\omega}\mathbf{Q}\neq0$. However, if $\mathbf{\nabla}$
is the covariant derivative operator acting on the sections of the tensor
bundle, then as we showed above, we have always
\begin{equation}
\mathbf{\nabla Q=(\nabla}_{{\mbox{\boldmath$\partial$}}_{\mu}}%
\mathbf{Q)\otimes}dx^{\nu}\mathbf{=0} \label{q12}%
\end{equation}

\section{ Comments on the\ `Evans Lemma'}

At page 440 of \textbf{\cite{0} \ }no distinction is made between the
connections \textbf{$\nabla$}$_{{\partial}_{\mu}}^{-},$\textbf{$\nabla$%
}$_{{\partial}_{\mu}}^{+}$ and \textbf{$\nabla$}$_{{\partial}_{\mu}}$. As,
e.g., in \cite{carroll}, all these three different connections are represented
by $D_{\mu}$ and the naive tetrad postulate is introduced by the equation
$D_{\mu}q_{\nu}^{\mathbf{a}}=0$, which as just showed is \ misleading if its
precise meaning is not well specified, which is just what happen in \cite{0}.
Then, author of \cite{0} states\textbf{ }that the\ `Evans Lemma' is a direct
consequence of the (\textit{naive)} \ `tetrad postulate'.

We shall now show that author of \cite{0} did a fatal flaw in the derivation
of his \ `Evans lemma'. Indeed, from a true equation, namely, Eq.(40E) that
should more correctly be written%
\begin{equation}
\mathbf{\nabla}_{\mu}^{+}V^{\mu}=\left(  \frac{\partial V^{\mu}}{\partial
x^{\mu}}+V^{\rho}\Gamma_{\mu\rho}^{\mu}\right)  , \tag{40E}%
\end{equation}
where the symbol \textbf{$\nabla$}$_{\mu}^{+}V^{\mu}$ has the precise meaning
discussed above, Remark \ref{dv} he inferred Eq.(41E), i.e., he
wrote:\footnote{That the symbols $\partial_{\mu}$ and $\partial^{\mu}$ used by
Evans are to be interpreted as meaning the basis vector fields
${\mbox{\boldmath$\partial$}}_{\mu}$ and ${\mbox{\boldmath$\partial$}}^{\mu}$
, as itis clear from Evans' Eq.(25E), one of the equations with correct
mathematical meaning in \cite{0}.}
\begin{equation}%
\begin{tabular}
[c]{|c|}\hline
\textbf{$\nabla$}$_{\mu}^{+}{\mbox{\boldmath$\partial$}}^{\mu}%
={\mbox{\boldmath$\partial$}}_{\mu}{\mbox{\boldmath$\partial$}}^{\mu}%
+\Gamma_{\mu\lambda}^{\mu}{\mbox{\boldmath$\partial$}}^{\mu}$\\\hline
\end{tabular}
\ \ \ \ \ \ \ \tag{41E}%
\end{equation}

This equation has \textit{no} mathematical meaning at all. Indeed, if the
symbol \textbf{$\nabla$}$_{\mu}^{+}$ is to have the precise mathematical
meaning disclosed in Section 4, then it can only be applied (with care) to
\textit{components} of vector fields (as, e.g., in Eq. (40E)), and not to
vector fields as it is the explicit case in Eq.(41E). If \textbf{$\nabla$%
}$_{\mu}^{+}$ is to be understood as really having the meaning of
\textbf{$\nabla$}$_{{\mbox{\boldmath$\partial$}}_{\mu}}^{+}$ then Eq. (41E) is
incorrect, because the correct equation in that case is, as recalled in
Eq.(\ref{14}) must be :%
\begin{equation}
\mathbf{\nabla}_{{\mbox{\boldmath$\partial$}}_{\mu}}^{+}%
{\mbox{\boldmath$\partial$}}^{\mu}=\Gamma_{\mu\alpha}^{\mu}%
{\mbox{\boldmath$\partial$}}^{\alpha}. \label{14bis}%
\end{equation}

Now, it is from the completely \textit{wrong} Eq.(41E), that author of
(\cite{0}) infers after a nonsense calculation (that we are not going to show
here) that the tetrad functions $q_{\mu}^{\mathbf{a}}:\varphi(U)\rightarrow
\mathbb{R}$ must satisfy his Eq.(49E), namely the \ `Evans Lemma'%

\begin{equation}%
\begin{tabular}
[c]{|c|}\hline
$\square q_{\mu}^{\mathbf{a}}=Rq_{\mu}^{\mathbf{a}},$\\\hline
\end{tabular}
\ \ \ \ \tag{49E}%
\end{equation}
where the symbol $\square$ in \cite{0} is defined as meaning $\square
={\mbox{\boldmath$\partial$}}_{\mu}{\mbox{\boldmath$\partial$}}^{\mu}$ and
called the D'Alembertian operator\footnote{Of course, in any case it is
\textit{not}, as well known, the covariant D'Alembertian operator on a general
Riemann-Cartan spacetime. \ Indeed, the covariant D'Alembertian operator is
given in Eq.(\ref{1792}a).} and it said that $R$ is the \textit{usual}
curvature scalar.

One more comment is in order. After arriving (illicitly) at Eq.(49E), author
of \cite{0} assumes the validity of Einstein's gravitational
equations\footnote{Einstein equations, by the way, are empirical equations and
have nothing to do with the \textit{foundations} of differential geometry.}
and write his `Evans field equations', which he claims to give an unified
theory of all fields...

That equations are giving by Eq.(2E) and are%

\begin{equation}%
\begin{tabular}
[c]{|c|}\hline
$(\square+\kappa T)q_{\mu}^{\mathbf{a}}=0,$\\\hline
\end{tabular}
\ \ \ \ \ \ \ \ \tag{2E}%
\end{equation}
where $\kappa$ is the gravitational constant and $T$ is the trace of the
energy-momentum tensor. We claim that Eq.(2E) is also \textit{wrong}. Since an
equation somewhat similar to Eq.(2E) appears also\footnote{Reference
\cite{kaniel} has been criticized in \cite{mughe}.}, e.g., in \cite{kaniel} it
is necessary to complete this paper by finding the correct equations satisfied
by the functions $\{q_{\nu}^{\mathbf{a}}\}$, at least for the case of a
Lorentzian spacetime. This will be done below, in two different ways. First we
use the wave equation satisfied by the tensor $\mathbf{Q.}$ Next we find the
correct equations satisfied by the \textit{tetrad fields }$\theta^{\mathbf{a}%
}$ representing a gravitational field in General Relativity, something that
was missing in the papers quoted above. Finally in Section 11 we describe the
Lagrangian formalism for the tetrad fields and derive the field equations from
a variational principle. In order to achieve that last goals we shall need to
introduced some mathematics of the theory of Clifford bundles as developed,
e.g., in \cite{10,14}. See also \cite{16} for details of the Clifford calculus
and some of the `tricks of the trade'.

\begin{remark}
Before leaving this section, we remark that \ since we already showed that the
identity tensor $\mathbf{Q}=q_{\nu}^{\mathbf{a}}\mathbf{e}_{\mathbf{a}}\otimes
dx^{\nu}\in\sec TU\otimes T^{\ast}U$ (Eq.(\ref{about Q}) \ is such that
$\mathbf{\nabla}_{{\mbox{\boldmath$\partial$}}_{\mu}}\mathbf{Q}=0$. It follows
immediately that in any Riemann-Cartan spacetime
\begin{equation}
g^{\nu\mu}\mathbf{\nabla}_{{\mbox{\boldmath$\partial$}}_{\nu}}\mathbf{\nabla
}_{{\mbox{\boldmath$\partial$}}_{\mu}}\mathbf{Q}=0 \label{q rc1}%
\end{equation}

\end{remark}

This can be called a wave equation for $\mathbf{Q=}q_{\nu}^{\mathbf{a}%
}\mathbf{e}_{\mathbf{a}}\otimes dx^{\nu}$, \ but it is indeed a very trivial
result. \ It cannot have any fundamental significance. Indeed, all encoded
differential geometry information is already encoded in the simply equation
$\mathbf{\nabla Q}=0$, which as we already know is an intrinsic writing of the
\ freshman identity (Eq.(\ref{18})) derived above.

\section{\ Clifford Bundles $\mathcal{C\ell}(T^{\ast}M)$ and $\mathcal{C\ell
}(TM)$}

In this section, we restrict ourselves, for simplicity to the case where
$(M,\mathbf{g},$\textbf{$\nabla$}$,\tau_{\mathbf{g}},\uparrow)$ refers to a
\textit{Lorentzian} spacetime as introduced in Section 2\footnote{The general
case of a Riemann-Cartan spacetime will be discussed elsewhere.}. This means
that \textbf{$\nabla$}\footnote{Of course, in what follows the connection
$\mathbf{\nabla}$ has the precise meaning presented in previous sections,but
\ for simplicity of notation, we shall use \ only the symbol $\mathbf{\nabla}%
$, instead of the more precise symbols $\mathbf{\nabla}^{+},\mathbf{\nabla
}^{-},\mathbf{\nabla}$.} is the Levi-Civita connection of $\mathbf{g}$,
i.e.\textit{\/}, \textbf{$\nabla$}$\mathbf{g}=0$, and $\mathbf{T}%
($\textbf{$\nabla$}$)=0$, but in general $\mathbf{R}($\textbf{$\nabla$}%
$)\neq0$. Recall that $\mathbf{R}$ and $\mathbf{T}$ denote respectively the
torsion and curvature tensors. Now, the Clifford bundle of differential forms
$\mathcal{C\ell}(T^{\ast}M)$ is the bundle of algebras $\mathcal{C\ell
}(T^{\ast}M)=\cup_{e\in M}\mathcal{C\ell}(T_{e}^{\ast}M)$, where $\forall e\in
M,\mathcal{C\ell}(T_{e}^{\ast}M)=\mathbb{R}_{1,3}$, the so-called
\emph{spacetime} \emph{algebra }(see, e.g.\emph{, }\cite{16}). Locally as a
linear space over the real field $\mathbb{R}$, $\mathcal{C\ell}(T_{e}^{\ast
}M)$ is isomorphic to the Cartan algebra $%
{\displaystyle\bigwedge}
(T_{e}^{\ast}M)$ of the cotangent space and $%
{\displaystyle\bigwedge}
T_{e}^{\ast}M=\bigoplus_{k=0}^{4}%
{\displaystyle\bigwedge}
{}^{k}T_{e}^{\ast}M$, where $%
{\displaystyle\bigwedge\nolimits^{k}}
T_{e}^{\ast}M$ is the $\binom{4}{k}$-dimensional space of $k$-forms. The
Cartan bundle $%
{\displaystyle\bigwedge}
T^{\ast}M=\cup_{e\in M}%
{\displaystyle\bigwedge}
T_{e}^{\ast}M$ can then be thought \cite{14} as \textquotedblleft
imbedded\textquotedblright\ in $\mathcal{C\ell}(T^{\ast}M)$. In this way
sections of $\mathcal{C\ell}(T^{\ast}M)$ can be represented as a sum of
nonhomogeneous differential forms. Let $\{\mathbf{e}_{\mathbf{a}}\}\in\sec
TM,(\mathbf{a}=0,1,2,3)$ be an orthonormal basis $\mathbf{g}(\mathbf{e}%
_{\mathbf{a}},\mathbf{e}_{\mathbf{b}})=\eta_{\mathbf{ab}}=\mathrm{diag}%
(1,-1,-1,-1)$ and let $\{\theta^{\mathbf{a}}\}\in\sec%
{\displaystyle\bigwedge\nolimits^{1}}
T^{\ast}M\hookrightarrow\sec\mathcal{C\ell}(T^{\ast}M)$ be the dual basis.
Moreover, we denote as in Section 2 by $g$ the metric in the cotangent bundle.

An analogous construction can be done for the tangent space. The corresponding
Clifford bundle is denoted $\mathcal{C\ell}(TM)$ and their sections are called
multivector fields. All formulas presented below for $\mathcal{C\ell}(T^{\ast
}M)$ have corresponding ones in $\mathcal{C\ell}(TM)$.

\begin{remark}
Let \ $\mathbf{V}$ be a real $n$-dimensional vector space equipped with a non
degenerate metric $\mathbf{G:V\times V\rightarrow}\mathbb{R}$ of signature
$(p,q)$, with $n=p+q$. Let $T\mathbf{V}=\bigoplus\nolimits_{r=0}^{\infty
}T^{r,0}\mathbf{V}$\ be the tensor algebra$.$ Let $\mathbf{I\subset
}T\mathbf{V}$ be the bilateral ideal generated by elements of the form
$a\otimes b+b\otimes a$, with $a,b\in\mathbf{V}$. Let $\mathbf{J}$ $\subset
T(\mathbf{V})$ be \ the bilateral \ ideals generated by elements of the form
$a\otimes b+b\otimes a-2\mathbf{G}(a,b)$. Then, we may define
\cite{bourbaki,crumeyrolle} the exterior algebra of \textbf{V }(denoted
$\bigwedge\mathbf{V}$) by the quotient set $T\mathbf{V}/\mathbf{I}$ and the
Clifford algebra of the pair \ $(\mathbf{V,G})$ \ (denoted $\mathbb{R}_{p,q}$)
by $\mathbb{R}_{p,q}=T\mathbf{V}/\mathbf{J}$. \ With these definitions, the
exterior product of $a,b\in\mathbf{V}$ is given by%
\begin{equation}
a\wedge b=\frac{1}{2}\left(  a\otimes b-b\otimes a\right)  .
\end{equation}
and the Clifford product of $a,b\in\mathbf{V}$ (denoted by juxtaposition of
symbols) satisfy the relation
\begin{equation}
ab+ba=2g(a,b),
\end{equation}
Moreover, we have
\begin{equation}
ab=g(a,b)+a\wedge b
\end{equation}
There exists another way for defining the Clifford product and the exterior
product. \ The algebraic structure of the alternative definition is of course,
equivalent to the one given above. However, the components of $p-$forms in a
given basis differ in the two cases. The interested reader may consult
\cite{fmr}.
\end{remark}

\subsection{Clifford product, scalar contraction and exterior products}

The fundamental \emph{Clifford product\footnote{If the reader need more detail
on the Clifford calculus of multivetors he may consult, e.g., \cite{16} and
the list of references therein.}} (in what follows to be denoted by
juxtaposition of symbols) is generated by $\theta^{\mathbf{a}}\theta
^{\mathbf{b}}+\theta^{\mathbf{b}}\theta^{\mathbf{a}}=2\eta^{\mathbf{ab}}$ and
if $\mathcal{C}\in\sec\mathcal{C\ell}(T^{\ast}M)$ we have \cite{14,16}%

\begin{equation}
\mathcal{C}=s+v_{\mathbf{a}}\theta^{\mathbf{a}}+\frac{1}{2!}b_{\mathbf{cd}%
}\theta^{\mathbf{c}}\theta^{\mathbf{d}}+\frac{1}{3!}a_{\mathbf{abc}}%
\theta^{\mathbf{a}}\theta^{\mathbf{b}}\theta^{\mathbf{c}}+p\theta^{\mathbf{5}%
}\;, \label{a.1}%
\end{equation}
where $\theta^{\mathbf{5}}=\theta^{\mathbf{0}}\theta^{\mathbf{1}}%
\theta^{\mathbf{2}}\theta^{\mathbf{3}}$ is the volume element and $s$,
$v_{\mathbf{a}}$, $b_{\mathbf{cd}}$, $a_{\mathbf{abc}}$, $p\in\sec%
{\displaystyle\bigwedge\nolimits^{0}}
T^{\ast}M\subset\sec\mathcal{C\ell}(T^{\ast}M)$.

Let $A_{r},\in\sec%
{\displaystyle\bigwedge\nolimits^{r}}
T^{\ast}M\hookrightarrow\sec\mathcal{C\ell}(T^{\ast}M),B_{s}\in\sec%
{\displaystyle\bigwedge\nolimits^{s}}
T^{\ast}M\hookrightarrow\sec\mathcal{C\ell}(T^{\ast}M)$. For $r=s=1$, we
define the \emph{scalar product} as follows:

For $a,b\in\sec%
{\displaystyle\bigwedge\nolimits^{1}}
T^{\ast}M\hookrightarrow\sec\mathcal{C\ell}(T^{\ast}M),$%
\begin{equation}
a\cdot b=\frac{1}{2}(ab+ba)=g(a,b). \label{a.2}%
\end{equation}
We also define the \emph{exterior product} ($\forall r,s=0,1,2,3)$ by
\begin{align}
A_{r}\wedge B_{s}  &  =\langle A_{r}B_{s}\rangle_{r+s},\nonumber\\
A_{r}\wedge B_{s}  &  =(-1)^{rs}B_{s}\wedge A_{r} \label{a.3}%
\end{align}
where $\langle\rangle_{k}$ is the component in the subspace $%
{\displaystyle\bigwedge\nolimits^{k}}
T^{\ast}M$ of the Clifford field. The exterior product is extended by
linearity to all sections of $\mathcal{C}\ell(T^{\ast}M).$

For $A_{r}=a_{1}\wedge...\wedge a_{r},B_{r}=b_{1}\wedge...\wedge b_{r}$, the
\textit{scalar product} is defined as
\begin{align}
A_{r}\cdot B_{r}  &  =(a_{1}\wedge...\wedge a_{r})\cdot(b_{1}\wedge...\wedge
b_{r})\nonumber\\
&  =\det\left[
\begin{array}
[c]{ccc}%
a_{1}\cdot b_{1} & \ldots & a_{1}\cdot b_{k}\\
\ldots & \ldots & \ldots\\
a_{k}\cdot b_{1} & \ldots & a_{k}\cdot b_{k}%
\end{array}
\right]  . \label{a.4}%
\end{align}

We agree that if $r=s=0$, the scalar product is simple the ordinary product in
the real field.

Also, if $r\neq s,$ $A_{r}\cdot B_{s}=0$ .

For $r\leq s,A_{r}=a_{1}\wedge...\wedge a_{r},B_{s}=b_{1}\wedge...\wedge
b_{s\text{ }}$we define the \textit{left contraction} by
\begin{equation}
\lrcorner:(A_{r},B_{s})\mapsto A_{r}\lrcorner B_{s}=%
{\displaystyle\sum\limits_{i_{1}<...<i_{r}}}
\epsilon_{1......s}^{i_{1}.....i_{s}}(a_{1}\wedge...\wedge a_{r}%
)\cdot(b_{i_{1}}\wedge...\wedge b_{i_{r}})^{\sim}b_{i_{r}+1}\wedge...\wedge
b_{i_{s}}, \label{a.5}%
\end{equation}
\ where $\sim$ denotes the reverse mapping (\emph{reversion})
\begin{equation}
\sim:\sec%
{\displaystyle\bigwedge\nolimits^{p}}
T^{\ast}M\ni a_{1}\wedge...\wedge a_{p}\mapsto a_{p}\wedge...\wedge a_{1},
\label{a.6}%
\end{equation}
and extended by linearity to all sections of $\mathcal{C\ell}(T^{\ast}M)$. We
agree that for $\alpha,\beta\in\sec%
{\displaystyle\bigwedge\nolimits^{0}}
T^{\ast}M$ the contraction is the ordinary (pointwise) product in the real
field and that if $\alpha\in\sec%
{\displaystyle\bigwedge\nolimits^{0}}
T^{\ast}M$, $\ A_{r}\in\sec%
{\displaystyle\bigwedge\nolimits^{r}}
T^{\ast}M,B_{s}\in\sec%
{\displaystyle\bigwedge\nolimits^{s}}
T^{\ast}M$ then $(\alpha A_{r})\lrcorner B_{s}=A_{r}\lrcorner(\alpha B_{s})$.
Left contraction is extended by linearity to all pairs of elements of sections
of $\mathcal{C\ell}(T^{\ast}M)$, i.e., for $A,B\in\sec\mathcal{C\ell}(T^{\ast
}M)$%

\begin{equation}
A\lrcorner B=\sum_{r,s}\langle A\rangle_{r}\lrcorner\langle B\rangle
_{s},\text{ }r\leq s. \label{a.7}%
\end{equation}

It is also necessary to introduce in $\mathcal{C\ell}(T^{\ast}M)$ the operator
of \emph{right contraction} denoted by $\llcorner$. The definition is obtained
from the one presenting the left contraction with the imposition that $r\geq
s$ and taking into account that now if $A_{r},\in\sec%
{\displaystyle\bigwedge\nolimits^{r}}
T^{\ast}M,B_{s}\in\sec%
{\displaystyle\bigwedge\nolimits^{s}}
T^{\ast}M$ then $A_{r}\llcorner(\alpha B_{s})=(\alpha A_{r})\llcorner B_{s}$.
Finally, note that%

\begin{equation}
A_{r}\lrcorner B_{r}=A_{r}\llcorner B_{r}=\tilde{A}_{r}\cdot B_{r}=A_{r}%
\cdot\tilde{B}_{r} \label{cont pscal}%
\end{equation}

\subsection{Some useful formulas}

The main formulas used in the Clifford calculus in the main text can be
obtained from the following ones, where $a\in\sec%
{\displaystyle\bigwedge\nolimits^{1}}
T^{\ast}M$ and $A_{r}\in\sec%
{\displaystyle\bigwedge\nolimits^{r}}
T^{\ast}M,B_{s}\in\sec%
{\displaystyle\bigwedge\nolimits^{s}}
T^{\ast}M$:
\begin{align}
aB_{s}  &  =a\lrcorner B_{s}+a\wedge B_{s},B_{s}a=B_{s}\llcorner a+B_{s}\wedge
a,\label{a.8}\\
a\lrcorner B_{s}  &  =\frac{1}{2}(aB_{s}-(-)^{s}B_{s}a),\nonumber\\
A_{r}\lrcorner B_{s}  &  =(-)^{r(s-1)}B_{s}\llcorner A_{r},\nonumber\\
a\wedge B_{s}  &  =\frac{1}{2}(aB_{s}+(-)^{s}B_{s}a),\nonumber\\
A_{r}B_{s}  &  =\langle A_{r}B_{s}\rangle_{|r-s|}+\langle A_{r}\lrcorner
B_{s}\rangle_{|r-s-2|}+...+\langle A_{r}B_{s}\rangle_{|r+s|}\nonumber\\
&  =\sum\limits_{k=0}^{m}\langle A_{r}B_{s}\rangle_{|r-s|+2k},\text{ }%
m=\frac{1}{2}(r+s-|r-s|). \label{1.201}%
\end{align}

\subsection{Hodge star operator}

Let $\star$ be the usual Hodge star operator $\star:%
{\displaystyle\bigwedge\nolimits^{k}}
T^{\ast}M\rightarrow%
{\displaystyle\bigwedge\nolimits^{4-k}}
T^{\ast}M$. If $B\in\sec%
{\displaystyle\bigwedge\nolimits^{k}}
T^{\ast}M$, $A\in\sec%
{\displaystyle\bigwedge\nolimits^{4-k}}
T^{\ast}M$ and $\tau\in\sec%
{\displaystyle\bigwedge\nolimits^{4}}
T^{\ast}M$ is the volume form, then $\star B$ is defined by
\[
A\wedge\star B=(A\cdot B)\tau.
\]

Then we can show that if $A_{p}\in\sec%
{\displaystyle\bigwedge\nolimits^{p}}
T^{\ast}M\hookrightarrow\sec\mathcal{C\!\ell}(T^{\ast}M)$ we have
\begin{equation}
\star A_{p}=\widetilde{A_{p}}\theta^{\mathbf{5}}. \label{a.hodge}%
\end{equation}
This equation is enough to prove very easily the following identities (which
are used below):%
\begin{align}
A_{r}\wedge\star B_{s}  &  =B_{s}\wedge\star A_{r};\hspace{0.15in}%
r=s,\nonumber\\
A_{r}\lrcorner\star B_{s}  &  =B_{s}\lrcorner\star A_{r};\hspace
{0.15in}r+s=4,\nonumber\\
A_{r}\wedge\star B_{s}  &  =(-1)^{r(s-1)}\star(\tilde{A}_{r}\lrcorner
B_{s});\hspace{0.15in}r\leq s,\nonumber\\
A_{r}\lrcorner\star B_{s}  &  =(-1)^{rs}\star(\tilde{A}_{r}\wedge
B_{s});\hspace{0.15in}r+s\leq4 \label{Aidentities}%
\end{align}

Let $d$ and $\delta$ be respectively the differential and Hodge codifferential
operators acting on sections of $%
{\displaystyle\bigwedge}
T^{\ast}M$. If $\omega_{p}\in\sec%
{\displaystyle\bigwedge\nolimits^{p}}
T^{\ast}M\hookrightarrow\sec\mathcal{C\ell}(T^{\ast}M)$, then $\delta
\omega_{p}=(-)^{p}\star^{-1}d\star\omega_{p}$, where $\star^{-1}%
\star=\mathrm{identity}$. When applied to a $p$-form we have%
\[
\star^{-1}=(-1)^{p(4-p)+1}\star\hspace{0.15in}.
\]

\subsection{Action of \textbf{$\nabla$}$_{\mathbf{e}_{\mathbf{a}}}$ on
Sections of $\mathcal{C\!\ell}(TM)$ and $\mathcal{C\!\ell}(T^{\ast}M)$}

Let \textbf{$\nabla$}$_{\mathbf{e}_{\mathbf{a}}}$ be a metrical compatible
covariant derivative operator acting on sections of the tensor bundle. It can
be easily shown (see, e.g., \cite{10}) that \textbf{$\nabla$}$_{\mathbf{e}%
_{\mathbf{a}}}$ is also a covariant derivative operator on the Clifford
bundles $\mathcal{C\!\ell}(TM)$ and $\mathcal{C\!\ell}(T^{\ast}M).$

Now, if \ $A_{p}\in\sec%
{\displaystyle\bigwedge\nolimits^{p}}
T^{\ast}M\hookrightarrow\sec\mathcal{C\!\ell}(M)$ \ we can show, very easily
by explicitly performing the calculations\footnote{A derivation of this
formula from the general theory of connections can be found in \cite{14}.}
that%
\begin{equation}
\mathbf{\nabla}_{\mathbf{e}_{\mathbf{a}}}A_{p}=\mathbf{e}_{\mathbf{a}}%
(A_{p})+\frac{1}{2}[\omega_{\mathbf{e}_{\mathbf{a}}},A_{p}], \label{der1}%
\end{equation}
where the $\omega_{\mathbf{e}_{\mathbf{a}}}\in\sec%
{\displaystyle\bigwedge\nolimits^{2}}
T^{\ast}M\hookrightarrow\sec\mathcal{C\!\ell}(M)$ may be called
\textit{Clifford} \textit{connection 2-forms. }They\textit{ }are given
by:\textit{ }%
\begin{equation}
\omega_{\mathbf{e}_{\mathbf{a}}}=\frac{1}{2}\omega_{\mathbf{a}}^{\mathbf{bc}%
}\theta_{\mathbf{b}}\theta_{\mathbf{c}}=\frac{1}{2}\omega_{\mathbf{bac}}%
\theta^{\mathbf{b}}\theta^{\mathbf{c}}=\frac{1}{2}\omega_{\mathbf{a}%
}^{\mathbf{bc}}\theta_{\mathbf{b}}\wedge\theta_{\mathbf{c}}, \label{der2}%
\end{equation}
where we use the (simplified) notation%
\begin{equation}
\mathbf{\nabla}_{\mathbf{e}_{\mathbf{a}}}\theta_{\mathbf{b}}=\omega
_{\mathbf{ab}}^{\mathbf{c}}\theta_{\mathbf{c}},\hspace{0.15in}\mathbf{\nabla
}_{\mathbf{e}_{\mathbf{a}}}\theta^{\mathbf{b}}=-\omega_{\mathbf{ac}%
}^{\mathbf{b}}\theta^{\mathbf{c}},\hspace{0.15cm}\omega_{\mathbf{a}%
}^{\mathbf{bc}}=-\omega_{\mathbf{a}}^{\mathbf{cb}} \label{der3}%
\end{equation}

\subsection{Dirac Operator, Differential and Codifferential}

\begin{definition}
The Dirac (like) operator acting on sections of $\mathcal{C\!\ell}(T^{\ast}M)$
is the invariant first order differential operator
\begin{equation}
{%
\mbox{\boldmath$\partial$}%
}=%
\mbox{\boldmath{$\theta$}}%
^{\mathbf{a}}\mathbf{\nabla}_{\mathbf{e}_{\mathbf{a}}}. \label{1.5}%
\end{equation}

\end{definition}

We can show (see, e.g., \cite{17}) that when \textbf{$\nabla$}$_{\mathbf{e}%
_{\mathbf{a}}}$ is the Levi-Civita covariant derivative operator (as assumed
here), the following important result holds:
\begin{equation}
{%
\mbox{\boldmath$\partial$}%
}={%
\mbox{\boldmath$\partial$}%
}\wedge\,+\,{%
\mbox{\boldmath$\partial$}%
}\lrcorner=d-\delta. \label{1.6}%
\end{equation}

\begin{definition}
The square of the Dirac operator $\,{%
\mbox{\boldmath$\partial$}%
}^{2}$ is called Hodge Laplacian.
\end{definition}

Some useful identities are:%
\begin{align}
dd  &  =\delta\delta=0,\nonumber\\
d{%
\mbox{\boldmath$\partial$}%
}^{2}  &  ={%
\mbox{\boldmath$\partial$}%
}^{2}d;\hspace{0.15in}\delta{%
\mbox{\boldmath$\partial$}%
}^{2}={%
\mbox{\boldmath$\partial$}%
}^{2}\delta,\nonumber\\
\delta\star &  =(-1)^{p+1}\star d;\hspace{0.15in}\star\delta=(-1)^{p}\star
d,\nonumber\\
d\delta\star &  =\star d\delta;\hspace{0.15in}\star d\delta=\delta
d\star;\hspace{0.15in}\star{%
\mbox{\boldmath$\partial$}%
}^{2}={%
\mbox{\boldmath$\partial$}%
}^{2}\star\label{A.identities1}%
\end{align}

\subsection{Covariant D'Alembertian, Ricci and Einstein Operators}

In this section we study in details the Hodge Laplacian and its decomposition
in the covariant D'Alembertian operator and the very important Ricci operator,
which do not have analogous in the standard presentation of differential
geometry in the Cartan and Hodge bundles, as given e.g., in \cite{choquet} .

Remembering that ${%
\mbox{\boldmath$\partial$}%
}=\theta^{\alpha}\,$\textbf{$\nabla$}$_{\mathbf{e}_{\alpha}}$, where
$\{\mathbf{e}_{\alpha}\}\in F(M)$ is an \textit{arbitrary} frame\footnote{This
means that it can be a cordiante basis or an orthonormal basis.} and
$\{\theta^{\alpha}\}$ its dual frame on the manifold $M$ and \textbf{$\nabla$}
is the Levi-Civita connection of the metric $\mathbf{g}$, such that
\begin{equation}
\mathbf{\nabla}_{\mathbf{e}_{\alpha}}\mathbf{e}_{\beta}=\mathbf{\gamma
}_{\alpha\beta}^{\mu}\mathbf{e}_{\mu},\text{ }\mathbf{\nabla}_{\mathbf{e}%
_{\alpha}}\theta^{\beta}=-\mathbf{\gamma}_{\alpha\mu}^{\beta}\theta^{\mu}%
\end{equation}
we have:
\begin{align}
{%
\mbox{\boldmath$\partial$}%
}^{2}  &  =(\theta^{\alpha}\mathbf{\nabla}_{\mathbf{e}_{\alpha}}%
)(\theta^{\beta}\mathbf{\nabla}_{\mathbf{e}_{\beta}})=\theta^{\alpha}%
(\theta^{\beta}\mathbf{\nabla}_{\mathbf{e}_{\alpha}}\mathbf{\nabla
}_{\mathbf{e}_{\beta}}+(\mathbf{\nabla}_{\mathbf{e}_{\alpha}}\theta^{\beta
})\mathbf{\nabla}_{\mathbf{e}_{\beta}})\nonumber\\
&  =g^{\alpha\beta}(\mathbf{\nabla}_{\mathbf{e}_{\alpha}}\mathbf{\nabla
}_{\mathbf{e}_{\beta}}-\mathbf{\gamma}_{\alpha\beta}^{\rho}\mathbf{\nabla
}_{\mathbf{e}_{\rho}})+\theta^{\alpha}\wedge\theta^{\beta}(\mathbf{\nabla
}_{\mathbf{e}_{\alpha}}\mathbf{\nabla}_{\mathbf{e}_{\beta}}-\mathbf{\gamma
}_{\alpha\beta}^{\rho}\mathbf{\nabla}_{\mathbf{e}_{\rho}}).
\end{align}
Next we introduce the operators:
\begin{equation}%
\begin{tabular}
[c]{c}%
(a)\\
(b)
\end{tabular}%
\begin{array}
[c]{ccl}%
\blacksquare={%
\mbox{\boldmath$\partial$}%
}\cdot{%
\mbox{\boldmath$\partial$}%
} & = & g^{\alpha\beta}(\mathbf{\nabla}_{\mathbf{e}_{\alpha}}\mathbf{\nabla
}_{\mathbf{e}_{\beta}}-\mathbf{\gamma}_{\alpha\beta}^{\rho}\mathbf{\nabla
}_{\mathbf{e}_{\rho}})\\
{%
\mbox{\boldmath$\partial$}%
}\wedge{%
\mbox{\boldmath$\partial$}%
} & = & \theta^{\alpha}\wedge\theta^{\beta}(\mathbf{\nabla}_{\mathbf{e}%
_{\alpha}}\mathbf{\nabla}_{\mathbf{e}_{\beta}}-\gamma_{\alpha\beta}^{\rho
}\mathbf{\nabla}_{\mathbf{e}_{\rho}}),
\end{array}
\label{1792}%
\end{equation}

\begin{definition}
We call $\blacksquare={%
\mbox{\boldmath$\partial$}%
}\cdot{%
\mbox{\boldmath$\partial$}%
}$ the covariant D'Alembertian operator and ${%
\mbox{\boldmath$\partial$}%
}\wedge{%
\mbox{\boldmath$\partial$}%
}$ the Ricci operator.
\end{definition}

The reason for the above names will become obvious through propositions
\ref{cov dalembert propos} and \ref{ricci proposition}.

Note that we can write:
\begin{equation}
{%
\mbox{\boldmath$\partial$}%
}^{2}={%
\mbox{\boldmath$\partial$}%
}\cdot{%
\mbox{\boldmath$\partial$}%
}+{%
\mbox{\boldmath$\partial$}%
}\wedge{%
\mbox{\boldmath$\partial$}%
} \label{1796}%
\end{equation}
or,%
\begin{align}
{%
\mbox{\boldmath$\partial$}%
}^{2}  &  =({%
\mbox{\boldmath$\partial$}%
}\lrcorner+{%
\mbox{\boldmath$\partial$}%
}\wedge)({%
\mbox{\boldmath$\partial$}%
}\lrcorner+{%
\mbox{\boldmath$\partial$}%
}\wedge)\nonumber\\
&  ={%
\mbox{\boldmath$\partial$}%
}\cdot{%
\mbox{\boldmath$\partial$}%
}\wedge+{%
\mbox{\boldmath$\partial$}%
}\wedge{%
\mbox{\boldmath$\partial$}%
}\lrcorner\label{1796a}\\
&  =-(d\delta+\delta d) \label{1796b}%
\end{align}

Before proceeding, let us calculate the commutator $\left[  \theta_{\alpha
},\theta_{\beta}\right]  $ and anticommutator $\{\theta_{\alpha},\theta
_{\beta}\}$. We have immediately%
\begin{equation}
\left[  \theta_{\alpha},\theta_{\beta}\right]  =c_{\alpha\beta}^{\rho}%
\theta_{\rho}, \label{726a}%
\end{equation}
where $c_{\alpha\beta}^{\rho}$ are the structure coefficients (see, e.g.,
\cite{choquet}) of the basis $\{\mathbf{e}_{\alpha}\}$, i.e., $\left[
\mathbf{e}_{\alpha},\mathbf{e}_{\beta}\right]  =c_{\alpha\beta}^{\rho
}\mathbf{e}_{\rho}.$

Also,
\begin{align}
\{\theta_{\alpha},\theta_{\beta}\}  &  =\mathbf{\nabla}_{\mathbf{e}_{\alpha}%
}\theta_{\beta}+\mathbf{\nabla}_{\mathbf{e}_{\beta}}\theta_{\alpha
},\nonumber\\
&  =(\mathbf{\gamma}_{\alpha\beta}^{\rho}+\mathbf{\gamma}_{\beta\alpha}^{\rho
})\theta_{\rho}\nonumber\\
&  =b_{\alpha\beta}^{\rho}\theta_{\rho}, \label{726}%
\end{align}
Eq.(\ref{726}) defines the coefficients $b_{\alpha\beta}^{\rho}$ which have a
very interesting geometrical meaning as discussed in \cite{soro}.

\begin{proposition}
The covariant D'Alembertian ${%
\mbox{\boldmath$\partial$}%
}\cdot{%
\mbox{\boldmath$\partial$}%
}$ operator can be written as:%
\begin{equation}
{%
\mbox{\boldmath$\partial$}%
}\cdot{%
\mbox{\boldmath$\partial$}%
}=\frac{1}{2}g^{\alpha\beta}\left[  \mathbf{\nabla}_{\mathbf{e}_{\alpha}%
}\mathbf{\nabla}_{\mathbf{e}_{\beta}}+\mathbf{\nabla}_{\mathbf{e}_{\beta}%
}\mathbf{\nabla}_{\mathbf{e}_{\alpha}}-b_{\alpha\beta}^{\rho}\mathbf{\nabla
}_{\mathbf{e}_{\rho}}\right]  . \label{sq1}%
\end{equation}

\end{proposition}

\begin{proof}
It is a simple computation left to the reader.
\end{proof}

\begin{proposition}
\label{cov dalembert propos}For every $r$-form field $\omega\in\sec
\bigwedge\nolimits^{r}M$, $\omega=\frac{1}{r!}\omega_{\alpha_{1}\ldots
\alpha_{r}}\theta^{\alpha_{1}}\wedge\ldots\wedge\theta^{\alpha_{r}}$, we
have:
\begin{equation}
({%
\mbox{\boldmath$\partial$}%
}\cdot{%
\mbox{\boldmath$\partial$}%
})\omega=\frac{1}{r!}g^{\alpha\beta}\mathbf{\nabla}_{\alpha}\mathbf{\nabla
}_{\beta}\omega_{\alpha_{1}\ldots\alpha_{r}}\theta^{\alpha_{1}}\wedge
\ldots\wedge\theta^{\alpha_{r}}, \label{1831}%
\end{equation}
where\/ \textbf{$\nabla$}$_{\alpha}$\textbf{$\nabla$}$_{\beta}\omega
_{\alpha_{1}\ldots\alpha_{r}}$ is to be calculate with the standard rule for
writing the covariant derivative of the components of a covector field.
\end{proposition}

\begin{proof}
We have \textbf{$\nabla$}$_{\mathbf{e}_{\beta}}\omega=\frac{1}{r!}%
$\textbf{$\nabla$}$_{\beta}\omega_{\alpha_{1}\ldots\alpha_{r}}\theta
^{\alpha_{1}}\wedge\ldots\wedge\theta^{\alpha_{r}}$, with \textbf{$\nabla$%
}$_{\beta}\omega_{\alpha_{1}\ldots\alpha_{r}}=(\mathbf{e}_{\beta}%
(\omega_{\alpha_{1}\ldots\alpha_{r}})-\mathbf{\gamma}_{\beta\alpha_{1}%
}^{\sigma}\omega_{\sigma\alpha_{2}\ldots\alpha_{r}}-\cdots-\mathbf{\gamma
}_{\beta\alpha_{r}}^{\sigma}\omega_{\alpha_{1}\ldots\alpha_{r-1}\sigma})$.
Therefore,
\begin{align*}
\mathbf{\nabla}_{\mathbf{e}_{\alpha}}\mathbf{\nabla}_{\mathbf{e}_{\beta}%
}\omega &  =\frac{1}{r!}(\mathbf{e}_{\alpha}(\mathbf{\nabla}_{\beta}%
\omega_{\alpha_{1}\ldots\alpha_{r}})-\mathbf{\gamma}_{\alpha\alpha_{1}%
}^{\sigma}\mathbf{\nabla}_{\beta}\omega_{\sigma\alpha_{2}\ldots\alpha_{r}%
}-\cdots\\
&  -\mathbf{\gamma}_{\alpha\alpha_{r}}^{\sigma}\mathbf{\nabla}_{\beta}%
\omega_{\alpha_{1}\ldots\alpha_{r-1}\sigma})\theta^{\alpha_{1}}\wedge
\ldots\wedge\theta^{\alpha_{r}}%
\end{align*}
and we conclude that:
\[
(\mathbf{\nabla}_{\mathbf{e}_{\alpha}}\mathbf{\nabla}_{\mathbf{e}_{\beta}%
}-\mathbf{\gamma}_{\alpha\beta}^{\rho}\mathbf{\nabla}_{e_{\rho}})\omega
=\frac{1}{r!}\mathbf{\nabla}_{\alpha}\mathbf{\nabla}_{\beta}\omega_{\alpha
_{1}\ldots\alpha_{r}}\theta^{\alpha_{1}}\wedge\ldots\wedge\theta^{\alpha_{r}%
}.
\]
Finally, multiplying this equation by $g^{\alpha\beta}$ and using the
Eq.(\ref{1792}a), we get the Eq.(\ref{1831}).\hfill\medskip
\end{proof}

The Ricci operator ${%
\mbox{\boldmath$\partial$}%
}\wedge{%
\mbox{\boldmath$\partial$}%
}$ can be written as:
\begin{equation}
{%
\mbox{\boldmath$\partial$}%
}\wedge{%
\mbox{\boldmath$\partial$}%
}=\frac{1}{2}\theta^{\alpha}\wedge\theta^{\beta}\left[  \mathbf{\nabla
}_{\mathbf{e}_{\alpha}}\mathbf{\nabla}_{\mathbf{e}_{\beta}}-\mathbf{\nabla
}_{\mathbf{e}_{\beta}}\mathbf{\nabla}_{\mathbf{e}_{\alpha}}-c_{\alpha\beta
}^{\rho}\mathbf{\nabla}_{\mathbf{e}_{\rho}}\right]  .
\end{equation}

\begin{proof}
It is a trivial exercise, left to the reader.
\end{proof}

Applying this operator to the 1-forms of the frame $\{\theta^{\mu}\}$, we
get:
\begin{equation}
({%
\mbox{\boldmath$\partial$}%
}\wedge{%
\mbox{\boldmath$\partial$}%
})\theta^{\mu}=-\frac{1}{2}R_{\rho}{}^{\mu}{}_{\alpha\beta}(\theta^{\alpha
}\wedge\theta^{\beta})\theta^{\rho}=-\mathcal{R}_{\rho}^{\mu}\theta^{\rho},
\end{equation}
where $R_{\rho}{}^{\mu}{}_{\alpha\beta}$ are the components of the Riemann
curvature tensor of the connection \textbf{$\nabla$}. We can write using the
first line in Eq.(\ref{a.8})
\begin{equation}
\mathcal{R}_{\rho}^{\mu}\theta^{\rho}=\mathcal{R}_{\rho}^{\mu}\llcorner
\theta^{\rho}+\mathcal{R}_{\rho}^{\mu}\wedge\theta^{\rho}.
\end{equation}
The second term in the r.h.s. of this equation is identically null because of
the Bianchi identity satisfied by the Riemann curvature tensor, as can be
easily verified. That result can be encoded in the equation:%
\begin{equation}
({%
\mbox{\boldmath$\partial$}%
}\wedge{%
\mbox{\boldmath$\partial$}%
)\wedge}\theta^{\mu}=0,
\end{equation}

For the term $\mathcal{R}_{\rho}^{\mu}\llcorner\theta^{\rho}$ we have (using
Eq.(\ref{a.5}) and~the third line in Eq.(\ref{a.8})):
\begin{align}
\mathcal{R}_{\rho}^{\mu}\llcorner\theta^{\rho}  &  =\frac{1}{2}R_{\rho}{}%
^{\mu}{}_{\alpha\beta}(\theta^{\alpha}\wedge\theta^{\beta})\llcorner
\theta^{\rho}\nonumber\\
&  =-\frac{1}{2}R_{\rho}{}^{\mu}{}_{\alpha\beta}(g^{\rho\alpha}\theta^{\beta
}-g^{\rho\beta}\theta^{\alpha})\nonumber\\
&  =-\mathring{g}^{\rho\alpha}R_{\rho}{}^{\mu}{}_{\alpha\beta}\theta^{\beta
}=-R_{\beta}^{\mu}\theta^{\beta},
\end{align}
where $R_{\beta}^{\mu}$ are the components of the Ricci tensor of the
Levi-Civita connection \ \textbf{$\nabla$} of $\mathbf{g}$. The above results
can be put in the form of the following

\begin{proposition}
\label{ricci proposition}
\begin{equation}
({%
\mbox{\boldmath$\partial$}%
}\wedge{%
\mbox{\boldmath$\partial$}%
})\theta^{\mu}=\mathcal{R}^{\mu}, \label{ricci equation}%
\end{equation}
where $\mathcal{R}^{\mu}=R_{\beta}^{\mu}\theta^{\beta}$ are the Ricci 1-forms
of the manifold.
\end{proposition}

The next proposition shows that the Ricci operator can be written in a purely
algebraic way:

\begin{proposition}
\label{ricci propos}The Ricci operator ${%
\mbox{\boldmath$\partial$}%
}\wedge{%
\mbox{\boldmath$\partial$}%
}$ satisfies the relation\textsl{: }
\begin{equation}
{%
\mbox{\boldmath$\partial$}%
}\wedge{%
\mbox{\boldmath$\partial$}%
}=\mathcal{R}^{\sigma}\wedge\theta_{\sigma}\lrcorner+\mathcal{R}^{\rho\sigma
}\wedge\theta_{\rho}\lrcorner\theta_{\sigma}\lrcorner, \label{1918}%
\end{equation}
where $\mathcal{R}^{\rho\sigma}=g^{\rho\mu}\mathcal{R}_{\mu}^{\sigma}=\frac
{1}{2}R^{\rho\sigma}{}_{\alpha\beta}\theta^{\alpha}\wedge\theta^{\beta}$ are
the curvature $2$-forms.
\end{proposition}

\begin{proof}
The Hodge Laplacian of an arbitrary $r$-form field $\omega=\frac{1}{r!}%
\omega_{\alpha_{1}\ldots\alpha_{r}}\theta^{\alpha_{1}}\wedge\ldots\wedge
\theta^{\alpha_{r}}$ is given by: (e.g., \cite{choquet}---recall that our
\textit{definition} differs by a sign from that given there) ${%
\mbox{\boldmath$\partial$}%
}^{2}\omega=\frac{1}{r!}({%
\mbox{\boldmath$\partial$}%
}^{2}\omega)_{\alpha_{1}\ldots\alpha_{r}}\theta^{\alpha_{1}}\wedge\ldots
\wedge\theta^{\alpha_{r}}$, with:
\begin{align}
({%
\mbox{\boldmath$\partial$}%
}^{2}\omega)_{\alpha_{1}\ldots\alpha_{r}}  &  =g^{\alpha\beta}\mathbf{\nabla
}_{\alpha}\mathbf{\nabla}_{\beta}\omega_{\alpha_{1}\ldots\alpha_{r}%
}\nonumber\\
&  -\sum_{p}(-1)^{p}R_{\alpha_{p}}^{\sigma}\omega_{\sigma\alpha_{1}%
\ldots\check{\alpha}_{p}\ldots\alpha_{r}}\nonumber\\
&  -2\sum_{%
\genfrac{}{}{0pt}{}{{p,q}}{{p<q}}%
}(-1)^{p+q}R^{\rho}{}_{\alpha_{q}}{}^{\sigma}{}_{\alpha_{p}}\omega_{\rho
\sigma\alpha_{1}\ldots\check{\alpha}_{p}\ldots\check{\alpha}_{q}\ldots
\alpha_{r}}, \label{hodge laplacian}%
\end{align}
where the notation $\check{\alpha}$ means that the index $\alpha$ was excluded
of the sequence.

The first term in the r.h.s. of this expression are the components of the
covariant D'Alembertian of the field $\omega$. Then,
\[
\mathcal{R}^{\sigma}\wedge\theta_{\sigma}\lrcorner\omega=-\frac{1}{r!}\left[
\sum_{p}(-1)^{p}R_{\alpha_{p}}^{\sigma}\omega_{\sigma\alpha_{1}\ldots
\check{\alpha}_{p}\ldots\alpha_{r}}\right]  \theta^{\alpha_{1}}\wedge
\ldots\wedge\theta^{\alpha_{r}}%
\]
and also,
\[
\mathcal{R}^{\rho\sigma}\wedge\theta_{\rho}\lrcorner\theta_{\sigma}%
\lrcorner\omega=-\frac{2}{r!}\left[  \sum_{%
\genfrac{}{}{0pt}{}{{p,q}}{{p<q}}%
}(-1)^{p+q}R^{\rho}{}_{\alpha_{q}}{}^{\sigma}{}_{\alpha_{p}}\omega_{\rho
\sigma\alpha_{1}\ldots\check{\alpha}_{p}\ldots\check{\alpha}_{q}\ldots
\alpha_{r}}\right]  \theta^{\alpha_{1}}\wedge\ldots\wedge\theta^{\alpha_{r}}.
\]

Hence, taking into account Eq.(\ref{1796}), we conclude that:
\begin{equation}
({%
\mbox{\boldmath$\partial$}%
}\wedge{%
\mbox{\boldmath$\partial$}%
})\omega=\mathcal{R}^{\sigma}\wedge\theta_{\sigma}\lrcorner\omega
+\mathcal{R}^{\rho\sigma}\wedge\theta_{\rho}\lrcorner\theta_{\sigma}%
\lrcorner\omega, \label{ricci bis}%
\end{equation}
for every $r$-form field $\omega$.\medskip\hfill
\end{proof}

Observe that applying the operator given by the second term in the r.h.s. of
Eq.(\ref{1918}) to the dual of the 1-forms $\theta^{\mu}$, we get:
\begin{align}
\mathcal{R}^{\rho\sigma}\wedge\theta_{\rho}\lrcorner\theta_{\sigma}%
\lrcorner\star\theta^{\mu}  &  =\mathcal{R}_{\rho\sigma}\wedge(\theta^{\rho
}\lrcorner(\theta^{\sigma}\lrcorner\star\theta^{\mu}))\nonumber\\
&  =-\mathcal{R}_{\rho\sigma}\wedge\star(\theta^{\rho}\wedge\theta^{\sigma
}\wedge\theta^{\mu})\label{440nbis}\\
&  =\star(\mathcal{R}_{\rho\sigma}\lrcorner(\theta^{\rho}\wedge\theta^{\sigma
}\wedge\theta^{\mu})),\nonumber
\end{align}
where we have used the Eqs.(\ref{Aidentities}). Then, recalling the definition
of the curvature forms and using the Eq.(\ref{a.5}), we conclude that:
\begin{equation}
\mathcal{R}^{\rho\sigma}\wedge\theta_{\rho}\lrcorner\theta_{\sigma}%
\lrcorner\star\theta^{\mu}=2\star(\mathcal{R}^{\mu}-\frac{1}{2}R\theta^{\mu
})=2\star\mathcal{G}^{\mu},
\end{equation}
where $R$ is the scalar curvature of the manifold and $\ $the $\mathcal{G}%
^{\mu}$ may be called the Einstein 1-form fields.

That observation motivates us to introduce the

\begin{definition}
The \textit{Einstein operator }of the manifold associated to the Levi-Civita
connection \textbf{$\nabla$} of $\mathbf{g}$ is the mapping
$\blacktriangledown$ $:\sec\mathcal{C\ell}(T^{\ast}M)\mathtt{\rightarrow}%
\sec\mathcal{C\ell}(T^{\ast}M)$ given by:
\begin{equation}
\blacktriangledown=\frac{1}{2}\star^{-1}(\mathcal{R}^{\rho\sigma}\wedge
\theta_{\rho}\lrcorner\theta_{\sigma}\lrcorner)\star.
\end{equation}
\ 
\end{definition}

Obviously, we have:
\begin{equation}
\blacktriangledown\theta^{\mu}=\mathcal{G}^{\mu}=\mathcal{R}^{\mu}-\frac{1}%
{2}R\theta^{\mu}.
\end{equation}
In addition, it is easy to verify that $\star^{-1}({%
\mbox{\boldmath$\partial$}%
}\wedge{%
\mbox{\boldmath$\partial$}%
})\star=-{%
\mbox{\boldmath$\partial$}%
}\wedge{%
\mbox{\boldmath$\partial$}%
}$ and $\star^{-1}(\mathcal{R}^{\sigma}\wedge\theta_{\sigma}\lrcorner
)\star=\mathcal{R}^{\sigma}\lrcorner\theta_{\sigma}\wedge$. Thus we can also
write the Einstein operator as:
\begin{equation}
\blacktriangledown=-\frac{1}{2}({%
\mbox{\boldmath$\partial$}%
}\wedge{%
\mbox{\boldmath$\partial$}%
}+\mathcal{R}^{\sigma}\lrcorner\theta_{\sigma}\wedge).
\end{equation}

Another important result is given by the following proposition:

\begin{proposition}
\textit{Let }$\omega_{\rho}^{\mu}$\textit{\ be the Levi-Civita connection
1-forms fields in an arbitrary moving frame }$\{\theta^{\mu}\}\in\sec
F(M)$\textit{\ on }$(M,$\textbf{$\nabla$}$,g)$\textit{. Then: }%
\begin{equation}%
\begin{tabular}
[c]{c}%
\textit{(a)}\\
\textit{(b)}%
\end{tabular}%
\begin{array}
[c]{rcl}%
({%
\mbox{\boldmath$\partial$}%
}\cdot{%
\mbox{\boldmath$\partial$}%
})\theta^{\mu} & = & -({%
\mbox{\boldmath$\partial$}%
}\cdot\omega_{\rho}^{\mu}-\omega_{\rho}^{\sigma}\cdot\omega_{\sigma}^{\mu
})\theta^{\rho}\\
({%
\mbox{\boldmath$\partial$}%
}\wedge{%
\mbox{\boldmath$\partial$}%
})\theta^{\mu} & = & -({%
\mbox{\boldmath$\partial$}%
}\wedge\omega_{\rho}^{\mu}-\omega_{\rho}^{\sigma}\wedge\omega_{\sigma}^{\mu
})\theta^{\rho},
\end{array}
\label{2017}%
\end{equation}
\textit{that is,\ }%
\begin{equation}
{%
\mbox{\boldmath$\partial$}%
}^{\hspace{0.1cm}2}\theta^{\mu}=-({%
\mbox{\boldmath$\partial$}%
}\omega_{\rho}^{\mu}-\omega_{\rho}^{\sigma}\omega_{\sigma}^{\mu})\theta^{\rho
}.
\end{equation}
\noindent\ \ \ 

\begin{proof}
We have%
\begin{align*}
{%
\mbox{\boldmath$\partial$}%
}\cdot\omega_{\rho}^{\mu}  &  =\theta^{\alpha}\cdot\mathbf{\nabla}_{e_{\alpha
}}(\mathbf{\gamma}_{\beta\rho}^{\mu}\theta^{\beta})\\
&  =\theta^{\alpha}\cdot(\mathbf{e}_{\alpha}(\mathbf{\gamma}_{\beta\rho}^{\mu
})\theta^{\beta}-\gamma_{\sigma\rho}^{\mu}\mathbf{\gamma}_{\alpha\beta
}^{\sigma}\theta^{\beta})\\
&  =g^{\alpha\beta}(\mathbf{e}_{\alpha}(\mathbf{\gamma}_{\beta\rho}^{\mu
})-\mathbf{\gamma}_{\sigma\rho}^{\mu}\mathbf{\gamma}_{\alpha\beta}^{\sigma})
\end{align*}
and $\omega_{\rho}^{\sigma}\cdot\omega_{\sigma}^{\mu}=(\mathbf{\gamma}%
_{\beta\rho}^{\sigma}\theta^{\beta})\cdot(\mathbf{\gamma}_{\alpha\sigma}^{\mu
}\theta^{\alpha})=g^{\beta\alpha}\mathbf{\gamma}_{\alpha\sigma}^{\mu
}\mathbf{\gamma}_{\beta\rho}^{\sigma}$. Then,%
\begin{align*}
&  -({%
\mbox{\boldmath$\partial$}%
}\cdot\omega_{\rho}^{\mu}-\omega_{\rho}^{\sigma}\cdot\omega_{\sigma}^{\mu
})\theta^{\rho}\\
&  =g^{\alpha\beta}(\mathbf{e}_{\alpha}(\mathbf{\gamma}_{\beta\rho}^{\mu
})-\mathbf{\gamma}_{\alpha\sigma}^{\mu}\mathbf{\gamma}_{\beta\rho}^{\sigma
}-\mathbf{\gamma}_{\alpha\beta}^{\sigma}\mathbf{\gamma}_{\sigma\rho}^{\mu
})\theta^{\rho}\\
&  =-\frac{1}{2}g^{\alpha\beta}(\mathbf{e}_{\alpha}(\mathbf{\gamma}_{\beta
\rho}^{\mu})+\mathbf{e}_{\beta}(\mathbf{\gamma}_{\alpha\rho}^{\mu
})-\mathbf{\gamma}_{\alpha\sigma}^{\mu}\mathbf{\gamma}_{\beta\rho}^{\sigma
}-\mathbf{\gamma}_{\beta\sigma}^{\mu}\mathbf{\gamma}_{\alpha\rho}^{\sigma
}-b_{\alpha\beta}^{\sigma}\mathbf{\gamma}_{\sigma\rho}^{\mu})\theta^{\rho}\\
&  =({%
\mbox{\boldmath$\partial$}%
}\cdot{%
\mbox{\boldmath$\partial$}%
})\theta^{\mu}.
\end{align*}

\end{proof}
\end{proposition}

\begin{proof}
Eq.(\ref{2017}b) is proved analogously.\hfill
\end{proof}

Now, for an \textit{orthonormal} coframe $\{\theta^{\mathbf{a}}\}$ we have
immediately, taken into account that \textbf{$\nabla$}$_{e_{a}}%
\mbox{\boldmath{$\theta$}}%
^{\mathbf{b}}=-\omega_{\mathbf{ac}}^{\mathbf{b}}%
\mbox{\boldmath{$\theta$}}%
^{\mathbf{c}}$,
\begin{align}
{%
\mbox{\boldmath$\partial$}%
}\cdot{%
\mbox{\boldmath$\partial$}%
}  &  =\eta^{\mathbf{ab}}(\mathbf{\nabla}_{\mathbf{e}_{\mathbf{a}}%
}\mathbf{\nabla}_{\mathbf{e}_{\mathbf{b}}}-\omega_{\mathbf{ab}}^{\mathbf{c}%
}\mathbf{\nabla}_{\mathbf{e}_{\mathbf{c}}}),\nonumber\\
{%
\mbox{\boldmath$\partial$}%
}\wedge{%
\mbox{\boldmath$\partial$}%
}  &  =\theta^{\mathbf{a}}\wedge\theta^{\mathbf{b}}(\mathbf{\nabla
}_{\mathbf{e}_{\mathbf{a}}}\mathbf{\nabla}_{\mathbf{e}_{\mathbf{b}}}%
-\omega_{\mathbf{ab}}^{\mathbf{c}}\mathbf{\nabla}_{\mathbf{e}_{\mathbf{c}}}).
\label{11.3}%
\end{align}

and\footnote{In \cite{18} there is an analogous equation, but there is a
misprint of a factor of 2.}%
\begin{equation}
({%
\mbox{\boldmath$\partial$}%
}\wedge{%
\mbox{\boldmath$\partial$}%
})\theta^{\mathbf{a}}=\mathcal{R}^{\mathbf{a}}, \label{11.5}%
\end{equation}

\section{ Equations for the Tetrad Fields\textbf{ }$\theta^{\mathbf{a}}$}

Here we want to \textit{recall }a not well known face\textit{ }of Einstein
equations, i.e., we show how to write the field equations for the tetrad
fields $%
\mbox{\boldmath{$\theta$}}%
^{\mathbf{a}}$ in such a way that the obtained equations are equivalent to
Einstein field equations. This is done in order to compare the correct
equations satisfied by those objects with equations proposed for those objects
that appeared in \textbf{\cite{0} }and also in other papers authored by Evans
(some quoted in the reference list).

\begin{proposition}
Let $\mathfrak{M}=(M,\mathbf{g},$\textbf{$\nabla$}$\mathbf{,\tau}_{\mathbf{g}%
},\mathbf{\uparrow})$ be a Lorentzian spacetime and also a spin manifold, and
suppose that $\mathbf{g}$ satisfies the classical Einstein's gravitational
equation, which reads in standard notation (and in natural units)%
\begin{equation}
\mathrm{Ricci}-\frac{1}{2}R\mathbf{g}=\mathcal{T}. \label{EGE}%
\end{equation}
Then, Eq.(\textbf{\ref{EGE}) }is equivalent to Eq.(\ref{11.1}) satisfied by
the fields $\theta^{\mathbf{a}}$ $(\mathbf{a}=0,1,2,3)$\ of a cotetrad
$\{\theta^{\mathbf{a}}\}$ on $\mathfrak{M}$. Also, under the same conditions
Eq.(\ref{11.1}) is equivalent to Einstein's equation.\footnote{Of course,
there are analogous equations for the $\mathbf{e}_{\mathbf{a}}$, where in that
case, the Dirac operator must be defined (in an obvious way) as acting on
sections of the Clifford bundle $\mathcal{C\ell(}TM)$ of non homogeneous
multivector fields. See, e.g., \cite{18}, but take notice that the equations
in \cite{18} have an (equivocated) extra factor of 2.}:
\begin{equation}
-({%
\mbox{\boldmath$\partial$}%
}\cdot{%
\mbox{\boldmath$\partial$}%
})\theta^{\mathbf{a}}+{%
\mbox{\boldmath$\partial$}%
}\wedge({%
\mbox{\boldmath$\partial$}%
}\cdot\theta^{\mathbf{a}})+{%
\mbox{\boldmath$\partial$}%
}\lrcorner({%
\mbox{\boldmath$\partial$}%
}\wedge\theta^{\mathbf{a}})=\mathcal{T}^{\mathbf{a}}-\frac{1}{2}%
T\theta^{\mathbf{a}}. \label{11.1}%
\end{equation}
In Eq.(\ref{EGE}) and Eq.(\ref{11.1}), \textrm{Ricci} is the Ricci tensor,
$\mathcal{T}$ is the energy momentum tensor (with components $T_{\mathbf{b}%
}^{\mathbf{a}}$), $R$ is the curvature scalar and\ $\mathcal{T}^{\mathbf{a}%
}=T_{\mathbf{b}}^{\mathbf{a}}\theta^{\mathbf{b}}\in\sec\bigwedge
\nolimits^{1}T^{\ast}M\hookrightarrow\sec\mathcal{C}\ell(T^{\ast}M)$ are the
energy momentum $1$-form fields and $T=T_{\mathbf{a}}^{\mathbf{a}%
}=-R=-R_{\mathbf{a}}^{\mathbf{a}}$.
\end{proposition}

\begin{proof}
\textit{We prove that Einstein's equations are equivalent to Eq.(\ref{11.1}).
The proof that Eq.(\ref{11.1}) is equivalent to Einstein's equation is left to
the reader. Einstein's equation reads in components relative to a tetrad}
$\{\mathbf{e}_{\mathbf{a}}\}\in\sec\mathrm{P}_{\mathrm{SO}_{1,3}^{e}}(M)$
\textit{and the cotetrad} $\{\theta^{\mathbf{a}}\}$, $\theta^{\mathbf{a}}%
\in\sec\bigwedge\nolimits^{1}TM\hookrightarrow\sec\mathcal{C\ell}(TM)$
\textit{as}:%
\begin{equation}
R_{\mathbf{b}}^{\mathbf{a}}-\frac{1}{2}\delta_{\mathbf{b}}^{\mathbf{a}%
}R=T_{\mathbf{b}}^{\mathbf{a}} \label{einstein}%
\end{equation}
\ \textit{Multiplying the above equation by} $\theta^{\mathbf{b}}$ \textit{and
summing we get},%
\begin{equation}
\mathcal{R}^{\mathbf{a}}-\frac{1}{2}R\theta^{\mathbf{a}}=\mathcal{T}%
^{\mathbf{a}} \label{einstein1}%
\end{equation}

\textit{Next we use in Eq}.\textit{(\ref{einstein1})} \textit{the
Eq.(\ref{11.5}), Eq.(\ref{1796}), Eq.(\ref{1796a})}, \textit{and that} $T=-R$
\textit{to write} \textit{Eq.(\ref{einstein1}) as}:%
\begin{equation}
-({%
\mbox{\boldmath$\partial$}%
}\cdot{%
\mbox{\boldmath$\partial$}%
v})\theta^{\mathbf{a}}+{%
\mbox{\boldmath$\partial$}%
}\wedge({%
\mbox{\boldmath$\partial$}%
}\cdot\theta^{\mathbf{a}})+{%
\mbox{\boldmath$\partial$}%
}\lrcorner({%
\mbox{\boldmath$\partial$}%
}\wedge\theta^{\mathbf{a}})=\mathcal{T}^{\mathbf{a}}-\frac{1}{2}%
T\theta^{\mathbf{a}},\label{correct}%
\end{equation}
and the proposition is proved.
\end{proof}

\bigskip Note that in a coordinate chart $\{x^{\mu}\}$\ of the maximal atlas
of $M$ covering $U\subset M$ , Eq.(\ref{einstein1}) can be written as
\begin{equation}
\mathcal{R}^{\mu}-\frac{1}{2}R\theta^{\mu}=\mathcal{T}^{\mu}, \label{11.6}%
\end{equation}
with $R^{\mu}=R_{\nu}^{\mu}dx^{\nu}$ and $T^{\mu}=T_{\nu}^{\mu}dx^{\nu}$,
$\theta^{\mu}=dx^{\mu}$. Eq.(\ref{11.6}) looks like an equation appearing in
some of Evans papers, but the meaning here is very different. From
Eq.(\ref{11.6}) we can show that an equation identical to Eq.(\ref{correct})
is also satisfied by the moving coordinate coframe $\{\theta^{\mu}=dx^{\mu
}\}.$If we suppose moreover that the coordinate functions are harmonic, i.e.,
$\delta\theta^{\mu}=-{%
\mbox{\boldmath$\partial$}%
}\theta^{\mu}=0$, Eq.(\ref{11.1}) becomes\footnote{A somewhat similar equation
with some (equivocated) extra factors of $2$ appears in \cite{18}.}%
\begin{equation}
\blacksquare\theta^{\mu}+\frac{1}{2}R\theta^{\mu}=-\mathcal{T}^{\mu},
\label{11.1BIS}%
\end{equation}

We recall that in \textbf{\cite{0}} it is wrongly derived that the equations
for $\theta^{\mathbf{a}}$, $\mathbf{a}=0,1,2,3$ are the equations
\footnote{Here we wrote the equation in units where $\kappa=1$. Note also that
in \cite{0} it is explicitly stated that the symbol $\square$ means
\ ${\mbox{\boldmath$\partial$}}_{\mu}{\mbox{\boldmath$\partial$}}^{\mu}$. It
is not to be confused with the covariant D'Alembertian, which in our paper is
represented by $\blacksquare$.
\par
{}}

\begin{center}%
\begin{equation}%
\begin{tabular}
[c]{|l|}\hline
$(\square-R(x))\theta^{\mathbf{a}}=0.$\\\hline
\end{tabular}
\tag{49E}%
\end{equation}

\end{center}

\begin{remark}
An equation looking similar to.(49E), namely,%
\begin{equation}
-{%
\mbox{\boldmath$\partial$}%
}^{2}\theta^{\mathbf{a}}+\lambda(x)\theta^{\mathbf{a}}=0 \label{11.7}%
\end{equation}
has been proposed in \cite{kaniel} as vacuum field equations for a theory of
the gravitational field not equivalent to General Relativity. Note that in
Eq.(\ref{11.7}) the wave equation is written with the Hodge Laplacian and
moreover $\lambda(x)\neq R(x).$Such a theory has been criticized in
\cite{mughe}, who point some particularizations\footnote{We shall discuss this
issue in another publication.} in the derivations of \cite{kaniel},but that
paper is really interesting. See also \cite{itin}.\ We shall discuss this
issue in another publication. However, even in \cite{mughe}, the wave
equations for the tetrad fields in General Relativity are not given.
\end{remark}

\subsection{Correct Equations for the $q_{\nu}^{\mathbf{a}}$ functions in a
Lorentzian Manifold}

First we obtain that equations for the functions $q_{\nu}^{\mathbf{a}}$ in a
Lorentzian manifold. This will be done using Eq.(\ref{q rc1}) for that
situation. We have:
\begin{align}
\mathbf{\nabla}_{{\mbox{\boldmath$\partial$}}_{\nu}}\mathbf{Q}  &
=\mathbf{\nabla}_{{\mbox{\boldmath$\partial$}}_{\nu}}(\mathbf{e}_{\mathbf{a}%
}\otimes\theta^{\mathbf{a}})\nonumber\\
&  =\mathbf{\nabla}_{{\mbox{\boldmath$\partial$}}_{\nu}}^{+}\mathbf{e}%
_{\mathbf{a}}\otimes\theta^{\mathbf{a}}+\mathbf{e}_{\mathbf{a}}\otimes
\mathbf{\nabla}_{{\mbox{\boldmath$\partial$}}_{\nu}}^{-}\theta^{\mathbf{a}}.
\label{q rc 2}%
\end{align}

Then,%
\begin{align}
&  g^{\mu\nu}\mathbf{\nabla}_{{\mbox{\boldmath$\partial$}}_{\mu}%
}\mathbf{\nabla}_{{\mbox{\boldmath$\partial$}}_{\nu}}\mathbf{Q}\nonumber\\
&  =\mathbf{\nabla}_{{\mbox{\boldmath$\partial$}}_{\mu}}^{+}\mathbf{\nabla
}_{{\mbox{\boldmath$\partial$}}_{\nu}}^{+}\mathbf{e}_{\mathbf{a}}\otimes
\theta^{\mathbf{a}}+\mathbf{\nabla}_{{\mbox{\boldmath$\partial$}}_{\nu}}%
^{+}\mathbf{e}_{\mathbf{a}}\otimes\mathbf{\nabla}%
_{{\mbox{\boldmath$\partial$}}_{\mu}}^{-}\theta^{\mathbf{a}}+\mathbf{\nabla
}_{{\mbox{\boldmath$\partial$}}_{\mu}}^{+}\mathbf{e}_{\mathbf{a}}%
\otimes\mathbf{\nabla}_{{\mbox{\boldmath$\partial$}}_{\nu}}^{-}\theta
^{\mathbf{a}}+\mathbf{e}_{\mathbf{a}}\otimes\mathbf{\nabla}%
_{{\mbox{\boldmath$\partial$}}_{\mu}}^{-}\mathbf{\nabla}%
_{{\mbox{\boldmath$\partial$}}_{\nu}}^{-}\theta^{\mathbf{a}}, \label{q  rc 2a}%
\end{align}
and%
\begin{align}
&  g^{\mu\nu}\mathbf{\nabla}_{{\mbox{\boldmath$\partial$}}_{\mu}%
}\mathbf{\nabla}_{{\mbox{\boldmath$\partial$}}_{\nu}}\mathbf{Q}\nonumber\\
&  =g^{\mu\nu}\mathbf{\nabla}_{{\mbox{\boldmath$\partial$}}_{\mu}}%
^{+}\mathbf{\nabla}_{{\mbox{\boldmath$\partial$}}_{\nu}}^{+}\mathbf{e}%
_{\mathbf{a}}\otimes\theta^{\mathbf{a}}+2g^{\mu\nu}\mathbf{\nabla
}_{{\mbox{\boldmath$\partial$}}_{\nu}}^{+}\mathbf{e}_{\mathbf{a}}%
\otimes\mathbf{\nabla}_{{\mbox{\boldmath$\partial$}}_{\mu}}^{-}\theta
^{\mathbf{a}}+\mathbf{e}_{\mathbf{a}}\otimes g^{\mu\nu}\mathbf{\nabla
}_{{\mbox{\boldmath$\partial$}}_{\mu}}^{-}\mathbf{\nabla}%
_{{\mbox{\boldmath$\partial$}}_{\nu}}^{-}\theta^{\mathbf{a}} \label{q rc 2b}%
\end{align}

Now, write the difference of Eq.(\ref{q rc 2b}) and the quantity $g^{\nu\mu
}\mathbf{\nabla}_{{\mbox{\boldmath$\partial$}}_{\nu}}\mathbf{\nabla
}_{{\mbox{\boldmath$\partial$}}_{\mu}}\mathbf{Q}$. This gives
\begin{equation}
g^{\mu\nu}\left[  \left(  \mathbf{\nabla}_{{\mbox{\boldmath$\partial$}}_{\mu}%
}^{+}\mathbf{\nabla}_{{\mbox{\boldmath$\partial$}}_{\nu}}^{+}-\mathbf{\nabla
}_{{\mbox{\boldmath$\partial$}}_{\nu}}^{+}\mathbf{\nabla}%
_{{\mbox{\boldmath$\partial$}}_{\mu}}^{+}\right)  \mathbf{e}_{\mathbf{a}%
}\right]  \otimes\theta^{\mathbf{a}}+\mathbf{e}_{\mathbf{a}}\otimes g^{\mu\nu
}\left[  \left(  \mathbf{\nabla}_{{\mbox{\boldmath$\partial$}}_{\mu}}%
^{-}\mathbf{\nabla}_{{\mbox{\boldmath$\partial$}}_{\nu}}^{-}-\mathbf{\nabla
}_{{\mbox{\boldmath$\partial$}}_{\nu}}^{-}\mathbf{\nabla}%
_{{\mbox{\boldmath$\partial$}}_{\mu}}^{-}\right)  \theta^{\mathbf{a}}\right]
=0 \label{q rc 3 0}%
\end{equation}

Now, recalling the operator identity \footnote{The operator identity given by
Eq.(\ref{epa}) is to be compared with the wrong Eq.(42E) and also with the
equation in line 11 of table 1 in \cite{0}.} (Eq.(\ref{1792}))%
\begin{equation}
g^{\mu\nu}\mathbf{\nabla}_{{\mbox{\boldmath$\partial$}}_{\mu}}^{-}%
\mathbf{\nabla}_{{\mbox{\boldmath$\partial$}}_{\nu}}^{-}={\blacksquare}%
+g^{\mu\nu}\Gamma_{\mu\nu}^{\rho}\mathbf{\nabla}_{{\mbox{\boldmath$\partial$}}%
\rho}^{-}, \label{epa}%
\end{equation}
and Eq.(\ref{1831}), we have
\begin{equation}
g^{\mu\nu}\mathbf{\nabla}_{{\mbox{\boldmath$\partial$}}_{\mu}}^{-}%
\mathbf{\nabla}_{{\mbox{\boldmath$\partial$}}_{\nu}}^{-}\theta^{\mathbf{a}}={%
\mbox{\boldmath$\partial$}%
\cdot%
\mbox{\boldmath$\partial$}%
}\theta^{\mathbf{a}}+g^{\mu\nu}\Gamma_{\mu\nu}^{\rho}\mathbf{\nabla
}_{{\mbox{\boldmath$\partial$}}\rho}^{-}\theta^{\mathbf{a}}. \label{q rc 3}%
\end{equation}
Also,%
\begin{equation}
g^{\mu\nu}\mathbf{\nabla}_{{\mbox{\boldmath$\partial$}}_{\mu}}^{-}%
\mathbf{\nabla}_{{\mbox{\boldmath$\partial$}}_{\nu}}^{-}\theta^{\mathbf{a}%
}=g^{\mu\nu}\left(  -{\mbox{\boldmath$\partial$}}_{\nu}\omega_{\mu\mathbf{b}%
}^{\mathbf{a}}+\omega_{\mu\mathbf{c}}^{\mathbf{a}}\omega_{\nu\mathbf{b}%
}^{\mathbf{c}}\right)  \theta^{\mathbf{b}}. \label{q rc 3a}%
\end{equation}
Also,%
\begin{align}
&  g^{\mu\nu}\left[  \left(  \mathbf{\nabla}_{{\mbox{\boldmath$\partial$}}%
_{\mu}}^{+}\mathbf{\nabla}_{{\mbox{\boldmath$\partial$}}_{\nu}}^{+}%
-\mathbf{\nabla}_{{\mbox{\boldmath$\partial$}}_{\nu}}^{+}\mathbf{\nabla
}_{{\mbox{\boldmath$\partial$}}_{\mu}}^{+}\right)  \mathbf{e}_{\mathbf{a}%
}\right] \nonumber\\
&  =g^{\mu\nu}R_{\mathbf{a}\text{ }\mu\nu}^{\text{ }\mathbf{b}}\mathbf{e}%
_{\mathbf{b}}=R_{\mathbf{a}}^{\mathbf{b}}\mathbf{e}_{\mathbf{b}}
\label{q rc 3b}%
\end{align}

Using Eqs. (\ref{q rc 3}), (\ref{q rc 3a}) and (\ref{q rc 3b}) in
Eq.(\ref{q rc 3 0}), we get%
\begin{equation}
g^{\alpha\beta}\mathbf{\nabla}_{{\alpha}}^{-}\mathbf{\nabla}_{{\beta}}%
^{-}q_{\mu}^{\mathbf{b}}+R_{\mathbf{a}}^{\mathbf{b}}q_{\mu}^{\mathbf{a}%
}-g^{\mu\nu}({\mbox{\boldmath$\partial$}}_{\nu}\omega_{\mu\mathbf{b}%
}^{\mathbf{a}}-\Gamma_{\mu\nu}^{\rho}\omega_{\rho\mathbf{a}}^{\mathbf{b}%
}-\omega_{\mu\mathbf{c}}^{\mathbf{a}}\omega_{\nu\mathbf{b}}^{\mathbf{c}%
})q_{\mu}^{\mathbf{a}}=0 \label{q rc 3c}%
\end{equation}

So, this is the \ `wave equation' satisfied by the functions $q_{\mu
}^{\mathbf{a}}$ in a Lorentzian manifold. It is to be compared with Eq.(2E)
found in \cite{0}, which it has been used there to derive the false `Evans
lemma' used by the author of \cite{0}. \ It is our opinion that as an wave
equation Eq.(\ref{q rc 3c}) has no utility. However, since as it is well
known, we can write the $\omega_{\mu\mathbf{b}}^{\mathbf{a}}$ and $g^{\mu\nu}$
in terms of the functions $q_{\mu}^{\mathbf{a}}$ and their inverses
$q_{\mathbf{b}}^{\nu}$. Doing that we can use Eq.(\ref{q rc 3c}) to write an
explicit expression (in the tetrad basis) for the components $R_{\mathbf{a}%
}^{\mathbf{b}}$ of the Ricci tensor in terms of the functions $q_{\mu
}^{\mathbf{a}}$ and $q_{\mathbf{b}}^{\nu}$. However, at the moment we cannot
see any advantage in writing such equation, for there are more efficient
methods to obtain the components of the Ricci tensor.

\subsection{Correct Equations for the $q_{\nu}^{\mathbf{a}}$ functions in
General Relativity}

Having obtained the correct equations for the tetrad fields $\theta
^{\mathbf{a}}$ in General Relativity (Eq.(\ref{correct})), we now derive the
corresponding equation for the $q_{\nu}^{\mathbf{a}}$ functions in a
Lorentzian spacetime representing a gravitational field

We first observe that
\begin{align}
&  {%
\mbox{\boldmath$\partial$}%
}\wedge({%
\mbox{\boldmath$\partial$}%
}\cdot\theta^{\mathbf{a}})+{%
\mbox{\boldmath$\partial$}%
}\lrcorner({%
\mbox{\boldmath$\partial$}%
}\wedge\theta^{\mathbf{a}})\nonumber\\
&  =\left[  -{\mbox{\boldmath$\partial$}}_{\mu}\left(  \omega_{\text{
\ }\mathbf{d}}^{\mathbf{a}\text{ }\mathbf{d}}\right)  q_{\mathbf{b}}^{\mu
}+q_{\mathbf{k}}^{\mu}{\mbox{\boldmath$\partial$}}_{\mu}\left(  \eta
^{\mathbf{kd}}\omega_{\mathbf{bd}}^{\mathbf{a}}-\omega_{\text{ \ }\mathbf{b}%
}^{\mathbf{a}\text{ }\mathbf{k}}\right)  \right]  \theta^{\mathbf{b}}
\label{l1}%
\end{align}
Next we define
\begin{equation}
\mathbf{K}_{\mathbf{a}}^{\mathbf{b}}=-\left[  -{\mbox{\boldmath$\partial$}}%
_{\mu}\left(  \omega_{\text{ \ }\mathbf{d}}^{\mathbf{b}\text{ }\mathbf{d}%
}\right)  q_{\mathbf{a}}^{\mu}+q_{\mathbf{k}}^{\mu}%
{\mbox{\boldmath$\partial$}}_{\mu}\left(  \eta^{\mathbf{kd}}\omega
_{\mathbf{ad}}^{\mathbf{b}}-\omega_{\text{ \ }\mathbf{a}}^{\mathbf{b}\text{
}\mathbf{k}}\right)  +T_{\mathbf{a}}^{\mathbf{b}}-\frac{1}{2}T\delta
_{\mathbf{a}}^{\mathbf{b}}\right]  \label{l2}%
\end{equation}
Using these results in Eq.(\ref{correct}) we get,%
\begin{equation}
g^{\alpha\beta}\mathbf{\nabla}_{{\alpha}}^{-}\mathbf{\nabla}_{\beta}^{-}%
q_{\mu}^{\mathbf{b}}+\mathbf{K}_{\mathbf{a}}^{\mathbf{b}}q_{\mu}^{\mathbf{a}%
}=0. \label{l3}%
\end{equation}

Comparing that equation with Eq.(\ref{q rc 3c}) we get the constraint
\begin{align}
R_{\mathbf{a}}^{\mathbf{b}}  &  -g^{\mu\nu}({\mbox{\boldmath$\partial$}}_{\nu
}\omega_{\mu\mathbf{b}}^{\mathbf{a}}-\Gamma_{\mu\nu}^{\rho}\omega
_{\rho\mathbf{a}}^{\mathbf{b}}-\omega_{\mu\mathbf{c}}^{\mathbf{a}}\omega
_{\nu\mathbf{b}}^{\mathbf{c}})+{\mbox{\boldmath$\partial$}}_{\mu}\left(
\omega_{\text{ \ }\mathbf{d}}^{\mathbf{b}\text{ }\mathbf{d}}\right)
q_{\mathbf{a}}^{\mu}+q_{\mathbf{k}}^{\mu}{\mbox{\boldmath$\partial$}}_{\mu
}\left(  \eta^{\mathbf{kd}}\omega_{\mathbf{ad}}^{\mathbf{b}}-\omega_{\text{
\ }\mathbf{a}}^{\mathbf{b}\text{ }\mathbf{k}}\right) \nonumber\\
&  =\frac{1}{2}T\delta_{\mathbf{a}}^{\mathbf{b}}-T_{\mathbf{a}}^{\mathbf{b}}.
\label{l4}%
\end{align}
which is a compatibility equation that must hold if the tetrad field equations
are to be equivalent to Einstein's equations.

\section{Correct Equation for the Electromagnetic Potential $A$}

In \cite{1,2,3} it is explicitly written several times that the
"electromagnetic potential" $\mathbf{A}$ of the "unified theory" (a 1-form
with values in a vector space) satisfies the following wave equation,
($\square={\mbox{\boldmath$\partial$}}_{\mu}{\mbox{\boldmath$\partial$}}^{\mu
}$)

\begin{center}%
\begin{tabular}
[c]{|c|}\hline
$(\square+T)\mathbf{A}=0.$\\\hline
\end{tabular}

\end{center}

Now, this equation cannot be correct even for the usual $U(1)$ gauge potential
of classical electrodynamics\footnote{Which must be one of the gauge
components of the gauge field.} $A\in\sec%
{\displaystyle\bigwedge\nolimits^{1}}
T^{\ast}M\subset\sec\mathcal{C\ell(}T^{\ast}M)$. To show that let us first
recall how to write electrodynamics in the Clifford bundle.

\subsection{Maxwell Equation}

Maxwell equations when can be written in the Clifford bundle formalism of
differential forms as single equation. Indeed, if $F\in\sec%
{\displaystyle\bigwedge\nolimits^{2}}
T^{\ast}M\subset\sec\mathcal{C\ell}(T^{\ast}M)$ is the electromagnetic field
and $J_{e}\in\sec%
{\displaystyle\bigwedge\nolimits^{1}}
T^{\ast}M\subset\sec\mathcal{C\ell}(T^{\ast}M)$ is the electromagnetic
current, we have Maxwell equation\footnote{Then, there is no misprint in the
title of this subsection.}:
\begin{equation}
{%
\mbox{\boldmath$\partial$}%
}F=J_{e}. \label{1.7}%
\end{equation}

Eq.(\ref{1.7}) is equivalent to the pair of equations%
\begin{align}
dF  &  =0,\label{1.8a}\\
\delta F  &  =-J_{e}. \label{1.8b}%
\end{align}

Eq.(\ref{1.8a}) is called the homogeneous equation and Eq.(\ref{1.8b}) is
called the nonhomogeneous equation. Note that it can be written also as:%
\begin{equation}
d\star F=-\star J_{e}. \label{1.9}%
\end{equation}

Now, in vacuum Maxwell equation reads
\begin{equation}
{%
\mbox{\boldmath$\partial$}%
}F=0, \label{11.9}%
\end{equation}
where $F={%
\mbox{\boldmath$\partial$}%
}A={%
\mbox{\boldmath$\partial$}%
}\wedge A=dA$, \textit{if} we work in the \textit{Lorenz} gauge ${%
\mbox{\boldmath$\partial$}%
}\cdot A={%
\mbox{\boldmath$\partial$}%
}\lrcorner A=-\delta A=0$. \ Now, since we have according to Eq.(\ref{1796b})
that ${%
\mbox{\boldmath$\partial$}%
}^{2}=-(d\delta+\delta d),$we get
\begin{equation}
{%
\mbox{\boldmath$\partial$}%
}^{2}A=0. \label{11.11}%
\end{equation}

Using Eq.(\ref{hodge laplacian}) (or Eq.(\ref{1796b}) coupled with
Eq.(\ref{1918})) and the coordinate basis introduced above we have,
\begin{equation}
({%
\mbox{\boldmath$\partial$}%
}^{2}A)_{\alpha}=g^{\mu\nu}\mathbf{\nabla}_{\mu}\mathbf{\nabla}_{\nu}%
A_{\alpha}+R_{\alpha}^{\nu}A_{\nu}. \label{11.12}%
\end{equation}
Then, we see that Eq.(\ref{11.11}) reads in components\footnote{Sometimes the
symbol $\square$ is used to denote the operator $D_{\alpha}D^{\alpha}$.
Eq.(\ref{11.13}) can be found, e.g., in Eddington's book \ \cite{19} on page
175.}%
\begin{equation}
\mathbf{\nabla}_{\alpha}\mathbf{\nabla}^{\alpha}A_{\mu}+R_{\mu}^{\nu}A_{\nu
}=0. \label{11.13}%
\end{equation}

Finally, we observe that in Einstein's theory, $R_{\mu}^{\nu}=0$ in vacuum,
and so in vacuum regions we end with:%
\begin{equation}
\mathbf{\nabla}_{\alpha}\mathbf{\nabla}^{\alpha}A_{\mu}=0. \label{11.13bis}%
\end{equation}

\section{Lagrangian Field Theory for the Tetrad Fields}

We show here how the Einstein-Hilbert Lagrangian (modulus an exact
differential) can be written in the suggestive form given by%

\begin{equation}
\mathcal{L}_{\mathtt{g}}=-\frac{1}{2}d\theta^{\mathbf{a}}\wedge\star
d\theta_{\mathbf{a}}+\frac{1}{2}\delta\theta^{\mathbf{a}}\wedge\star
\delta\theta_{\mathbf{a}}+\frac{1}{4}\left(  d\theta^{\mathbf{a}}\wedge
\theta_{\mathbf{a}}\right)  \wedge\star\left(  d\theta^{\mathbf{b}}%
\wedge\theta_{\mathbf{b}}\right)  . \label{8.1}%
\end{equation}
Here, $\mathtt{\mathbf{g}}=\eta_{\mathbf{ab}}\theta^{\mathbf{a}}\otimes
\theta^{\mathbf{b}}$ and
\begin{equation}
\theta^{\mathbf{a}}\theta^{\mathbf{b}}+\theta^{\mathbf{b}}\theta^{\mathbf{a}%
}=2\eta^{\mathbf{ab}}. \label{8.2}%
\end{equation}
Now, the classical Einstein-Hilbert Lagrangian density in appropriate
(geometrical) units is given by
\begin{equation}
\mathcal{L}_{EH}=\frac{1}{2}R\tau_{\mathtt{\mathbf{g}}}=\frac{1}{2}%
R\theta^{\mathbf{5}}, \label{8.6}%
\end{equation}
where $R=\eta^{\mathbf{cd}}R_{\mathbf{cd}}$ is the scalar curvature. We
observe that we can write $\mathcal{L}_{EH}$ as
\begin{align}
\mathcal{L}_{EH}  &  =\frac{1}{2}\mathcal{R}_{\mathbf{cd}}\wedge\star
(\theta^{\mathbf{c}}\wedge\theta^{\mathbf{d}})\label{8.7}\\
&  =\frac{1}{2}\mathcal{R}_{\mathbf{cd}}\wedge(\theta^{\mathbf{d}}\wedge
\theta^{\mathbf{c}})\theta^{\mathbf{5}}%
\end{align}
Indeed, we have immediately that
\begin{align}
\mathcal{R}_{\mathbf{cd}}\wedge\star(\theta^{\mathbf{c}}\wedge\theta
^{\mathbf{d}})  &  =(\theta^{\mathbf{c}}\wedge\theta^{\mathbf{d}})\wedge
\star\mathcal{R}_{\mathbf{cd}}=-(\theta^{\mathbf{c}})\wedge\star
(\theta^{\mathbf{d}}\lrcorner\mathcal{R}_{\mathbf{cd}})\nonumber\\
&  =-\star\lbrack\theta^{\mathbf{c}}\lrcorner(\theta^{\mathbf{d}}%
\lrcorner\mathcal{R}_{\mathbf{cd}})], \label{8.8}%
\end{align}
and since%
\begin{align}
\theta^{\mathbf{d}}\lrcorner\mathcal{R}_{\mathbf{cd}}  &  =\frac{1}%
{2}R_{\mathbf{cdab}}\theta^{\mathbf{d}}\lrcorner(\theta^{\mathbf{a}}%
\wedge\theta^{\mathbf{b}})=\frac{1}{2}R_{\mathbf{cdab}}(\eta^{\mathbf{da}%
}\theta^{\mathbf{b}}-\eta^{\mathbf{db}}\theta^{\mathbf{a}})\nonumber\\
&  =-R_{\mathbf{ca}}\theta^{\mathbf{b}}=-\mathcal{R}_{\mathbf{c,}} \label{8.9}%
\end{align}
it follows that $-\theta^{\mathbf{c}}\lrcorner(\theta^{\mathbf{d}}%
\lrcorner\mathcal{R}_{\mathbf{cd}})=\theta^{\mathbf{c}}\cdot\mathcal{R}%
_{c}=R.$

Now, taking into account that $\mathcal{R}_{\mathbf{cd}}=d\omega_{\mathbf{cd}%
}+\omega_{\mathbf{ca}}\wedge\omega_{\mathbf{d}}^{\mathbf{a}}$, we can obtain
the free Einstein's field equations $\star\mathcal{G}_{\mathbf{a}}=0$ by
varying \ the Einstein-Hilbert action $\int\mathcal{L}_{EH}$ with respect to
the fields $\theta^{\mathbf{a}}$ and $\omega_{\mathbf{ca}}$. Indeed, after a
very long calculation (see Appendix B) which requires the notion of derivative
of multivector functions and functionals
\cite{fmr,fmr2,fmr3,fmr4,fmr5,fmr6,fmr7} we get%
\begin{equation}%
\mbox{\boldmath{$\delta$}}%
\mathcal{L}_{EH}=-\frac{1}{2}d\left[  \star\left(  \theta^{\mathbf{c}}%
\wedge\theta^{\mathbf{d}}\right)  \wedge%
\mbox{\boldmath{$\delta$}}%
\omega_{\mathbf{cd}}\right]  +%
\mbox{\boldmath{$\delta$}}%
\theta^{\mathbf{a}}\wedge\left[  \frac{1}{2}\star(\theta^{\mathbf{c}}%
\wedge\theta^{\mathbf{d}}\wedge\theta_{\mathbf{a}})\right]  \wedge
\mathcal{R}_{\mathbf{cd}}. \label{8.10}%
\end{equation}

Now, taking into account that
\begin{equation}
-\frac{1}{2}\left[  \star(\theta^{\mathbf{c}}\wedge\theta^{\mathbf{d}}%
\wedge\theta_{\mathbf{a}})\right]  \wedge\mathcal{R}_{\mathbf{cd}}%
=\star\mathcal{G}_{\mathbf{a}}=\star(\mathcal{R}_{\mathbf{a}}-\frac{1}%
{2}R\theta_{\mathbf{a}}). \label{8.11}%
\end{equation}
Of course, in order to obtain Einstein's equations in the presence of matter
we have to vary the total Lagrangian density $\mathcal{L}=\mathcal{L}%
_{EH}+\mathcal{L}_{m}$, where we explicitly suppose that $\mathcal{L}%
_{m}(\theta^{\mathbf{a}},d\theta^{\mathbf{a}},\phi^{A},d\phi^{A})$, the matter
Lagrangian of a set of fields $\phi^{A}$ (which may be Clifford or spinor
fields, \ the latter fields, also represented in each spin frame as a sum of
non homogeneous differential forms, as explained in \cite{10}) does not depend
explicitly on the $\omega_{\mathbf{cd}}$. In that case, we have%
\begin{equation}%
\mbox{\boldmath{$\delta$}}%
\mathcal{L=}-\frac{1}{2}d\left[  \star\left(  \theta^{\mathbf{c}}\wedge
\theta^{\mathbf{d}}\right)  \wedge%
\mbox{\boldmath{$\delta$}}%
\omega_{\mathbf{cd}}\right]  +%
\mbox{\boldmath{$\delta$}}%
\theta^{\mathbf{a}}\wedge\left\{  \left[  \frac{1}{2}\star(\theta^{\mathbf{c}%
}\wedge\theta^{\mathbf{d}}\wedge\theta_{\mathbf{a}})\right]  \wedge
\mathcal{R}_{\mathbf{cd}}+\star T_{\mathbf{a}}\right\}  , \label{8.12}%
\end{equation}
and the field equations results in%
\begin{equation}
\star\mathcal{G}_{\mathbf{a}}=\star T_{\mathbf{a}}, \label{8.13}%
\end{equation}
but this equation as we know, gives by use of the Ricci, \ Einstein, covariant
D'Alembertian and the Hodge Laplacian, directly the equations for the tetrad fields.

\begin{remark}
We observe that $\mathcal{L}_{g}$ is the first order Lagrangian density (first
introduced by Einstein) written in \textit{intrinsic} form. Indeed, the dual
of Eq.(\ref{8.8}), i.e., $[\theta^{\mathbf{c}}\lrcorner(\theta^{\mathbf{d}%
}\lrcorner\mathcal{R}_{\mathbf{cd}})]$ is given by
\begin{equation}
\lbrack\theta^{\mathbf{c}}\lrcorner(\theta^{\mathbf{d}}\lrcorner
\mathcal{R}_{\mathbf{cd}})]=\theta^{\mathbf{c}}\lrcorner(\theta^{\mathbf{d}%
}\lrcorner d\omega_{\mathbf{cd}})+\theta^{\mathbf{a}}\lrcorner\theta
^{\mathbf{b}}\lrcorner\left(  \omega_{\mathbf{ac}}\wedge\omega_{\mathbf{b}%
}^{\mathbf{c}}\right)  . \label{8.16}%
\end{equation}
Writing $\omega_{\mathbf{b}}^{\mathbf{a}}=\omega_{\mathbf{bc}}^{\mathbf{a}%
}\theta^{\mathbf{c}}$ we verify that
\begin{equation}
\theta^{\mathbf{a}}\lrcorner\theta^{\mathbf{b}}\lrcorner\left(  \omega
_{\mathbf{ac}}\wedge\omega_{\mathbf{b}}^{\mathbf{c}}\right)  =\eta
^{\mathbf{bk}}\left(  \omega_{\mathbf{kc}}^{\mathbf{d}}\omega_{\mathbf{db}%
}^{\mathbf{c}}-\omega_{\mathbf{dc}}^{\mathbf{d}}\omega_{\mathbf{kb}%
}^{\mathbf{c}}\right)  , \label{8.17}%
\end{equation}
and moreover,
\begin{equation}
\star\lbrack\theta^{\mathbf{c}}\lrcorner(\theta^{\mathbf{d}}\lrcorner
d\omega_{\mathbf{cd}})]=-d\left(  \theta^{\mathbf{a}}\wedge\star
d\theta_{\mathbf{a}}\right)  . \label{8.18}%
\end{equation}
Now, since
\begin{equation}
\omega^{\mathbf{cd}}=\frac{1}{2}\left[  \theta^{\mathbf{d}}\lrcorner
d\theta^{\mathbf{c}}-\theta^{\mathbf{c}}\lrcorner d\theta^{\mathbf{d}}%
+\theta^{\mathbf{c}}\lrcorner\left(  \theta^{\mathbf{d}}\lrcorner
d\theta_{\mathbf{a}}\right)  \theta^{\mathbf{a}}\right]  , \label{8.19}%
\end{equation}
using Eq.(\ref{8.19}) in Eq.(\ref{8.12}) we get,%
\begin{align}
\mathcal{L}_{\mathtt{g}}  &  =-\frac{1}{2}\tau_{\mathtt{\mathbf{g}}}%
\theta^{\mathbf{a}}\lrcorner\theta^{\mathbf{b}}\lrcorner\{\frac{1}{2}%
[\theta_{\mathbf{a}}\lrcorner d\theta_{\mathbf{c}}+\theta_{\mathbf{c}%
}\lrcorner d\theta_{\mathbf{a}}+\theta_{\mathbf{a}}\lrcorner(\theta
_{\mathbf{c}}\lrcorner d\theta_{\mathbf{k}})\theta^{\mathbf{k}}]\nonumber\\
&  \wedge\frac{1}{2}[\theta_{\mathbf{b}}\lrcorner d\theta^{\mathbf{c}}%
+\theta^{\mathbf{c}}\lrcorner d\theta_{\mathbf{b}}+\theta^{\mathbf{c}%
}\lrcorner(\theta_{\mathbf{b}}\lrcorner d\theta^{\mathbf{l}})\theta
_{\mathbf{l}}]\}, \label{8.20}%
\end{align}
which after some algebraic manipulations reduces to Eq.(\ref{8.1}), i.e.,%
\begin{equation}
\mathcal{L}_{g}=-\frac{1}{2}d\theta^{\mathbf{a}}\wedge\star d\theta
_{\mathbf{a}}+\frac{1}{2}\delta\theta^{\mathbf{a}}\wedge\star\delta
\theta_{\mathbf{a}}+\frac{1}{4}\left(  d\theta^{\mathbf{a}}\wedge
\theta_{\mathbf{a}}\right)  \wedge\star\left(  d\theta^{\mathbf{b}}%
\wedge\theta_{\mathbf{b}}\right)  . \label{8.1bis}%
\end{equation}

\end{remark}

The Lagrangian density $\mathcal{L}_{g}$ looks like the Lagrangians of gauge
theories. The first term is of the Yang-Mills type. The second term, will be
called the gauge fixing term, since as can be verified $\delta\theta
^{\mathbf{a}}=0$ is equivalent to the harmonic gauge as we already observed
above. The third term is the auto-interacting term, responsible for the
nonlinearity of Einstein's equations. Lagrangians of this type have been
discussed by some authors, see, e.g., \cite{wallner}, where no use of the
Clifford bundle formalism is used. In \cite{17} $\mathcal{L}_{g}$ has been
used to give a theory of the gravitational field in Minkowski spacetime, by
writing $\star$ in terms of the Hodge dual associated \ to a constant
Minkowski metric defined in the world manifold $M$ which is supposed
diffeomorphic to $\mathbb{R}^{4}$

\section{Conclusions}

We discussed in details in this paper the genesis of an ambiguous statement
called `tetrad postulate', \ which should be more precisely called naive
tetrad postulate. We show that if the naive `tetrad postulate' is not used in
a very \textit{special} context-- where it has a precise meaning as a correct
mathematical statement--namely, that the Eq.(\ref{TETRAD POSTULATE})
$\mathbf{\nabla Q=0}$ \ \ is satisfied (an intrinsic expression of the
\textit{obvious} freshman identity given in Eq.(\ref{18})) it may produce some
serious misunderstandings. We give explicit examples of such misunderstandings
appearing in many books and articles.

We presented moreover a detailed derivation\footnote{These equations already
appeared in \cite{14,17}, but the necessary theorems (proved in this report)
needed to prove them have not been given there.} (including all the necessary
mathematical theorems) of the correct differential equations satisfied by the
(\textit{co})\textit{tetrad} fields $\theta^{\mathbf{a}}=q_{\mu}^{\mathbf{a}%
}dx^{\mu}$ on a Lorentzian manifold, modelling a gravitational field in
General Relativity. This has been done using modern mathematical tools, namely
the theory of \textit{Clifford bundles} and the theory of the square of the
Dirac operator. The correct equations are to be compared with the ones given,
e.g., in (\cite{1,2,3,4,5,kaniel}) and which also \ appears as Eq.(49E) in
\textbf{\cite{0}}. We derive also the tetrad equations in General Relativity
from a variational principle.

\ The functions $q_{\mu}^{\mathbf{a}}$ \ appearing as components of the
tetrads $\theta^{\mathbf{a}}$ \ in a coordinate basis, appear also as
components of the tensor $\mathbf{Q}=q_{\mu}^{\mathbf{a}}\mathbf{e}%
_{\mathbf{a}}\otimes dx^{\mu}$ (see Eq.(\ref{sachs 1})) that satisfies
\textit{trivially} in any general Riemann-Cartan spacetime a second order
differential equation, namely Eq.(\ref{q rc1}). From that equation, we derived
\ for the particular case of a general Lorentzian spacetime a \ `wave
equation' for the functions $q_{\mu}^{\mathbf{a}}$. Since a wave equation for
the functions $q_{\mu}^{\mathbf{a}}$ \ can also be derived from the correct
equations satisfied by the $\theta^{\mathbf{a}}$\ in General Relativity, by
comparing both equations we obtained a constraint equation (Eq.(\ref{l4})).
That equation couples the functions $q_{\mu}^{\mathbf{a}}$, the components of
the Ricci tensor and the components of the energy-momentum tensor and its
trace. \ 

In a series of papers \cite{0,1,2,3,4,5,20} \ (to quote some of them)
a\ `unified field theory' is proposed. In \cite{0} it is claimed that such
'unified theory' follows from a so called\ `Evans Lemma' of differential
geometry. We proved that as presented in \cite{0} `Evans Lemma' \ is a
\textit{false} statement. Then it follows that the \ `unified field theory' is
wrong. \ Before closing, it is eventually worth to give additional pertinent
comments concerning some other statements in \cite{0}.

At page 442 of \textbf{\cite{0}}, concerning his discovery of the \ \ `Evans
Lemma', i.e., the wrong Eq.(2E), the author said:

`The Lemma is an identity of differential geometry, and so is comparable in
generality and power to the well-known Poincar\'{e} Lemma [14]. In other
words, new theorems of topology can be developed from the Evans Lemma in
analogy with topological theorems [2,14] from Poincar\'{e} Lemma.'

Well, a \textit{correct} corollary (not lemma, please)] of Eq.(\ref{18}),
which in intrinsic form reads \ $\mathbf{\nabla Q}=0$, is simply our
Eq.(\ref{q rc1}), $g^{\nu\mu}\mathbf{\nabla}_{{\mbox{\boldmath$\partial$}}%
_{\nu}}\mathbf{\nabla}_{{\mbox{\boldmath$\partial$}}_{\mu}}\mathbf{Q}=0$. The
author of \cite{0} derived a wrong equation from the components $q_{\mu
}^{\mathbf{a}}$ of \textbf{Q} and dubbed this equation Evans lemma of
differential geometry. So, we leave to the reader to judge if such a
triviality has the same status of the Poincar\'{e} lemma.

Note that we did not comment on \ many other errors \ in \cite{0} and in
particular on Section 3 of that paper. But we emphasize that they are subtle
confusions there as some of the ones we have enough patience to
describe\ above. Those confusions are of the same caliber as the following on
that we can find in \cite{20} and which according to our view is a very
convincing proof of the sloppiness of \cite{0,1,2,3,4,5} and other papers from
that author and collaborators. Indeed, e.g., in \cite{20}, Evans and his
coauthor Clements\footnote{At the time of publication, a Ph.D. student at
Oxford University.} try to identify Sachs supposed\footnote{On this issue see
\cite{13,14}.} `electromagnetic' field (which Sachs believes to follow from
his `unified' theory) with a supposed existing longitudinal electromagnetic
field predicted by Evans \ `theory', the so-called \textbf{B}(3) mentioned
several times in \textbf{\cite{0}} and the other papers we quoted. Well, on
\cite{20} we can read at the beginning of section 1.1:

\textquotedblleft The antisymmetrized form of special relativity [1] has
spacetime metric given by the enlarged structure%
\begin{equation}
\eta^{\mu\nu}=\frac{1}{2}\left(  \sigma^{\mu}\sigma^{\nu\ast}+\sigma^{\nu
}\sigma^{\mu\ast}\right)  , \tag{1.1.}%
\end{equation}
where $\sigma^{\mu}$ are the Pauli matrices satisfying a Clifford algebra
\[
\{\sigma^{\mu},\sigma^{\nu}\}=2\delta^{\mu\nu},
\]
which are represented by
\begin{equation}
\sigma^{0}=\left(
\begin{array}
[c]{cc}%
1 & 0\\
0 & 1
\end{array}
\right)  ,\sigma^{1}=\left(
\begin{array}
[c]{cc}%
0 & 1\\
1 & 0
\end{array}
\right)  ,\sigma^{2}=\left(
\begin{array}
[c]{cc}%
0 & -i\\
i & 0
\end{array}
\right)  ,\sigma^{3}=\left(
\begin{array}
[c]{cc}%
1 & 0\\
0 & -1
\end{array}
\right)  . \tag{1.2}%
\end{equation}
The $\ast$ operator denotes quaternion conjugation, which translates to a
spatial parity transformation.\textquotedblright

Well, we comment as follows: the $\ast$ is not really defined anywhere in
\cite{20}. If it refers to a spatial parity operation, we infer that
$\sigma^{0\ast}=\sigma^{0}$ and \ $\sigma^{i\ast}=-\sigma^{i}$. Also,
$\eta^{\mu\nu}$ is not defined, but Eq.(3.5) of \cite{20} makes us to infer
that $\eta^{\mu\nu}=$ \textrm{diag}$(1,-1,-1,-1)$. In that case Eq.(1.1) above
(with the first member understood as $\eta^{\mu\nu}\sigma^{0}$) is true but
the equation $\{\sigma^{\mu},\sigma^{\nu}\}=2\delta^{\mu\nu}$ is false. Enough
is to see that \ $\{\sigma^{0},\sigma^{i}\}=2\sigma^{i}$ $\neq2\delta^{0i}$. \ 

\ We left to the reader who fells expert enough on Mathematics matters to set
the final judgment.\medskip

\textbf{Acknowledgement}: Authors are grateful to Dr.. R. Rocha for a careful
reading of the manuscript and to Drs. R. A. Mosna, E. Capelas de Oliveira, J.
Vaz Jr. and Professor G. W. Bruhn for very useful discussions. Of course, we
are the only responsible for eventual errors and will be glad of being
informed of any one, if they are found, since we are only interested in truth
and beauty. And,

\textit{Beauty is truth, truth}

beauty. \textit{That is all ye know}

\textit{on Earth, and all ye need to know.}

\textit{J. Keats}

\appendix

\section{ Counterexample to the \ Naive `Tetrad Postulate'}

(i) Consider the structure $(\mathring{S}^{2},g,$\textbf{$\nabla$}$)$, where
the manifold $\mathring{S}^{2}$ $=\{S^{2}\backslash$north pole$\}\subset
\mathbb{R}^{3}$ is an sphere of radius $R$ excluding the north pole, $g\in\sec
T_{0}^{2}\mathring{S}^{2}$ is a metric field for $\mathring{S}^{2}$, the
natural one that it inherits from euclidean space $\mathbb{R}^{3}$, and
\textbf{$\nabla$} is the Levi-Civita connection on $\mathring{S}^{2}$, which
may be understood as $\mathbf{\nabla}^{+},\mathbf{\nabla}^{-}$ or
$\mathbf{\nabla}$ in each appropriate case.

(ii) Introduce the usual spherical coordinate functions $(x^{1},x^{2}%
)=(\vartheta,\varphi)$, $0<\vartheta<\pi$, $0<\varphi<2\pi$, which covers all
the open set $U$ which is $\mathring{S}^{2}$ with the exclusion of a
semi-circle uniting the north and south poles.

(iii) Introduce first \textbf{coordinate bases }
\begin{equation}
\{{\mbox{\boldmath$\partial$}}_{\mu}\},\{\theta^{\mu}=dx^{\mu}\} \label{1x}%
\end{equation}
for $TU$ and $T^{\ast}U$.

(iv) Then,
\begin{equation}
g=R^{2}dx^{1}\otimes dx^{1}+R^{2}\sin^{2}x^{1}dx^{2}\otimes dx^{2} \label{2x}%
\end{equation}

(v) Introduce now the \textbf{orthonormal bases }$\{\mathbf{e}_{\mathbf{a}%
}\},\{%
\mbox{\boldmath{$\theta$}}%
^{\mathbf{a}}\}$ for $TU$ and $T^{\ast}U$ with%
\begin{align}
\mathbf{e}_{\mathbf{1}}  &  =\frac{1}{R}{\mbox{\boldmath$\partial$}}%
_{1}\text{, }\mathbf{e}_{\mathbf{2}}=\frac{1}{R\sin x^{1}}%
{\mbox{\boldmath$\partial$}}_{2},\label{3x}\\%
\mbox{\boldmath{$\theta$}}%
^{\mathbf{1}}  &  =Rdx^{1}\text{, }%
\mbox{\boldmath{$\theta$}}%
^{\mathbf{2}}=R\sin x^{1}dx^{2}. \label{4x}%
\end{align}
We immediately get that%
\begin{align*}
\lbrack\mathbf{e}_{\mathbf{i}},\mathbf{e}_{\mathbf{j}}]  &  =c_{\mathbf{ij}%
}^{\mathbf{k}}\mathbf{e}_{\mathbf{k}},\\
c_{\mathbf{12}}^{\mathbf{2}}  &  =-c_{\mathbf{21}}^{\mathbf{2}}=-\cot
x^{\mathbf{1}}%
\end{align*}

(vi) Writing%
\begin{equation}
\mathbf{e}_{\mathbf{a}}=q_{\mathbf{a}}^{\mu}{\mbox{\boldmath$\partial$}}_{\mu
},%
\mbox{\boldmath{$\theta$}}%
^{\mathbf{a}}=q_{\mu}^{\mathbf{a}}dx^{\mu}, \label{5x}%
\end{equation}
we read from Eq.(\ref{3x}) and Eq.(\ref{4x}),%
\begin{align}
q_{\mathbf{1}}^{1}  &  =\frac{1}{R}\text{, }q_{\mathbf{1}}^{2}=0,\label{6x}\\
q_{\mathbf{2}}^{1}  &  =0\text{, }q_{\mathbf{2}}^{2}=\frac{1}{R\sin x^{1}%
},\label{7x}\\
q_{1}^{\mathbf{1}}  &  =R\text{, }q_{2}^{\mathbf{1}}=0,\label{8x}\\
q_{1}^{\mathbf{2}}  &  =0\text{, }q_{2}^{\mathbf{2}}=R\sin x^{1}. \label{9x}%
\end{align}

(vii) Christoffel symbols. Before proceeding we put for \textit{simplicity}
$R=1$. Then, the non zero Christoffel symbols are:
\begin{align}
\mathbf{\nabla}_{{\mbox{\boldmath$\partial$}}_{\mu}}^{+}%
{\mbox{\boldmath$\partial$}}_{\nu}  &  =\Gamma_{\mu\nu}^{\rho}%
{\mbox{\boldmath$\partial$}}_{\rho},\nonumber\\
\Gamma_{22}^{1}  &  =\Gamma_{\varphi\varphi}^{\vartheta}=-\cos\vartheta
\sin\vartheta\text{, }\Gamma_{21}^{2}=\Gamma_{\theta\varphi}^{\varphi}%
=\Gamma_{12}^{2}=\Gamma_{\varphi\theta}^{\varphi}=\cot\vartheta. \label{10x}%
\end{align}

(viii) Then we have, e.g.,
\begin{align}
\mathbf{\nabla}_{{\mbox{\boldmath$\partial$}}_{2}}^{-}\theta^{2}  &  =\cot
x^{1}\theta^{1}=\cot\vartheta\theta^{1}\label{11x}\\
\text{ }\mathbf{\nabla}_{{\mbox{\boldmath$\partial$}}_{2}}^{-}\theta^{1}  &
=\cos x^{1}\sin x^{1}\theta^{2}=\cos\vartheta\sin\vartheta\theta^{2}\\
\mathbf{\nabla}_{{\mbox{\boldmath$\partial$}}_{1}}^{-}\theta^{2}  &  =-\cot
x^{1}\theta^{2}=-\cot\vartheta\theta^{2},\\
\mathbf{\nabla}_{{\mbox{\boldmath$\partial$}}_{1}}^{-}\theta^{1}  &  =0
\end{align}

(ix) We also have, e.g.,
\begin{align}
\mathbf{\nabla}_{{\mbox{\boldmath$\partial$}}_{2}}^{-}%
\mbox{\boldmath{$\theta$}}%
^{\mathbf{2}}  &  =\mathbf{\nabla}_{{\mbox{\boldmath$\partial$}}_{2}}%
^{-}\left(  q_{\mu}^{\mathbf{2}}\theta^{\mu}\right)  =\mathbf{\nabla
}_{{\mbox{\boldmath$\partial$}}_{2}}\left(  q_{\mu}^{\mathbf{2}}dx^{\mu
}\right) \nonumber\\
&  =\mathbf{\nabla}_{{\mbox{\boldmath$\partial$}}_{2}}^{-}\left(  \sin
x^{1}dx^{2}\right)  =\sin x^{1}\mathbf{\nabla}_{{\mbox{\boldmath$\partial$}}%
_{2}}^{-}dx^{2}=-\cos x^{1}dx^{1}\nonumber\\
&  =(\mathbf{\nabla}_{2}^{-}q_{\mu}^{\mathbf{2}})dx^{\mu}. \label{14x}%
\end{align}

Then, the symbols \textbf{$\nabla$}$_{2}^{-}q_{1}^{\mathbf{2}}$ and
\textbf{$\nabla$}$_{2}^{-}q_{2}^{\mathbf{2}}$ are according to Eq.(\ref{19})
\begin{align}
\mathbf{\nabla}_{2}^{-}q_{1}^{\mathbf{2}}  &  =-\cos x^{1}\neq0,\nonumber\\
\mathbf{\nabla}_{2}^{-}q_{2}^{\mathbf{2}}  &  =0. \label{14ax}%
\end{align}

This seems strange, but is correct, because of the \textit{definition} of the
symbols \textbf{$\nabla$}$_{\mu}^{-}q_{\nu}^{\mathbf{a}}$ (see Eq.(\ref{16})
and Eq.(\ref{17})) . Now, even if $q_{1}^{\mathbf{2}}=0$, and $q_{2}%
^{\mathbf{2}}=\sin x^{1}$, we get,%
\begin{align}
\mathbf{\nabla}_{1}^{-}q_{2}^{\mathbf{2}}  &  =\frac{\partial}{\partial x^{1}%
}q_{2}^{\mathbf{2}}-\Gamma_{12}^{1}q_{1}^{\mathbf{2}}-\Gamma_{12}^{2}%
q_{2}^{\mathbf{2}}=-\Gamma_{21}^{2}q_{2}^{\mathbf{2}}=\cos x^{1}-\cos
x^{1}=0,\nonumber\\
\mathbf{\nabla}_{2}^{-}q_{2}^{\mathbf{2}}  &  =\frac{\partial}{\partial x^{2}%
}q_{2}^{\mathbf{2}}-\Gamma_{22}^{1}q_{1}^{\mathbf{2}}-\Gamma_{22}^{2}%
q_{2}^{\mathbf{2}}=\frac{\partial}{\partial x^{2}}(\sin x^{1})-(-\sin
x^{1}\cos x^{1})(0)-(0)(\sin x^{1})=0. \label{16x}%
\end{align}

For future reference we note also that%
\begin{align}
\mathbf{\nabla}_{1}^{-}q_{1}^{\mathbf{1}}  &  =0\text{, }\mathbf{\nabla}%
_{1}^{-}q_{2}^{\mathbf{1}}=0\text{, }\mathbf{\nabla}_{1}^{-}q_{1}^{\mathbf{2}%
}=0\text{, }\label{16xx}\\
\mathbf{\nabla}_{2}^{-}q_{1}^{\mathbf{1}}  &  =0\text{, }\mathbf{\nabla}%
_{2}^{-}q_{2}^{\mathbf{1}}=\cos x^{1}\sin x^{1}\text{, }\mathbf{\nabla}%
_{2}^{-}q_{1}^{\mathbf{2}}=-\cos x^{1}%
\end{align}

So, in definitive we exhibit a \textit{counterexample} to the \textit{naive}%
\ \ `tetrad postulate' (when \textbf{$\nabla$}$_{\mu}^{-}q_{\nu}^{\mathbf{a}}$
is written \textbf{$\nabla$}$_{\mu}q_{\nu}^{\mathbf{a}}$ and interpreted by an
equation Eq.(\ref{19})), because we just found, e.g., that \textbf{$\nabla$%
}$_{2}^{-}q_{1}^{\mathbf{2}}=-\cos x^{1}\neq0.$

Note that in our example, if it happened that all the symbols \textbf{$\nabla
$}$_{\mu}^{-}q_{\nu}^{\mathbf{a}}=0,$ it would result that \textbf{$\nabla$%
}$_{\mathbf{e}_{\mathbf{b}}}\mathbf{e}_{\mathbf{a}}=0,$ for $\mathbf{a,b}%
=1,2$. In that case the Riemann curvature tensor of \textbf{$\nabla$} would be
\textit{null} and the torsion tensor would be \textit{non null}. But this
would be a contradiction, because in that case \textbf{$\nabla$} would not be
the Levi-Civita connection as supposed.

Suppose now that we calculate the symbols \textbf{$\nabla$}$_{\mu}^{+}q_{\nu
}^{\mathbf{a}}$ for our problem

We get,
\begin{equation}
\mathbf{\nabla}_{1}^{+}q_{2}^{\mathbf{1}}=0,\mathbf{\nabla}_{1}^{+}%
q_{2}^{\mathbf{2}}=\cos\vartheta\mathbf{,\nabla}_{2}^{+}q_{2}^{\mathbf{1}%
}=0,\mathbf{\nabla}_{2}^{+}q_{1}^{\mathbf{2}}=\cos\vartheta, \label{torsion5}%
\end{equation}
and the torsion tensor is zero, as it may be. However, if we forget about the
necessary distinction of symbols and use the symbols \textbf{$\nabla$}$_{\mu
}^{-}q_{\nu}^{\mathbf{a}}$ to calculate the torsion tensor we would get the
wrong result.%
\begin{equation}
T_{12}^{\mathbf{2}}=\cos\vartheta\text{, }T_{21}^{\mathbf{2}}=-\cos\vartheta.
\label{torsion 6}%
\end{equation}

Of course, we can define for the manifold $\mathring{S}_{2}$ \ introduced
above a metric compatible teleparallel connection $\overset{c}{\mathbf{\nabla
}}$ (the so-called navigator or Columbus connection \cite{9}), by imposing
that
\begin{equation}
\overset{c}{\mathbf{\nabla}^{+}}_{\mathbf{e}_{\mathbf{a}}}\mathbf{e}%
_{\mathbf{b}}=0,\text{ }\mathbf{a,b=1,2} \label{torsion 6 bis}%
\end{equation}
This corresponds to the following transport law. A vector is parallel
transported along a curve $C$ if the angle between the vector and the latitude
line intersecting the curve $C$ is kept constant. For that
\ \textit{particular} connection the statement $\overset{c}{\mathbf{\nabla
}\text{ }^{-}}_{1}q_{\nu}^{\mathbf{a}}=0$\ is correct. However, we may verify
that for that connection the $\overset{c}{\mathbf{\nabla}\text{ }^{+}}_{\mu
}q_{\nu}^{\mathbf{a}}$ are not all null and the torsion is \textit{not} null,
for we have $T_{\mathbf{12}}^{\mathbf{2}}=-T_{\mathbf{12}}^{\mathbf{2}}%
=\cot\vartheta$ And of course, should\ we use naively the always true equation
$\overset{c}{\mathbf{\nabla}}_{\mu}q_{\nu}^{\mathbf{a}}=0$ , and use
$\overset{c}{\mathbf{\nabla}}_{\mu}q_{\nu}^{\mathbf{a}}$ instead $\overset
{c}{\mathbf{\nabla}\text{ }^{+}}_{\mu}q_{\nu}^{\mathbf{a}}$ of to calculate
the components of torsion tensor we would obtain that it would be null, a contradiction.

\section{Variation of $\mathcal{L}_{EH}$}

Given a Lagrangian density $\mathcal{L}_{\wedge}(\phi)=\mathcal{L}_{\wedge
}\mathcal{(}x,\phi,d\phi)$ for a homogenous field $\phi\in\sec\bigwedge
^{r}T^{\ast}M\hookrightarrow\sec\mathcal{C}\ell(M,\mathtt{g})$ the functional
derivative (or Euler Lagrange functional) of $\mathcal{L}_{\wedge}$ is the
functional $\mathcal{\star}\mathbf{\Sigma}$, with $\mathcal{\star
}\mathbf{\Sigma}\mathcal{(\phi)=}\frac{%
\mbox{\boldmath{$\delta$}}%
\mathcal{L}_{\wedge}}{%
\mbox{\boldmath{$\delta$}}%
\phi}(\phi)\in\sec\bigwedge^{4-r}T^{\ast}M\hookrightarrow\sec\mathcal{C}%
\ell(M,\mathtt{g})$ such that\
\begin{subequations}
\begin{align}%
\mbox{\boldmath{$\delta$}}%
\mathcal{L}_{\wedge}(\phi) &  =%
\mbox{\boldmath{$\delta$}}%
\phi\wedge\frac{\partial\mathcal{L}_{\wedge}(\phi)}{\partial\phi}+%
\mbox{\boldmath{$\delta$}}%
(d\phi)\wedge\frac{\partial\mathcal{L}_{\wedge}(\phi)}{\partial d\phi
}\nonumber\\
&  =%
\mbox{\boldmath{$\delta$}}%
\phi\wedge\frac{\partial\mathcal{L}_{\wedge}(\phi)}{\partial\phi}+d(%
\mbox{\boldmath{$\delta$}}%
\phi)\wedge\frac{\partial\mathcal{L}_{\wedge}(\phi)}{\partial d\phi
}\nonumber\\
&  =%
\mbox{\boldmath{$\delta$}}%
\phi\wedge\left(  \frac{\partial\mathcal{L}_{\wedge}(\phi)}{\partial\phi
}-(-1)^{r}d\left(  \frac{\partial\mathcal{L}_{\wedge}(\phi)}{\partial d\phi
}\right)  \right)  +d\left(
\mbox{\boldmath{$\delta$}}%
\phi\wedge\frac{\partial\mathcal{L}_{\wedge}(\phi)}{\partial d\phi}\right)
\nonumber\\
&  =%
\mbox{\boldmath{$\delta$}}%
\phi\wedge\frac{%
\mbox{\boldmath{$\delta$}}%
\mathcal{L}_{\wedge}}{%
\mbox{\boldmath{$\delta$}}%
\phi}+d\left(
\mbox{\boldmath{$\delta$}}%
\phi\wedge\frac{\partial\mathcal{L}_{\wedge}(\phi)}{\partial d\phi}\right)
\nonumber\\
&  =%
\mbox{\boldmath{$\delta$}}%
\phi\wedge\mathcal{\star}\mathbf{\Sigma}\mathcal{(\phi)}+d\left(
\mbox{\boldmath{$\delta$}}%
\phi\wedge\frac{\partial\mathcal{L}_{\wedge}(\phi)}{\partial d\phi}\right)
,\label{7.1a}%
\end{align}%
\end{subequations}
\begin{equation}
\mathcal{\star}\mathbf{\Sigma}\mathcal{(\phi)}=\frac{\partial\mathcal{L}%
_{\wedge}(\phi)}{\partial\phi}-(-1)^{r}d\left(  \frac{\partial\mathcal{L}%
_{\wedge}(\phi)}{\partial d\phi}\right)  \label{7.1bb}%
\end{equation}

\begin{definition}
The terms $\frac{\partial\mathcal{L}_{\wedge}}{\partial\phi}$ and
$\frac{\partial\mathcal{L}_{\wedge}}{\partial d\phi}$ are called in what
follows algebraic derivatives of $\mathcal{L}_{\wedge}$\ \emph{\footnote{This
terminology was originally introduced in \cite{thiwal}. The exterior product
\ $%
\mbox{\boldmath{$\delta$}}%
\phi\wedge\frac{\partial}{\partial\phi}$ is a particular instance of the
$A\wedge\frac{\partial}{\partial\phi}$ directional derivatives introduced in
the multiform calculus developed in \cite{fmr,fmr2,fmr3,fmr4,fmr5,fmr6,fmr7}
with $%
\mbox{\boldmath{$\delta$}}%
\phi=A$.}} and
\end{definition}

For our present problem, we are fortunate, since we only need to know the
following rule \cite{thirring,wallner} (besides, of course a series of
identities of the Clifford bundle formalism that we summarized in Section 8):
Given two action functionals depending, say, only of $\phi\sec\bigwedge
^{r}T^{\ast}M$ , $\mathcal{F(\phi)\in}\sec\bigwedge^{p}T^{\ast}M$ and
$\mathcal{K(\phi)\in}\sec\bigwedge^{q}T^{\ast}M,$%
\begin{equation}
\frac{\partial}{\partial\phi}[\mathcal{F(\phi)\wedge K(\phi)}]=\frac{\partial
}{\partial\phi}\mathcal{F(\phi)\wedge K(\phi)}+(-1)^{pr}\mathcal{F(\phi
)\wedge}\frac{\partial}{\partial\phi}\mathcal{K(\phi)}.\label{7.2}%
\end{equation}

With these preliminaries, we can find the algebraic derivatives $\frac
{\partial\mathcal{L}_{EH}}{\partial\theta^{\mathbf{d}}}$ and $\frac
{\partial\mathcal{L}_{EH}}{\partial d\theta^{\mathbf{d}}}$ of Einstein-Hilbert
Lagrangian density $\mathcal{L}_{EH}$, necessary to obtain its variation. We
know that
\begin{equation}
\mathcal{L}_{EH}=-d(\theta^{\mathbf{a}}\wedge\star d\theta_{\mathbf{a}}%
)-\frac{1}{2}d\theta^{\mathbf{a}}\wedge\star d\theta_{\mathbf{a}}+\frac{1}%
{2}\delta\theta^{\mathbf{a}}\wedge\star\delta\theta_{\mathbf{a}}+\frac{1}%
{4}\left(  d\theta^{\mathbf{a}}\wedge\theta_{\mathbf{a}}\right)  \wedge
\star\left(  d\theta^{\mathbf{b}}\wedge\theta_{\mathbf{b}}\right)
\label{7.ex0}%
\end{equation}

Before proceeding we must take into account that for any $\phi\in\sec
\bigwedge^{r}T^{\ast}M\hookrightarrow\sec\mathcal{C}\ell(M,\mathtt{g})$, it
holds as can be easily verified
\begin{align}
\lbrack%
\mbox{\boldmath{$\delta$}}%
,\star]\phi &  =%
\mbox{\boldmath{$\delta$}}%
\star\phi-\star%
\mbox{\boldmath{$\delta$}}%
\phi\label{7.ex00}\\
&  =%
\mbox{\boldmath{$\delta$}}%
\theta^{\mathbf{a}}\wedge\left(  \theta_{\mathbf{a}}\lrcorner\star\phi\right)
-\star\left[
\mbox{\boldmath{$\delta$}}%
\theta^{\mathbf{a}}\wedge\left(  \theta_{\mathbf{a}}\lrcorner\phi\right)
\right] \nonumber
\end{align}
Then since if $\phi=\theta_{\mathbf{c}}$ any variation induced by a local
Lorentz rotation or by an arbitrary diffeomorphism must be a constrained
variation of the Lorentz type \cite{quintino}, i.e., $%
\mbox{\boldmath{$\delta$}}%
\theta_{\mathbf{c}}$ $=\chi_{\mathbf{cd}}\theta^{\mathbf{d}}$, $\chi
_{\mathbf{cd}}=-\chi_{\mathbf{dc}}$. If that is the case, we have that for any
product of $1$-forms $\theta^{\mathbf{a}}\wedge...\wedge\theta^{\mathbf{d}}$
since $\star\left(  \theta^{\mathbf{a}}\wedge...\theta^{\mathbf{d}}\right)
=\widetilde{\left(  \theta^{\mathbf{a}}\wedge...\wedge\theta^{\mathbf{d}%
}\right)  }\lrcorner(\theta^{\mathbf{0}}\wedge\theta^{\mathbf{1}}\wedge
\theta^{\mathbf{2}}\wedge\theta^{\mathbf{3}})$ \ it holds that
\begin{equation}%
\mbox{\boldmath{$\delta$}}%
\star\left(  \theta^{\mathbf{a}}\wedge...\wedge\theta^{\mathbf{d}}\right)  =%
\mbox{\boldmath{$\delta$}}%
\theta_{\mathbf{c}}\wedge\left[  \theta^{\mathbf{c}}\lrcorner\star\left(
\theta^{\mathbf{a}}\wedge...\wedge\theta^{\mathbf{d}}\right)  \right]  =%
\mbox{\boldmath{$\delta$}}%
\theta_{\mathbf{c}}\wedge\star\left(  \theta^{\mathbf{a}}\wedge...\wedge
\theta^{\mathbf{d}}\wedge\theta^{\mathbf{c}}\right)  . \label{7.ex000}%
\end{equation}

The first term in Eq.(\ref{7.ex0}), $\mathcal{L}^{(1)}=-d(\theta^{\mathbf{a}%
}\wedge\star d\theta_{\mathbf{a}})$ is an exact differential and so did not
contribute for the variation of the action. The variation of the second term,
$\mathcal{L}^{(2)}=-\frac{1}{2}d\theta^{\mathbf{a}}\wedge\star d\theta
_{\mathbf{a}}$ is calculated as follows. We have
\begin{equation}
\theta^{\mathbf{b}}\wedge\theta^{\mathbf{c}}\wedge\star d\theta_{\mathbf{a}%
}=d\theta_{\mathbf{a}}\wedge\star(\theta^{\mathbf{b}}\wedge\theta^{\mathbf{c}%
}). \label{7.ex1}%
\end{equation}
Then, writing \ $\theta^{\mathbf{bc}}=\theta^{\mathbf{b}}\wedge\theta
^{\mathbf{c}}$ we have
\begin{align}
\theta^{\mathbf{bc}}\wedge%
\mbox{\boldmath{$\delta$}}%
\star d\theta_{\mathbf{a}}  &  =%
\mbox{\boldmath{$\delta$}}%
d\theta_{\mathbf{a}}\wedge\star\theta^{\mathbf{bc}}+d\theta_{\mathbf{a}}\wedge%
\mbox{\boldmath{$\delta$}}%
\star\theta^{\mathbf{bc}}-%
\mbox{\boldmath{$\delta$}}%
\theta^{\mathbf{bc}}\wedge\star d\theta_{\mathbf{a}}\nonumber\\
&  =%
\mbox{\boldmath{$\delta$}}%
d\theta_{\mathbf{a}}\wedge\star\theta^{\mathbf{bc}}+d\theta_{\mathbf{a}}\wedge%
\mbox{\boldmath{$\delta$}}%
\theta^{\mathbf{d}}\wedge(\theta_{\mathbf{d}}\lrcorner\star\theta
^{\mathbf{bc}})-%
\mbox{\boldmath{$\delta$}}%
\theta^{\mathbf{d}}\wedge(\theta_{\mathbf{d}}\lrcorner\theta^{\mathbf{bc}%
})\wedge\star d\theta_{\mathbf{a}}\nonumber\\
=  &
\mbox{\boldmath{$\delta$}}%
d\theta_{\mathbf{a}}\wedge\star\theta^{\mathbf{bc}}+%
\mbox{\boldmath{$\delta$}}%
\theta^{\mathbf{d}}\wedge\lbrack d\theta_{\mathbf{a}}\wedge(\theta
_{\mathbf{d}}\lrcorner\star\theta^{\mathbf{bc}})-(\theta_{\mathbf{a}}%
\lrcorner\theta^{\mathbf{bc}})\wedge\star d\theta_{\mathbf{a}}]. \label{7.ex2}%
\end{align}
Multiplying Eq.(\ref{7.ex2}) by $\frac{1}{2!}(d\theta_{\mathbf{a}%
})_{\mathbf{bc}}=-\frac{1}{2}\eta_{\mathbf{ad}}c_{\mathbf{bc}}^{\mathbf{d}}$
we get%
\begin{equation}
d\theta^{\mathbf{a}}\wedge%
\mbox{\boldmath{$\delta$}}%
\star d\theta_{\mathbf{a}}=%
\mbox{\boldmath{$\delta$}}%
d\theta^{\mathbf{a}}\wedge\lbrack d\theta_{\mathbf{a}}\wedge(\theta
_{\mathbf{d}}\lrcorner\star d\theta^{\mathbf{a}})-(\theta_{\mathbf{d}%
}\lrcorner d\theta^{\mathbf{a}})\wedge\star d\theta_{\mathbf{a}}. \label{7.b2}%
\end{equation}
Taking into account that $%
\mbox{\boldmath{$\delta$}}%
(d\theta^{\mathbf{a}}\wedge\star d\theta_{\mathbf{a}})=%
\mbox{\boldmath{$\delta$}}%
d\theta^{\mathbf{a}}\wedge\star d\theta_{\mathbf{a}}+d\theta^{\mathbf{a}%
}\wedge%
\mbox{\boldmath{$\delta$}}%
\star d\theta_{\mathbf{a}})$ we have%
\begin{equation}%
\mbox{\boldmath{$\delta$}}%
\mathcal{L}^{(2)}=-%
\mbox{\boldmath{$\delta$}}%
d\theta^{\mathbf{d}}\wedge\star d\theta^{\mathbf{d}}+\frac{1}{2}%
\mbox{\boldmath{$\delta$}}%
\theta^{\mathbf{d}}\wedge\left[  \theta_{\mathbf{d}}\lrcorner d\theta
^{\mathbf{a}}\wedge\star d\theta_{\mathbf{a}}-d\theta_{\mathbf{a}}%
\wedge(\theta_{\mathbf{d}}\lrcorner\star d\theta^{\mathbf{a}})\right]
\label{7.b3}%
\end{equation}
From Eq.(\ref{7.b3}) it follows that the algebraic derivatives of
$\mathcal{L}^{(2)}$ relative to $\theta^{\mathbf{d}}$ and $d\theta
^{\mathbf{d}}$are:%
\begin{align}
\frac{\partial\mathcal{L}^{(2)}}{\partial\theta^{\mathbf{d}}}  &  =\frac{1}%
{2}\left[  \theta_{\mathbf{d}}\lrcorner d\theta^{\mathbf{a}}\wedge\star
d\theta_{\mathbf{a}}-d\theta_{\mathbf{a}}\wedge(\theta_{\mathbf{d}}%
\lrcorner\star d\theta^{\mathbf{a}})\right]  ,\nonumber\\
\frac{\partial\mathcal{L}^{(2)}}{\partial d\theta^{\mathbf{d}}}  &  =-\star
d\theta_{\mathbf{d}} \label{7.b5}%
\end{align}

The variation of the third term in $\mathcal{L}_{EH}$, $\mathcal{L}%
^{(3)}=\frac{1}{2}\delta\theta^{\mathbf{a}}\wedge\star\delta\theta
_{\mathbf{a}}=-\frac{1}{2}\star d\star\theta^{\mathbf{a}}\wedge d\star
\theta_{\mathbf{a}}$ is calculated as follows. First, we observe that
$\theta^{\mathbf{bcrs}}\wedge\star d\star\theta^{\mathbf{a}}=d\star
\theta^{\mathbf{a}}\wedge\star\theta^{\mathbf{bcrs}}$. Then
\begin{align}
\theta^{\mathbf{bcrs}}\wedge%
\mbox{\boldmath{$\delta$}}%
\star d\star\theta_{\mathbf{a}}  &  =%
\mbox{\boldmath{$\delta$}}%
d\star\theta_{\mathbf{a}}\wedge\star\theta^{\mathbf{bcrs}}+d\star
\theta_{\mathbf{a}}\wedge%
\mbox{\boldmath{$\delta$}}%
\star\theta^{\mathbf{bcrs}}-%
\mbox{\boldmath{$\delta$}}%
\theta^{\mathbf{bcrs}}\wedge\star d\star\theta_{\mathbf{a}}\nonumber\\
&  =d%
\mbox{\boldmath{$\delta$}}%
\star\theta_{\mathbf{a}}\wedge\star\theta^{\mathbf{bcrs}}-%
\mbox{\boldmath{$\delta$}}%
\theta^{\mathbf{d}}\wedge(\theta_{\mathbf{d}}\lrcorner\theta^{\mathbf{bcrs}%
})\wedge\star d\star\theta_{\mathbf{a}} \label{7.b6}%
\end{align}

Multiplying Eq.(\ref{7.b6}) by the coefficients $\frac{1}{4!}(d\star
\theta^{\mathbf{a}})_{\mathbf{bcrs}}$ we get%
\begin{equation}
d\star\theta^{\mathbf{a}}\wedge%
\mbox{\boldmath{$\delta$}}%
\star d\star\theta_{\mathbf{a}}=%
\mbox{\boldmath{$\delta$}}%
d\star\theta_{\mathbf{a}}\wedge\star d\star\theta^{\mathbf{a}}-%
\mbox{\boldmath{$\delta$}}%
\theta^{\mathbf{d}}\wedge\left(  \theta_{\mathbf{d}}\lrcorner d\star
\theta^{\mathbf{a}}\right)  \wedge\star d\star\theta_{\mathbf{a}} \label{7.b7}%
\end{equation}
The first member of the right hand side of Eq.(\ref{7.b7} ) gives%
\begin{align}%
\mbox{\boldmath{$\delta$}}%
d\star\theta_{\mathbf{a}}\wedge\star d\star\theta^{\mathbf{a}}  &  =d%
\mbox{\boldmath{$\delta$}}%
\star\theta^{\mathbf{a}}\wedge\star d\star\theta_{\mathbf{a}}\nonumber\\
&  =d\left[
\mbox{\boldmath{$\delta$}}%
\theta^{\mathbf{d}}\wedge(\theta_{\mathbf{d}}\lrcorner\star\theta^{\mathbf{a}%
})\right]  \wedge\star d\star\theta_{\mathbf{a}}\nonumber\\
=  &
\mbox{\boldmath{$\delta$}}%
d\theta^{\mathbf{d}}\wedge\left(  \theta_{\mathbf{d}}\lrcorner\star
\theta^{\mathbf{a}}\right)  \wedge\star d\star\theta_{\mathbf{a}}-%
\mbox{\boldmath{$\delta$}}%
\theta^{\mathbf{d}}\wedge d\left(  \theta_{\mathbf{d}}\lrcorner\star
\theta^{\mathbf{a}}\right)  \wedge\star d\star\theta_{\mathbf{a}},
\label{7.b66}%
\end{align}
and recalling that $%
\mbox{\boldmath{$\delta$}}%
\left(  d\star\theta^{\mathbf{a}}\wedge\star d\star\theta_{\mathbf{a}}\right)
=%
\mbox{\boldmath{$\delta$}}%
d\star\theta^{\mathbf{a}}\wedge\star d\star\theta_{\mathbf{a}}+d\star
\theta^{\mathbf{a}}\wedge%
\mbox{\boldmath{$\delta$}}%
\star d\star\theta_{\mathbf{a}}$ we get%
\begin{align}%
\mbox{\boldmath{$\delta$}}%
\mathcal{L}^{(3)}  &  =-%
\mbox{\boldmath{$\delta$}}%
d\theta^{\mathbf{d}}\wedge\left(  \theta_{\mathbf{d}}\lrcorner\star
\theta^{\mathbf{a}}\right)  \wedge\star d\star\theta_{\mathbf{a}}\nonumber\\
&  +%
\mbox{\boldmath{$\delta$}}%
\theta^{\mathbf{d}}\wedge\frac{1}{2}\left[  d\left(  \theta_{\mathbf{d}%
}\lrcorner\star\theta^{\mathbf{a}}\right)  \wedge\star d\star\theta
_{\mathbf{a}}+\left(  \theta_{\mathbf{d}}\lrcorner d\star\theta^{\mathbf{a}%
}\right)  \wedge\star d\star\theta_{\mathbf{a}}\right]  . \label{7.b9}%
\end{align}
Then,%
\begin{align}
\frac{\partial\mathcal{L}^{(3)}}{\partial\theta^{\mathbf{d}}}  &  =\frac{1}%
{2}d\left(  \theta_{\mathbf{d}}\lrcorner\star\theta^{\mathbf{a}}\right)
\wedge\star d\star\theta_{\mathbf{a}}+\frac{1}{2}\left(  \theta_{\mathbf{d}%
}\lrcorner d\star\theta^{\mathbf{a}}\right)  \wedge\star d\star\theta
_{\mathbf{a}}\nonumber\\
\frac{\partial\mathcal{L}^{(3)}}{\partial d\theta^{\mathbf{d}}}  &  =-\left(
\theta_{\mathbf{d}}\lrcorner\star\theta^{\mathbf{a}}\right)  \wedge\star
d\star\theta_{\mathbf{a}} \label{7.b10}%
\end{align}

The variation of the fourth term of \ $\mathcal{L}_{EH}$, $\mathcal{L}%
^{(4)}=\frac{1}{4}\left(  d\theta^{\mathbf{a}}\wedge\theta_{\mathbf{a}%
}\right)  \wedge\star\left(  d\theta^{\mathbf{b}}\wedge\theta_{\mathbf{b}%
}\right)  $ is done as follows. First, we observe that since $\theta
^{\mathbf{bcr}}\wedge\star\left(  d\theta^{\mathbf{a}}\wedge\theta
_{\mathbf{a}}\right)  =d\theta^{\mathbf{a}}\wedge\theta_{\mathbf{a}}%
\wedge\star\theta^{\mathbf{bcr}}$ we can write
\begin{align}
\theta^{\mathbf{bcr}}\wedge%
\mbox{\boldmath{$\delta$}}%
\star\left(  d\theta^{\mathbf{a}}\wedge\theta_{\mathbf{a}}\right)   &  =%
\mbox{\boldmath{$\delta$}}%
d\theta^{\mathbf{d}}\wedge\theta_{\mathbf{d}}\wedge\star\theta^{\mathbf{bcr}}+%
\mbox{\boldmath{$\delta$}}%
\theta^{\mathbf{d}}\wedge d\theta_{\mathbf{d}}\wedge\star\theta^{\mathbf{bcr}%
}\nonumber\\
&  -%
\mbox{\boldmath{$\delta$}}%
\theta^{\mathbf{d}}\wedge d\theta^{\mathbf{a}}\wedge\theta_{\mathbf{a}}%
\wedge\left(  \theta_{\mathbf{d}}\lrcorner\star\theta^{\mathbf{bcr}}\right)
\nonumber\\
&  -%
\mbox{\boldmath{$\delta$}}%
\theta^{\mathbf{d}}\wedge\left(  \theta_{\mathbf{d}}\lrcorner\theta
^{\mathbf{bcr}}\right)  \wedge\star\left(  d\theta^{\mathbf{a}}\wedge
\theta_{\mathbf{a}}\right)  . \label{7.b12}%
\end{align}
Multiplying Eq.(\ref{7.b12}) by $\frac{1}{3!}\left(  d\theta^{\mathbf{e}%
}\wedge\theta_{\mathbf{e}}\right)  _{\mathbf{ars}}$ we get%
\begin{align}
d\theta^{\mathbf{e}}\wedge\theta_{\mathbf{e}}\wedge%
\mbox{\boldmath{$\delta$}}%
\star\left(  d\theta^{\mathbf{a}}\wedge\theta_{\mathbf{a}}\right)   &  =%
\mbox{\boldmath{$\delta$}}%
d\theta^{\mathbf{d}}\wedge\theta_{\mathbf{d}}\wedge\star\left(  d\theta
^{\mathbf{a}}\wedge\theta_{\mathbf{a}}\right) \nonumber\\
&  +%
\mbox{\boldmath{$\delta$}}%
\theta^{\mathbf{d}}\wedge d\theta_{\mathbf{d}}\wedge\star\left(
d\theta^{\mathbf{a}}\wedge\theta_{\mathbf{a}}\right) \nonumber\\
&  -%
\mbox{\boldmath{$\delta$}}%
\theta^{\mathbf{d}}\wedge d\theta^{\mathbf{a}}\wedge\theta_{\mathbf{a}}%
\wedge\left[  \theta_{\mathbf{d}}\lrcorner\star\left(  d\theta^{\mathbf{e}%
}\wedge\theta_{\mathbf{e}}\right)  \right] \nonumber\\
&  -%
\mbox{\boldmath{$\delta$}}%
\theta^{\mathbf{d}}\wedge\left[  \theta_{\mathbf{d}}\lrcorner\left(
d\theta^{\mathbf{e}}\wedge\theta_{\mathbf{e}}\right)  \wedge\star\left(
d\theta^{\mathbf{a}}\wedge\theta_{\mathbf{a}}\right)  \right]  .
\label{7.b12bis}%
\end{align}
Then, since
\begin{align}%
\mbox{\boldmath{$\delta$}}%
\left[  d\theta^{\mathbf{e}}\wedge\theta_{\mathbf{e}}\wedge\star\left(
d\theta^{\mathbf{a}}\wedge\theta_{\mathbf{a}}\right)  \right]   &  =%
\mbox{\boldmath{$\delta$}}%
d\theta^{\mathbf{e}}\wedge\theta_{\mathbf{e}}\wedge\star\left(  d\theta
^{\mathbf{a}}\wedge\theta_{\mathbf{a}}\right) \nonumber\\
&  +%
\mbox{\boldmath{$\delta$}}%
\theta^{\mathbf{e}}\wedge d\theta_{\mathbf{e}}\wedge\star\left(
d\theta^{\mathbf{a}}\wedge\theta_{\mathbf{a}}\right) \nonumber\\
&  +d\theta^{\mathbf{e}}\wedge\theta_{\mathbf{e}}\wedge%
\mbox{\boldmath{$\delta$}}%
\star\left(  d\theta^{\mathbf{a}}\wedge\theta_{\mathbf{a}}\right)  ,
\label{7.12biss}%
\end{align}
it follows that%
\begin{align}%
\mbox{\boldmath{$\delta$}}%
\mathcal{L}^{(4)}  &  =\frac{1}{2}%
\mbox{\boldmath{$\delta$}}%
d\theta^{\mathbf{d}}\wedge\theta_{\mathbf{d}}\wedge\star\left(  d\theta
^{\mathbf{a}}\wedge\theta_{\mathbf{a}}\right) \nonumber\\
&  +\frac{1}{2}%
\mbox{\boldmath{$\delta$}}%
\theta^{\mathbf{d}}\wedge d\theta_{\mathbf{d}}\wedge\star\left(
d\theta^{\mathbf{a}}\wedge\theta_{\mathbf{a}}\right) \nonumber\\
&  -\frac{1}{4}%
\mbox{\boldmath{$\delta$}}%
\theta^{\mathbf{d}}\wedge d\theta^{\mathbf{a}}\wedge\theta_{\mathbf{a}}%
\wedge\left[  \theta_{\mathbf{d}}\lrcorner\star\left(  d\theta^{\mathbf{e}%
}\wedge\theta_{\mathbf{e}}\right)  \right] \nonumber\\
&  -\frac{1}{4}%
\mbox{\boldmath{$\delta$}}%
\theta^{\mathbf{d}}\wedge\left[  \theta_{\mathbf{d}}\lrcorner\left(
d\theta^{\mathbf{e}}\wedge\theta_{\mathbf{e}}\right)  \right]  \wedge
\star\left(  d\theta^{\mathbf{a}}\wedge\theta_{\mathbf{a}}\right)  .
\label{7.b13}%
\end{align}

Then,
\begin{align}
\frac{\partial\mathcal{L}^{(4)}}{\partial\theta^{\mathbf{d}}}  &  =\frac{1}%
{2}d\theta_{\mathbf{d}}\wedge\star\left(  d\theta^{\mathbf{a}}\wedge
\theta_{\mathbf{a}}\right)  -\frac{1}{4}d\theta^{\mathbf{a}}\wedge
\theta_{\mathbf{a}}\wedge\left[  \theta_{\mathbf{d}}\lrcorner\star\left(
d\theta^{\mathbf{e}}\wedge\theta_{\mathbf{e}}\right)  \right] \nonumber\\
&  -\frac{1}{4}\left[  \theta_{\mathbf{d}}\lrcorner\left(  d\theta
^{\mathbf{e}}\wedge\theta_{\mathbf{e}}\right)  \right]  \wedge\star\left(
d\theta^{\mathbf{a}}\wedge\theta_{\mathbf{a}}\right)  ,\\
\frac{\partial\mathcal{L}^{(4)}}{\partial d\theta^{\mathbf{d}}}  &  =\frac
{1}{2}\theta_{\mathbf{d}}\wedge\star\left(  d\theta^{\mathbf{a}}\wedge
\theta_{\mathbf{a}}\right)  . \label{7.b14}%
\end{align}

Finally, disregarding the contribution of the exact differential and
collecting all terms in Eqs.(\ref{7.b5}), (\ref{7.b10}) and (\ref{7.b14}) we
get:%
\begin{align}
\frac{\partial\mathcal{L}}{\partial\theta^{\mathbf{d}}}  &  =\frac{1}%
{2}[(\theta_{\mathbf{d}}\lrcorner d\theta^{\mathbf{a}})\wedge\star
d\theta_{\mathbf{a}}-d\theta^{\mathbf{a}}\wedge(\theta_{\mathbf{d}}%
\lrcorner\star d\theta_{\mathbf{a}})]\nonumber\\
&  +\frac{1}{2}d\left(  \theta_{\mathbf{d}}\lrcorner\star\theta^{\mathbf{a}%
}\right)  \wedge\star d\star\theta_{\mathbf{a}}+\frac{1}{2}\left(
\theta_{\mathbf{d}}\lrcorner d\star\theta^{\mathbf{a}}\right)  \wedge\star
d\star\theta_{\mathbf{a}}\nonumber\\
&  +\frac{1}{2}d\theta_{\mathbf{d}}\wedge\star\left(  d\theta^{\mathbf{a}%
}\wedge\theta_{\mathbf{a}}\right)  -\frac{1}{4}d\theta^{\mathbf{a}}%
\wedge\theta_{\mathbf{a}}\wedge\left[  \theta_{\mathbf{d}}\lrcorner
\star\left(  d\theta^{\mathbf{e}}\wedge\theta_{\mathbf{e}}\right)  \right]
\nonumber\\
&  -\frac{1}{4}\left[  \theta_{\mathbf{d}}\lrcorner\left(  d\theta
^{\mathbf{e}}\wedge\theta_{\mathbf{e}}\right)  \right]  \wedge\star\left(
d\theta^{\mathbf{a}}\wedge\theta_{\mathbf{a}}\right)  , \label{7.b15}%
\end{align}
and%
\begin{equation}
\frac{\partial\mathcal{L}}{\partial d\theta^{\mathbf{d}}}=-\star
d\theta_{\mathbf{d}}-\left(  \theta_{\mathbf{d}}\lrcorner\star\theta
^{\mathbf{a}}\right)  \wedge\star d\star\theta_{\mathbf{a}}+\frac{1}{2}%
\theta_{\mathbf{d}}\wedge\star\left(  d\theta^{\mathbf{a}}\wedge
\theta_{\mathbf{a}}\right)  . \label{7.b16}%
\end{equation}

Collecting all these terms we arrive at the Euler Lagrange equation,%
\begin{equation}
\frac{\partial\mathcal{L}}{\partial\theta^{\mathbf{a}}}+d\left(
\frac{\partial\mathcal{L}}{\partial d\theta^{\mathbf{a}}}\right)  =\star
t^{\mathbf{a}}+d\star\mathcal{S}^{\mathbf{a}}=-(\star\mathcal{R}^{\mathbf{a}%
}-\frac{1}{2}R\star\theta^{\mathbf{a}}) \label{7.b17}%
\end{equation}
where $\mathcal{S}^{\mathbf{a}}$ are the superpotentials and $\star
t^{\mathbf{a}}$ the (pseudo) energy-momentum $1$-forms of the gravitational
field (see, e.g.,\cite{14,thirring})
\begin{align}
\star\mathcal{S}^{\mathbf{c}}  &  =\frac{1}{2}%
\mbox{\boldmath{$\omega$}}%
_{\mathbf{ab}}\wedge\star(\theta^{\mathbf{a}}\wedge\theta^{\mathbf{b}}%
\wedge\theta^{\mathbf{c}})\in\sec\bigwedge\nolimits^{2}T^{\ast}%
M\hookrightarrow\mathcal{C\ell}\left(  T^{\ast}M\right)  ,\nonumber\\
\star t_{\mathbf{\ }}^{\mathbf{c}}  &  =-\frac{1}{2}%
\mbox{\boldmath{$\omega$}}%
_{\mathbf{ab}}\wedge\lbrack%
\mbox{\boldmath{$\omega$}}%
_{\mathbf{d}}^{\mathbf{c}}\star(\theta^{\mathbf{a}}\wedge\theta^{\mathbf{b}%
}\wedge\theta^{\mathbf{d}})+%
\mbox{\boldmath{$\omega$}}%
_{\mathbf{d}}^{\mathbf{b}}\star(\theta^{\mathbf{a}}\wedge\theta^{\mathbf{d}%
}\wedge\theta^{\mathbf{c}})]\label{7.10.17}\\
&  \in\sec\bigwedge\nolimits^{3}T^{\ast}M\hookrightarrow\mathcal{C\ell}\left(
T^{\ast}M\right)  .\nonumber
\end{align}

\section{A Note on a Reply to a Previous Version of this Paper}

After taking notice of a preliminary version of our paper posted at the arXiv
[\texttt{math-ph/0411085}] the author of \cite{0} posted a paper entitled:
\textit{Refutation of Rodrigues}: \textit{The Correctness of Differential
Geometry}\footnote{Posted at
{\small http://www.aias.us/Comments/comments01062005b.html}} (called simply
\textit{reply}, in what follows).\ As anticipated he said that he interpreted
correctly the tetrad postulate, since for him a tetrad is a vector valued
1-form, although he did \textit{not} explain if by this statement he means the
pullback $\mathbf{\theta}$ of the soldering form $\overset{\blacktriangle
}{\mathbf{\theta}}$ (see Eq.(\ref{q7})), or the $(1-1)$ tensor \textbf{Q }(see
Eq.(\ref{Q})). He then claims that his conclusions concerning the proof of his
\ `Evans lemma' are correct. However a look to his \textit{reply }reveals,
that he did not really grasp what is going on. Indeed, the way he introduces
into the game the symbols $q_{\nu}^{\mathbf{a}}$ in equation (6) of the reply,
is as a \ `matrix' connecting the components of a vector field $V$ \ from the
coordinate basis $\{{\mbox{\boldmath$\partial$}}_{\mu}\}$ to the orthonormal
basis. He explicitly wrote: $V^{\mathbf{a}}=q_{\mu}^{\mathbf{a}}V^{\mu}$. This
immediately requires that for each $\nu$, the $q_{\nu}^{\mathbf{a}}$ are the
components of the coordinate vector field ${\mbox{\boldmath$\partial$}}_{\nu}$
in the orthonormal basis $\{\mathbf{e}_{\mathbf{a}}\}$. To obtain the
covariant derivative of ${\mbox{\boldmath$\partial$}}_{\nu}$ in the direction
of the vector field ${\mbox{\boldmath$\partial$}}_{\mu}$ \ we need (as
discussed above) use the covariant derivative $\mathbf{\nabla}%
_{{\mbox{\boldmath$\partial$}}_{\mu}}^{+}$ which acts on the sections of $TM$.
Once we do this, as showed in detail in Section 4, we arrive at the
calculation of \ \textbf{$\nabla$}$_{\mu}^{+}q_{\nu}^{\mathbf{a}}$. And, in
general \textbf{$\nabla$}$_{\mu}^{+}q_{\nu}^{\mathbf{a}}\neq0$. The only licit
way of obtaining \ the \ `tetrad postulate' ( and in this case it is a
proposition, as we showed in the main text) is by calculation of the covariant
derivative of the tensor field $\mathbf{Q}$ in the direction of the vector
field ${\mbox{\boldmath$\partial$}}_{\mu}$, i.e., $\mathbf{\nabla
}_{{\mbox{\boldmath$\partial$}}_{\mu}}^{+}\mathbf{Q}$. This has not be done in
\cite{0} nor has it been done in the \textit{reply}. Thus, we state here: the
would be proof of the
\'{}%
tetrad postulate' offered in the \textit{reply} is unfortunately one more
example of wishful thinking.

Besides that, the \textit{reply}, the author of \cite{0}, did not
address\ himself to the other strong criticisms we done to his work (and which
already appeared in the preliminary version of this paper) as, e.g.,

(i) our statement and demonstration that his proof of the \ `Evans lemma' is a
nonsequitur, that his Eq.(41) is completely meaningless,

(ii) our statement that the tetrad differential equations of his paper are wrong,

(ii) our statement that the sequence of calculations done by him in paper
written sometime ago with collaborator (at that time Ph.D. student at Oxford)
and that we reproduced in the conclusions our paper, shows that
(unfortunately) he effectively does not know what a Clifford algebra is and
worse, does not know how to multiply $2\times2$ matrices.

These statements are \textit{sad} \textit{facts} that cannot be hidden
anymore, and \ so cannot be considered as \textit{ad Hominen} attack, contrary
to many of the arguments that author of \cite{0} used in his reply against one
of us.

Also, it must be registered here that instead of directing himself to the
mathematical questions, the author of \cite{0} preferred to suggest to his
readers that we must succumb under the weight of authorities. Indeed, he said
that we are contradicting authors like, e.g.,\ Carroll, Greene, Wheeler and
Witten. What the author of \cite{0} forgot is that a\ name does not mean
authority in science. In the formal sciences a valid argument must fulfil the
rules of logic. What we did was simply to find serious ambiguities in a
statement that some authors called
\'{}%
tetrad postulate', and the bad use made of that statement in some papers.

So, whereas it is true that we criticize some writings of the above authors
(and some others, quoted in the references), we express here our admiration
and respect for all of them, and also to any honest researcher that has at
least enough humility to recognize errors. We are sure that our comments have
been fair, educated and constructive. Besides that we think that our
clarification of the necessity to explicitly distinguish the different
covariant derivatives acting on different associate vector associate to the
principal bundles $F(M)$ (and $P_{\mathrm{SO}_{1,3}^{e}}(M)$) will be welcome.

And to end, we must say that we agree with at least one statement of the
\textit{reply}, namely: that \textit{differential geometry is correct}.
However, the use that author of \cite{0} made of this notable theory in his
many papers is \textit{not} correct. Certainly, the reader that knows enough
Mathematics and had enough patience to arrive here already knows that.

\end{document}